\documentclass[twocolumn,twocolappendix]{aastex63}

\usepackage{comment}
\usepackage{cases}

\accepted{\today}
\submitjournal{ApJS}

\shorttitle{ACA Band8 U/LIRGs Survey}
\shortauthors{Michiyama et al.}
\graphicspath{{./}{figures/}}
\begin{document}
\title{An ACA Survey of [\ion{C}{1}]~$^3P_1$--$^3P_0$,  CO~$J$=4--3, and Dust Continuum in Nearby U/LIRGs}

\correspondingauthor{Tomonari Michiyama}
\email{t.michiyama.astr@gmail.com}
\author[0000-0003-2475-7983]{Tomonari Michiyama}
\affiliation{Kavli Institute for Astronomy and Astrophysics, Peking University, 5 Yiheyuan Road, Haidian District, Beijing 100871, P.R.China}
\affiliation{Department of Earth and Space Science, Osaka University, 1-1 Machikaneyama, Toyonaka, Osaka 560-0043, Japan}
\affiliation{National Astronomical Observatory of Japan, National Institutes of Natural Sciences, 2-21-1 Osawa, Mitaka, Tokyo, 181-8588}

\author{Toshiki Saito}
\affiliation{Max-Planck-Institut f\"ur Astronomie, K\"onigstuhl 17, 69117 Heidelberg, Germany}
\affiliation{Department of Physics, General Studies, College of Engineering, Nihon University, 1 Nakagawara, Tokusada, Tamuramachi, Koriyama, Fukushima, 963-8642, Japan}
\affiliation{National Astronomical Observatory of Japan, National Institutes of Natural Sciences, 2-21-1 Osawa, Mitaka, Tokyo, 181-8588}

\author{Ken-ichi Tadaki}
\affiliation{National Astronomical Observatory of Japan, National Institutes of Natural Sciences, 2-21-1 Osawa, Mitaka, Tokyo, 181-8588}

\author{Junko Ueda}
\affiliation{National Astronomical Observatory of Japan, National Institutes of Natural Sciences, 2-21-1 Osawa, Mitaka, Tokyo, 181-8588}

\author{Ming-Yang Zhuang}
\affiliation{Kavli Institute for Astronomy and Astrophysics, Peking University, 5 Yiheyuan Road, Haidian District, Beijing 100871, P.R.China}
\affiliation{Department of Astronomy, School of Physics, Peking University, Beijing 100871, China}

\author{Juan Molina}
\affiliation{Kavli Institute for Astronomy and Astrophysics, Peking University, 5 Yiheyuan Road, Haidian District, Beijing 100871, P.R.China}

\author{Bumhyun Lee}
\affiliation{Kavli Institute for Astronomy and Astrophysics, Peking University, 5 Yiheyuan Road, Haidian District, Beijing 100871, P.R.China}

\author{Ran Wang}
\affiliation{Kavli Institute for Astronomy and Astrophysics, Peking University, 5 Yiheyuan Road, Haidian District, Beijing 100871, P.R.China}
\affiliation{Department of Astronomy, School of Physics, Peking University, Beijing 100871, China}

\author{Alberto Bolatto}
\affiliation{Department of Astronomy and Laboratory for Millimeter-Wave Astronomy, University of Maryland, College Park, MD 20742, USA}

\author{Daisuke Iono}
\affiliation{National Astronomical Observatory of Japan, National Institutes of Natural Sciences, 2-21-1 Osawa, Mitaka, Tokyo, 181-8588}
\affiliation{Department of Astronomical Science, The Graduate University for Advanced Studies, SOKENDAI, 2-21-1 Osawa, Mitaka, Tokyo 181-8588}

\author{Kouichiro Nakanishi}
\affiliation{National Astronomical Observatory of Japan, National Institutes of Natural Sciences, 2-21-1 Osawa, Mitaka, Tokyo, 181-8588}
\affiliation{Department of Astronomical Science, The Graduate University for Advanced Studies, SOKENDAI, 2-21-1 Osawa, Mitaka, Tokyo 181-8588}

\author{Takuma Izumi}
\affiliation{National Astronomical Observatory of Japan, National Institutes of Natural Sciences, 2-21-1 Osawa, Mitaka, Tokyo, 181-8588}
\affiliation{Department of Astronomical Science, The Graduate University for Advanced Studies, SOKENDAI, 2-21-1 Osawa, Mitaka, Tokyo 181-8588}

\author{Takuji Yamashita}
\affiliation{National Astronomical Observatory of Japan, National Institutes of Natural Sciences, 2-21-1 Osawa, Mitaka, Tokyo, 181-8588}

\author{Luis C. Ho}
\affiliation{Kavli Institute for Astronomy and Astrophysics, Peking University, 5 Yiheyuan Road, Haidian District, Beijing 100871, P.R.China}
\affiliation{Department of Astronomy, School of Physics, Peking University, Beijing 100871, China}



\begin{abstract}
We present the results of surveying [\ion{C}{1}]~$^3P_1$--$^3P_0$,  $^{12}$CO~$J$=4--3, and 630~$\mu$m dust continuum emission for 36 nearby ultra/luminous infrared galaxies (U/LIRGs) using the Band 8 receiver mounted on the Atacama Compact Array (ACA) of the Atacama Large Millimeter/submillimeter Array. 
We describe the survey, observations, data reduction, and results; the main results are as follows.
(i) We confirmed that [\ion{C}{1}]~$^3P_1$--$^3P_0$ has a linear relationship with both the $^{12}$CO~$J$=4--3 and 630~$\mu$m continuum. 
(ii) In NGC~6052 and NGC~7679, $^{12}$CO~$J$=4--3 was detected but [\ion{C}{1}]~$^3P_1$--$^3P_0$ was not detected with a [\ion{C}{1}]~$^3P_1$--$^3P_0$/ $^{12}$CO~$J$=4--3 ratio of $\lesssim0.08$. Two possible scenarios of weak [\ion{C}{1}]~$^3P_1$--$^3P_0$ emission are C$^0$-poor/CO-rich environments or an environment with an extremely large [\ion{C}{1}]~$^3P_1$--$^3P_0$ missing flux.
(iii) There is no clear evidence showing that galaxy mergers, AGNs, and dust temperatures control the ratios of [\ion{C}{1}]~$^3P_1$--$^3P_0$/ $^{12}$CO~$J$=4--3 and  $L'_{\rm [CI](1-0)}/L_{\rm 630\mu m}$.
(iv) We compare our nearby U/LIRGs with high-z galaxies, such as galaxies on the star formation main sequence (MS) at z$\sim1$ and submillimeter galaxies (SMGs) at $z=2-4$.
We found that the mean value for the [\ion{C}{1}]~$^3P_1$--$^3P_0$/ $^{12}$CO~$J$=4--3 ratio of U/LIRGs is similar to that of SMGs but smaller than that of galaxies on the MS. 
\end{abstract}



\section{Introduction} \label{sec:intro}
The  molecular hydrogen gas mass ($M_{\rm H_2}$) of galaxies is a fundamental observable quantity that is directly linked to galaxy evolution.
As cold H$_2$  does not radiate strong emission, astronomers use carbon monoxide (CO\footnote{We use the term ``CO" to infer $^{12}$C$^{16}$O throughout this paper.}) as a tracer of the extragalactic cold molecular gas mass \citep[e.g.,][]{Bolatto_2013}. 
The $J$=1--0 ground rotational transition of CO  (hereafter CO~(1--0)) has low excitation energy and the emission falls in the transparent atmospheric window (at a frequency of 115~GHz and a wavelength of 2.6~mm).
For high-z galaxies, the observed frequency of low-$J$ CO lines is shifted out of the main high-transparency atmospheric window (i.e., $<80$~GHz) and we cannot use powerful telescopes -- such as  the Atacama Large Millimeter/submillimeter Array (ALMA) -- to observe low-$J$ CO lines until the lowest frequency band of ALMA (band~1) becomes available in the near future. 
Therefore, detecting low-$J$ CO lines requires long integration radio observation, even for bright objects \citep[e.g.,][]{Carilli_2013}. 
We thus need to develop alternative methods to measure $M_{\rm H_2}$, which can be observed at a higher frequency (shorter wavelength) than CO~(1--0) emission.

One possible method is to observe the mid- and high-$J$ CO lines ($J_{\rm upp}\geqq$4) \citep[e.g.,][]{Solomon_2005, Carilli_2013}.
For example, CO~(4--3) line emits at a frequency of 460~GHz (corresponding to a  wavelength of 650~$\mu$m). 
However, higher-$J$ CO lines are sensitive to highly excited gases (i.e., warm and dense) that would occupy a minor fraction of the total molecular mass.
The assumption of the line ratios, that is, CO~(4--3)/CO~(1--0), makes non-negligible systematic errors to measure the bulk of the molecular gas using higher-$J$ CO lines \citep[e.g.,][]{Carilli_2013}.
Therefore, astronomers have developed various alternative methods.

One proposed method is to use the lower forbidden $^3P$ fine structure line of atomic carbon -- i.e., [\ion{C}{1}]~$^3P_1$--$^3P_0$, hereafter referred to as [\ion{C}{1}] ~(1--0) -- which  emits at a frequency of 492~GHz (corresponding to a wavelength of 609~$\mu$m). 
For example, [\ion{C}{1}] observations in Galactic clouds have shown that [\ion{C}{1}] distributions coincide with those of CO \citep{Ojha_2001, Oka_2001, Ikeda_2002, Kramer_2008, Burton_2015, Izumi_2021}, suggesting that [\ion{C}{1}] ~(1--0) luminosity can trace the bulk of molecular gas.
Besides, \citet{Israel_2001, Israel_2002, Israel_2003} reported bright [\ion{C}{1}] emission at the center of nearby galaxies.
\citet{Papa_2004_theory} proposed that [\ion{C}{1}] is superior to CO for measuring $M_{\rm H_2}$, especially in the UV intense diffuse region ($\sim10^{2}-10^{3}$~cm$^{-3}$)  and/or metal-poor environments due to the dissociation of CO. Subsequently, \citet{Papa_2004_ULIRG} applied the [\ion{C}{1}] method for two Ultra/ infrared luminous galaxies: U/LIRGs -- Arp~220 and NGC~1614 -- and demonstrated that the molecular gas mass inferred from [\ion{C}{1}]~(1--0) luminosity is consistent with that from CO observations. 
Many studies have investigated the [\ion{C}{1}] properties of galaxies using the data obtained by the Herschel Space Observatory \citep[e.g.,][]{Israel_2015, Kame_2016, Jiao_2017, Jiao_2019, Crocker_2019}. Theoretical modeling suggests that [\ion{C}{1}] is a better tracer of $M_{\rm H_2}$ than CO~(1--0) in several environments, such as metal-poor environments or those with a high-cosmic rays, and active galactic nuclei (AGNs) \citep[e.g.,][]{Bisbas_2015, Papadopoulos_2018, Clark_2019}.
More recently, high-frequency receivers (e.g., Band 8 and 10 receivers) 
on ALMA  have also enabled investigation of the spatial distribution of atomic carbon in nearby galaxies 
\citep[e.g.,][]{Krips_2016, Cicone_2018, Izumi_2018, Miyamoto_2018, Salak_2019, Izumi_2020, Saito_2020}.
In addition,
[\ion{C}{1}] measurements of bright high-z galaxies, 
such as gravitational lensing galaxies and submillimeter galaxies (SMGs)
\citep[e.g.,][]{Walter_2011, Alaghband-Zadeh_2013, Bothwell_2017, Yang_2017, Andreani_2018, Caameras_2018, Harrington_2018, Tadaki_2018, Dannerbauer_2019, Jin_2019, Cortzen_2020},
and galaxies on main sequence (MS; e.g., \citealt[][]{Popping_2017, Talia_2018, Valentino_2018, Bourne_2019,Valentino_2020, Lee_2021}) are increasingly available.

The other proposed method uses the submillimeter dust continuum emission \citep{Eales_2012, Magdis_2012, Scoville_2014}.
\citet{Scoville_2016} derived the same empirical calibration constant from 350~GHz ($\lambda=850~\mu$m) continuum flux to the interstellar
medium (ISM) mass among local star-forming galaxies, molecular clouds of the Milky Way, and high-redshift SMGs; this suggests that the Rayleigh-Jeans tail of the dust emission can be used as an accurate and very efficient probe of the ISM in galaxies.

Before applying these new methods to measure $M_{\rm H_2}$ in high-z galaxies, it is necessary to investigate the experimental relationship between CO,   [\ion{C}{1}], and the dust continuum in well-studied nearby galaxies.
For example, \citet{Michiyama_2020} discovered a  [\ion{C}{1}]-faint CO-bright gas-rich galaxy (NGC~6052), implying a large galaxy-to-galaxy variation with respect to the [\ion{C}{1}]/ CO ratio and the importance of a systematic comparison between CO and [\ion{C}{1}].
Here, we expand the analysis of \citet{Michiyama_2020} by a survey of [\ion{C}{1}]~(1--0),  CO~(4--3), and dust continuum in U/LIRGs, using the Band 8 receiver mounted on the Atacama Compact (Morita) Array (ACA).
This project can improve the sensitivity and resolution of legacy Herschel results in terms of the higher sensitivity and the higher resolution (i.e., Hershel’s beam is $\sim45\arcsec$ and the sensitivity is $\sim0.5$~mJy whereas ACA’s beam is $\sim3\arcsec$ and the sensitivity is $\sim0.05$ ~mJy). 

In Section~\ref{sec:Survey}, we describe the survey design, supplemental data, observation, data reduction and data analysis.
In Section~\ref{sec:result}, we summarize the results of the survey.
In Section~\ref{Sec:Discussion}, we investigate the correlation between [\ion{C}{1}]~(1--0), CO~(4--3), and the dust continuum. In addition, the details of [\ion{C}{1}]-faint CO-bright gas-rich galaxy, NGC~7679, are explained.
We investigated whether AGNs, mergers, and dust temperatures control the ratio of [\ion{C}{1}]/CO. Finally, we compare our nearby U/LIRGs and high-z galaxies ($z\sim1$ galaxies on MS and $z=2-4$ SMGs).
Section~\ref{sec:summary} provides a summary of the study.
We use the cosmological parameters for a $\Lambda$CDM, $H_0=67.8$, $\Omega_m=0.308$, and $\Omega_\Lambda=0.692$ \citep{Planck_2016}.

\section{Survey} \label{sec:Survey}
\subsection{Sample Selection}
The ACA Band~8 U/LIRG survey was designed to observe [\ion{C}{1}]~(1--0), CO~(4--3), and dust continuum emission for nearby ultraluminous infrared galaxies 
(ULIRGs: $10^{12}~L_{\odot}<L_{\rm TIR}<10^{13}~L_{\odot}$)
and luminous infrared galaxies
(LIRGs: $10^{11}~L_{\odot}<L_{\rm TIR}<10^{12}~L_{\odot}$),
where the $L_{\rm TIR}$ is the total infrared luminosity integrated from 8-1000~$\mu$m.
We observed 36\footnote{We observed 39 U/LIRGs, however, we do not investigate UGC~02338, IRAS~05442+1732, and NGC~7771. In the case of UGC~02338 and IRAS~05442+1732, the target information were wrong due to artificial mistake. For NGC~7771, the final images were uncertain due to technical issues and it was considered as QA2 fail from observatory.} U/LIRGs.
We selected 12ULIRGs\footnote{NGC~6240 satisfies our criteria but was not observed due to duplication issue.} from the IRAS Revised Bright Galaxy Sample \citep{Sanders_2003}, whose declinations were below 20 degrees and exhibited no strong atmospheric absorption around [\ion{C}{1}]~(1--0) or CO~(4--3).
We selected  24LIRGs, from which CO~(1--0) was detected in literature by cross-matching the CO~(1--0) samples compiled by \citet{Kame_2016} and the references therein.
The list of targets is listed in Table~\ref{tab:target}. 

\subsection{Supplemental data}
\begin{figure}[!htbp]
\begin{center}
\includegraphics[angle=0,scale=0.25]{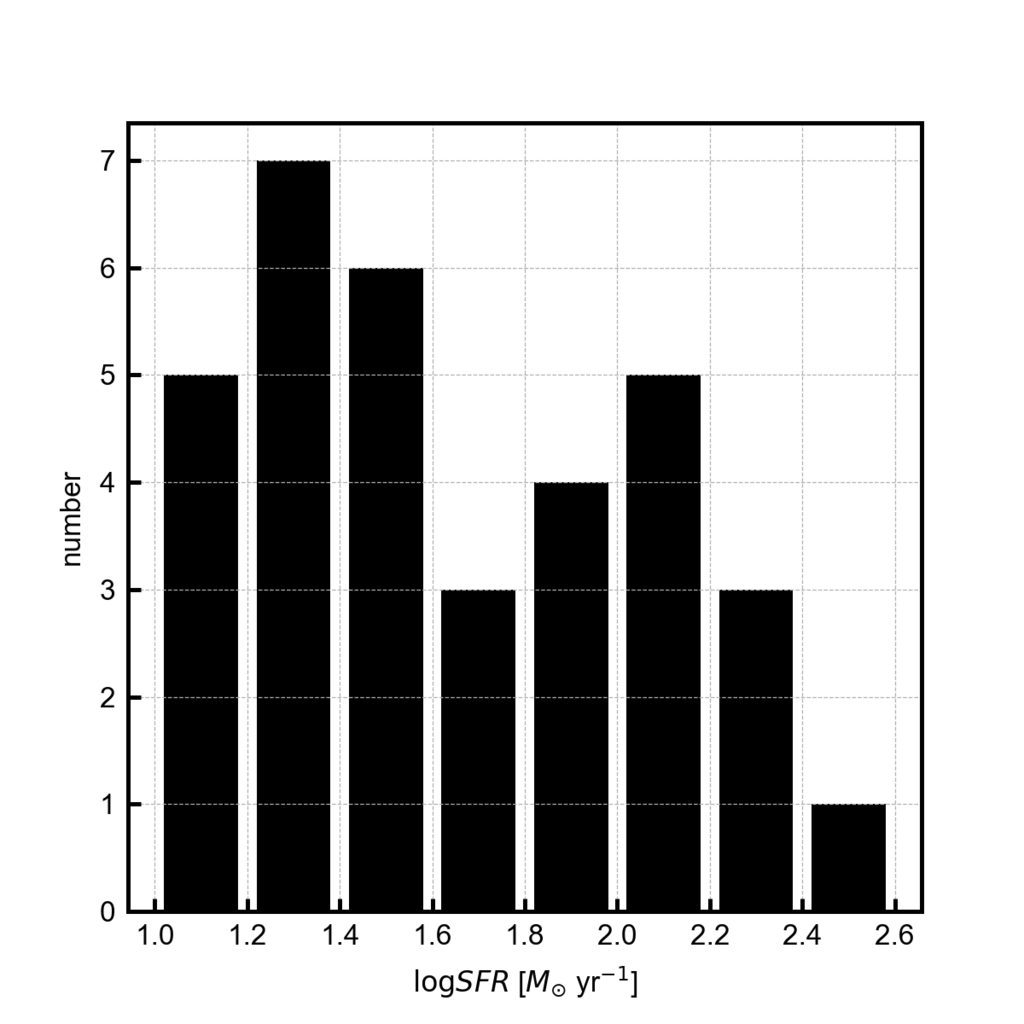}
\caption{A histogram of SFR for our ACA Band~8 U/LIRGs survey. \label{fig:hist}}
\end{center}
\end{figure}

Several galaxies were observed using the Spectral and
Photometric Imaging Receiver Fourier Transform Spectrometer
(SPIRE/FTS) mounted on the Herschel Space Observatory \citep{Kame_2016, Lu_2017}, which provides the total [\ion{C}{1}]~(1--0) and CO~(4--3) flux.
The galaxy-integrated global properties such as SFR, dust mass ($M_{\rm dust}$), dust temperature ($T_{\rm dust}$), and $L_{\rm TIR}$ were estimated based on continuum spectrum fitting (SED fitting) by \citet{Shangguan_2019}\footnote{$T_{\rm dust}$ and  $L_{\rm TIR}$ is estimated with the same method explained in \citet{Shangguan_2019} in private communication.}. 
The CO~(1--0) velocity integrated flux obtained by single dish telescopes was compiled by \citet{Kame_2016}, which corresponds to a 45\arcsec aperture after aperture correction.
The supplemental information is summarized in Table~\ref{tab:SED}\\
A histogram of the SFRs of the sample sources is presented in Figure~\ref{fig:hist}. \\

\subsection{Observation}\label{sec:Observation}
Single pointing mode ACA observations were performed during the cycle~6 season (ALMA project ID: 2018.1.00994.S, PI: T.Michiyama).
For NGC~232, IRAS F10565+2448, and ESO 069-IG006, the fainter pairs of the interacting system were also observed.
Among the 36 U/LIRGs, only CO~(4--3) observations were completed in NGC 5990 and IRAS~19542+1110 and only [\ion{C}{1}]~(1--0) observations were completed in UGC~02982.
The details -- e.g., date, typical system temperature, and the number of antennas -- are summarized in Table~\ref{tab:obs}, which are based on QA0 reports provided by the ALMA observatory. 
The detailed information in each observation run is shown in the weblog provided by the ALMA observatory.

Most important are the small field of view (FoV) and small maximum recovery scale (MRS) because the observation was a single-point snapshot ($\sim$5 min integration) without a total power array. 
The FoV is defined as:
\begin{equation}\label{eq:FoV}
{\rm FoV} = 1.13\frac{\lambda_{\rm sky}}{D_{\rm ant}},
\end{equation}
where $\lambda_{\rm sky}$ is the observed wavelength and $D_{\rm ant}=7$~m is the antenna aperture. 
For example, the FoV are $\sim$ 20\farcs1 at 480~GHz.
For the nearest galaxy NGC~7552 and the most distant galaxy IRAS F12112+0305, 20\farcs1 corresponds to 2.2~kpc and 29~kpc, respectively.
The FoVs are shown in Figure~\ref{fig:SED}.
The typical baseline coverage is 8--47~m, which corresponds to an MRS of $10\arcsec$--$14\arcsec$, Table~\ref{tab:obs}).
To estimate the possible recovered flux (RF) as a function of the source size, we used the $\tt{simobs}$ task in CASA.  The RF was 50~\% when we assumed a 2D gaussian distribution with the FWHM=7\arcsec. If the FWHM is $>12\arcsec$, the recovered flux is $<10\%$; this means that our observation may not recover the total flux in the case of an extended gas and dust structure. 

\subsection{Data Reduction and Imaging}
The calibrated visibility data were obtained 
by running the standard pipeline calibration scripts 
provided by ALMA using $\tt{CASA}$ \citep{McMullin_2007}. 
We used the various $\tt{CASA}$ versions which are specified in each observation.
The imaging processes were conducted by $\tt{tclean}$ in $\tt{CASA}$ (version 5.4.0-70)
with a velocity resolution of 30~km\,s$^{-1}$ for line emission, a cell size of 0$\farcs$5, an image size of 80 pixels, and Briggs weighting with a robust parameter of 0.5. 
Clean masks for each channel were manually selected.
The continuum was subtracted in the $u–v$ domain using $\tt{uvcontsub}$ in the $\tt{CASA}$ task after manually inspecting the emission-free frequencies.
Table~\ref{tab:obs_res} summarizes the parameters of  the reduced data (i.e., the synthesized beam, sensitivity, and the integrated velocity ranges).
The sensitivity was calculated as the rms value of the residual image after the $\tt{tclean}$ task. The velocity range for the moment maps was manually selected by checking the datacube. 
The velocity integrated intensity map (mom0), velocity field map (mom1),  velocity dispersion map (mom2), and continuum maps, and the spatially integrated spectrum are shown in Figure~\ref{fig:mom}.

\section{Results}\label{sec:result}
We detected 
CO~(4--3) emission from 36 out of 39 observed galaxies,  [\ion{C}{1}]~(1--0) from 32 out of 38 galaxies, and continuum emission from 35 out of 41 galaxies.
CO~(4--3) emission was not detected in NGC~232-3, IRAS~F10565+2448-2, and ESO~069-IG006-2, which are fainter galaxy pairs. 
In the case of NGC~6052, NGC~7679, and ESO~467-G027, [\ion{C}{1}]~(1--0) and continuum emission were not detected while the CO~(4--3) emission line was detected. 
The details are mentioned in \citet{Michiyama_2020} for NGC~6052 and Section~\ref{sec:ND} for NGC~7679.
ESO~467-G027 also shows weak [\ion{C}{1}]~(1--0) but the spatial and velocity distribution of the CO~(4--3) line is uncertain (Figure~\ref{fig:mom}) owing to the observation condition. Therefore, we used it just as a candidate for [\ion{C}{1}]-faint galaxies. More sensitive observations are needed to investigate ESO 467-G027 in detail.
In the case of IC~4280, both CO~(4--3) and [\ion{C}{1}]~(1--0) were detected but continuum emission were not.

\subsection{Flux}\label{sec:Flux}
The total velocity integrated line flux ($S_{\rm line}{\Delta v}$ in Jy~km\,s$^{-1}$), based on [\ion{C}{1}]~(1--0) and CO~(4--3) moment~0  maps, is derived by using the``${\tt imfit}$" task in ${\tt CASA}$, which identifies the source automatically, conducts 2D Gaussian fitting, and calculates the total flux and the statistical errors ($\sigma_{\rm imfit}$). 
The ${\tt imfit}$ task was used on the moment~0 map before primary beam correction. To estimate the correction factor due to the primary beam response, we also measured the flux  (by the ${\tt imstat}$ task in ${\tt CASA}$) using the map before/after primary beam correction with the same aperture (1.5 $\times$ larger than the FWHM size derived from the ${\tt imfit}$ task). The ratio between the two flux is used as a primary beam correction factor of total flux.
The continuum flux density is also derived from the ``${\tt imfit}$" task in the same manner.
A summary of the line flux and continuum flux density is presented in Table~\ref{tab:flux_line} and \ref{tab:flux_cont}. 
The full width at half maximum (FWHM) was derived from a single Gaussian fitting of the spectrum shown in Figure~\ref{fig:mom}.
The errors in Table~\ref{tab:flux_line} are calculated by:
\begin{eqnarray} \label{eq:err}
 \sigma(S_{\rm line}{\Delta v)}&=&\sqrt{\sigma_{\rm imfit}^2 + (0.3S_{\rm line}{\Delta v})^2}f
\end{eqnarray}
for lines and 
\begin{eqnarray} \label{eq:err_cont}
 \sigma(S_{\rm cont})&=\sqrt{\sigma_{\rm imfit}^2 + (0.3S_{\rm cont})^2}
\end{eqnarray}
for continuum emission.
A factor of 0.3 indicates a systematic error for absolute flux calibrations. 
The absolute flux accuracy is expected to be 10~\%  for Band~8 by ALMA main array observations\footnote{\url{https://almascience.nao.ac.jp/documents-and-tools/cycle6/alma-proposers-guide}}. However, the actual performance of the flux calibration has not been explored for the ACA standalone mode in Band~8 by the ALMA observatory.
We adopt a conservative systematic error value of 30~\%.
When the ${\tt imfit}$ does not work correctly due to side lobes, we manually selected the aperture (e.g., NGC~5990, ESO~148-IG002, IRAS~09022-3615, IRAS~F14378-3651, ESO~148-IG002); the details are explained in Section~\ref{sec:individual}. 

In the case of non-detection, the $3\sigma$ upper limit of the integrated line flux can be derived from
\begin{eqnarray} \label{eq:err}
S_{\rm line}{\Delta v}&<&3\times\sigma_{\rm ch} \Delta V_{\rm ch} \sqrt{\frac{N_{\rm ch}N_{\rm S}}{ N_{\rm B}}} 
\end{eqnarray}
for lines and
\begin{eqnarray} \label{eq:err_cont}
S_{\rm cont}&<&3\times\sigma_{\rm cont} \sqrt{\frac{N_{\rm S}}{ N_{\rm B}}} 
\end{eqnarray}
for continuum emission,
where 
$\sigma_{\rm ch}$ is the RMS level of the channel map, $\Delta V_{\rm ch}$ is the velocity resolution,
$N_{\rm ch} ={\rm FWHM}/\Delta V_{\rm ch}$,
$N_{\rm S} $ is the the number of pixels for the source size, and
$N_{\rm B}$ is the number of pixels for the synthesized beam. 
The equations are from \citet{Hainline_2004}. 
If CO~(4--3) is detected and [\ion{C}{1}]~(1--0) is not, we assume that $N_{\rm ch}$ and $N_{\rm S}$ of [\ion{C}{1}]~(1--0) are the same as CO~(4--3), where  $N_{\rm ch}$ and $N_{\rm S}$ are calculated based on the FWHM value of the velocity width and source size measured in Table~\ref{tab:flux_line}. 
If both CO~(4--3) and [\ion{C}{1}]~(1--0) are not detected, we assume that the FWHM of line width is 150~km~s$^{-1}$ and $N_{\rm S}=3\times N_{\rm B}$. However, these upper limits were not included in the discussion section.
We note that  some [\ion{C}{1}]~(1--0) emission is on the edge of the spectral window because of the incorrect redshift entered in observation requests in some galaxies (e.g., IRAS~F05189-2524). In this case, we assume the same line profile as CO~(4--3) to calculate the total  [\ion{C}{1}]~(1--0) flux. 

\subsection{Missing Flux}\label{sec:MissingFlux}
As explained in Section~\ref{sec:Observation}, the ACA single-point observation without the total power array may not cover the total flux. 
In Table~\ref{tab:flux_line}, the SPIRE/FTS velocity integrated [\ion{C}{1}]~(1--0) and CO~(4--3) fluxes from \citet{Kame_2016} are listed if accessible. 
The typical ACA RF is approximately $\sim 59~\%$ and $\sim 69~\%$ for [\ion{C}{1}]~(1--0) and CO~(4--3), respectively.
The CO~(4-3) and  [\ion{C}{1}]~(1--0) lines are close enough in frequency that they are both observed with similar uv-coverage, tracing similar spatial scales in the sample galaxies.  Therefore, we argue that the resultant line ratios between  CO~(4-3) and  [\ion{C}{1}]~(1--0) provided here are a reasonable representation of the line ratios, despite the missing flux of order 30-40\%.
The extrapolated total dust continuum flux densities are listed in Table~\ref{tab:flux_cont}, whenever the dust mass was estimated in \citet{Shangguan_2019}\footnote{Dust continuum fluxes are provided by Jinyi Shangguan in private communication.}.
We found that the typically RF for continuum emission was 46\% and 52\%, for the 609 and 650~$\mu$m continuum flux density, respectively. 
The RF for individual galaxies is derived from the ratio between the continuum flux density by the ACA and continuum spectrum fitting.
The typical RF 
suggests that almost half of the continuum flux density comes outside the FoV or from a structure that is larger than MRS. 
We note that the SPIRE beam is 45\farcs5 at the CO~(4--3) transition but the ACA FoV is 23\farcs5 (shown as magenta circles in Figure~\ref{fig:SED}).
Therefore, a large missing flux might be observed, especially, for galaxies with small luminosity distances (e.g., $D<100$~Mpc).
The individual notes for galaxies that have a large missing flux (e.g., IC~4280) are given in Section~\ref{sec:individual}. 

\subsection{Luminosity}
The luminosities for the lines are measured using
\begin{eqnarray}\label{L1}
&&L_{\rm line}~[L_{\odot}]  \nonumber \\
&& =1.04\times10^{-3}~S_{\rm line}\Delta v~ \nu_{\rm rest} ~(1+z)^{-1}~D_{\rm L}^2
\end{eqnarray}
and
\begin{eqnarray}\label{L2}
&&L'_{\rm line}~[{\rm K~km~s^{-1}~pc^2}]  \nonumber \\
&&= 3.25\times10^{7}~S_{\rm line}\Delta v~ \nu_{\rm obs}^{-2} ~(1+z)^{-3}~D_{\rm L}^2,
\end{eqnarray}
where $\nu_{\rm rest}$ is the rest frequency of the line emission in GHz, 
$z$ is the redshift, $D_{\rm L}$ is the luminosity distance in Mpc, 
and $ \nu_{\rm obs}=\nu_{\rm rest}/(1+z)$ is \added{the} sky (observed) frequency in GHz \citep{Solomon_2005}. 
We used $L'_{\rm line}$ to calculate the molecular gas mass (Section~\ref{appendix:Mgas}) 
and $L_{\rm line}$ for the photodissociation region (PDR) modeling (Section~\ref{sec:highz}).

The ACA continuum flux density at the rest frames of 609~$\mu$m and 650~$\mu$m ($S_{609}$ and $S_{650}$) are converted to the 630~$\mu$m specific luminosity $L_{\nu({\rm 630\mu m})}$ based on a formulation in \citet{Scoville_2016},
\begin{eqnarray}\label{Lcont}
&&L_{\nu({\rm 630\mu m})}~[{\rm erg~s^{-1}~Hz^{-1}}]  \nonumber \\
&&=(1.19\times10^{27}) \times \left(\frac{\nu_{630 {\rm \mu m}}}{\nu_{\rm obs}(1+z)}\right)^{3.8} \times \frac{D_{\rm L}^2}{(1+z)}\nonumber \\
&&~~\times \frac{\Gamma_{\rm RJ}(T_{\rm dust},\nu_{630 {\rm \mu m}},0)}{\Gamma_{\rm RJ}(T_{\rm dust},\nu_{\rm obs},z)} \times S_{\nu}.
\end{eqnarray}
where $S_{\nu}$ is $(S_{609}+S_{650})/2$ in Jy and $\nu_{\rm obs}$ is the average of the red-shifted sky frequency\footnote{The red-shifted sky frequencies are labeled in  Figure~\ref{fig:mom}(d) and (h).}. 
If either $S_{609}$ or $S_{650}$ were detected, we used the flux density and frequency of the detected signal.
We assume the dust temperature of $T_{\rm dust}=25$~K for simple analyses.
$\Gamma_{\rm RJ}$ is the correction for departure in the rest frame of the Planck function from Rayleigh–Jeans given by
\begin{equation}\label{eq:RJ}
\Gamma_{\rm RJ}(T_{\rm dust},\nu,z)=\frac{h\nu(1+z)/kT_{\rm dust}}{e^{h\nu(1+z)/kT_{\rm dust}}-1}.
\end{equation}
The derived luminosities are listed in Table~\ref{tab:luminosity}. 
The ratios among the derived  luminosities are listed in Table~\ref{tab:ratio_appendix}.

\subsection{Molecular gas mass}\label{sec:Mgas}
We supplementally estimated $M_{\rm H2}$ from CO~(1--0), [\ion{C}{1}]~(1--0), CO~(4--3), and  the continuum, respectively (See detail in Section~\ref{appendix:Mgas} and Table~\ref{tab:Mgas}).
We assumed a simple luminosity-to-$M_{\rm H2}$ conversion factor.
We note that  the main purpose of this study is not to investigate the effect of the systematic uncertainties introduced by the various assumptions when estimating $M_{\rm H2}$. 

\section{Discussion}\label{Sec:Discussion}
\begin{figure*}[!htbp]
\begin{center}
\includegraphics[angle=0,scale=0.6]{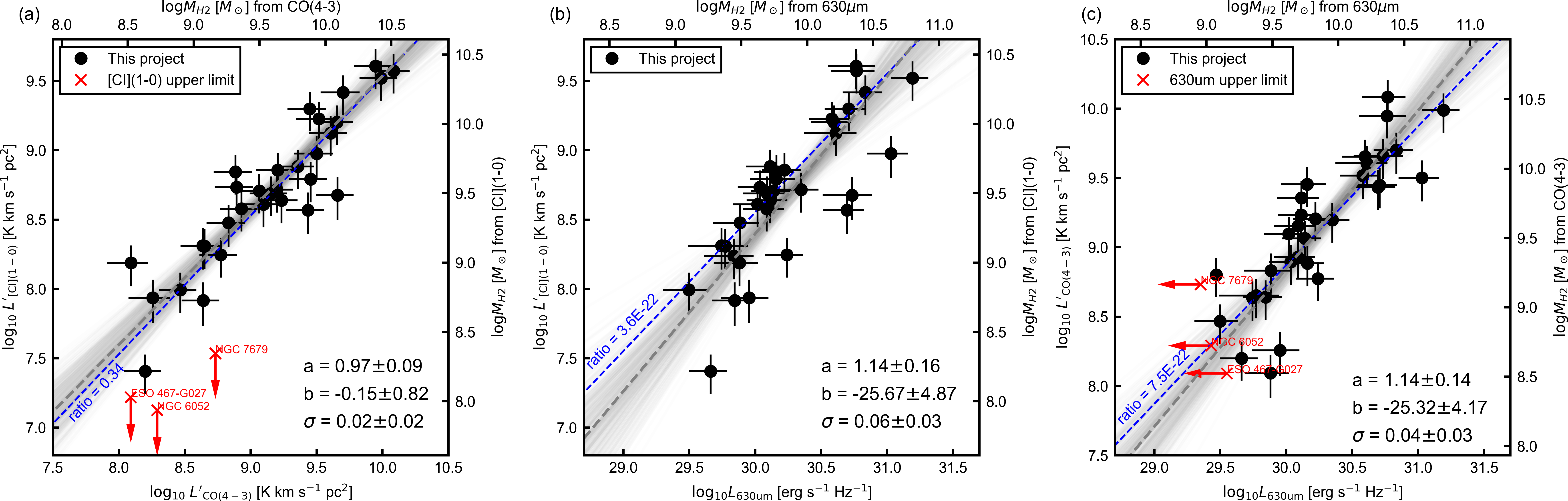}
\caption{The relation between luminosities of (a) CO~(4--3) and [\ion{C}{1}]~(1--0), (b)   630~$\mu$m and [\ion{C}{1}]~(1--0), and (c)  630~$\mu$m and CO~(4--3) obtained by this project. 
The top and right-hand axes of each panel supplementally display the data in terms of
$M_{\rm H2}$ estimated in Section~\ref{appendix:Mgas}.
In each panel, the blue-dashed line indicates the median of the ratio between both quantities. The median is calculated including non-detection galaxies.
The linear regression lines (grey lines) for detected galaxies are calculated by the \textsc{Python} package {\tt linmix} developed by \citet{Kelly_2007}.
The grey-dashed indicates the best fit.
The grey translucent lines indicate samples from the posterior distribution.
The $a$ and $b$ values for the fitting results ($y=ax+b$) and  the intrinsic scatter ($\sigma$) are shown at the right bottom corner.
The red arrows indicate the upper limits in the case of [\ion{C}{1}]~(1--0) and 630~$\mu$m non-detection.\label{fig:CI-CO43-cont}}
\end{center}
\end{figure*}

Figure~\ref{fig:CI-CO43-cont} shows the relationship between [\ion{C}{1}]~(1--0), CO~(4--3), and 630~$\mu$m continuum luminosities.
The linear regression lines (gray lines) for detected galaxies were calculated using the \textsc{Python} package of {\tt linmix} developed by \citet{Kelly_2007}.
Minimum, maximum, average, median, and standard errors of the  $L_{\rm [CI](1-0)}/L_{\rm CO(4-3)}$, $L_{\rm 630\mu m}/L'_{\rm [CI](1-0)}$, and $L_{\rm 630\mu m}/L'_{\rm CO(4-3)}$ in our targets are shown in Table~\ref{tab:ratio}.
The correlations are naturally explained because $L'_{\rm [CI](1-0)}$, $L'_{\rm CO(4-3)}$, and $L_{\rm 630\mu m}$ could be used to measure M$_{\rm H2}$ (Section~\ref{appendix:Mgas}).
The scatter for the correlations could reflect the systematic uncertainties introduced by the various assumptions when measuring $M_{\rm H2}$.
Figure~\ref{fig:CI-CO43-cont} (a) shows that the upper limits for the non-detections, i.e., $L'_{\rm [CI](1-0)}/L'_{\rm CO(4-3)} <0.08$ in NGC~6052 and  $L'_{\rm [CI](1-0)}/L'_{\rm CO(4-3)}<0.08$ for NGC~7679 is $> 4$ times smaller than the typical $L'_{\rm [CI](1-0)}/L'_{\rm CO(4-3)}$ ratio of 0.34 (see Section~\ref{sec:ND} in detail).
In Section \ref{Sec:CI-ratio}, we further investigate a diverse range of galaxies in terms of $L'_{\rm [CI](1-0)}/L'_{\rm CO(4-3)}$ and $L'_{\rm [CI](1-0)}/L_{\rm 630\mu m}$.
\begin{deluxetable}{lllllll}
\tablenum{1}
\tablecaption{Summary of statistics for the ratios between [\ion{C}{1}]~(1--0), CO~(4--3) and 630~$\mu$m \label{tab:ratio}}
\tablewidth{0pt}
\tablehead{
\colhead{}  &
\colhead{min,max}  &
\colhead{ave} &
\colhead{med} &
\colhead{SE} & 
\colhead{$N$}
}
\startdata
$L_{\rm [CI](1-0)}/L_{\rm CO(4-3)}$  & 0.1,1.5 & 0.5 & 0.4 & 0.05 & 31\\
$F^{\rm p}_{\rm [CI](1-0)}/F^{\rm p}_{\rm CO(4-3)}$  & 0.2,1.1 & 0.4 & 0.4 & 0.05 & 20\\
$L'_{\rm [CI](1-0)}/L_{\rm 630um}$$\times10^{-22}$ & 0.6,7 & 3.3 & 3.6 & 0.3 & 32\\
$F^{\rm p}_{\rm [CI](1-0)}/S_{609}$ & 1.3,37 & 9.8 & 7.7 & 1.6 & 27\\
$L'_{\rm CO(4-3)}/L_{\rm 630um}$$\times10^{-22}$  & 1.6,21 & 9.1 & 8.1 & 0.9 & 32\\
$F^{\rm p}_{\rm CO(4-3)}/S_{650}$  & 4.9,63 & 31 & 26 & 3.0 & 32\\

\enddata
\tablecomments{the minimum (min), maximum (max), average (ave), median (med), and standard error (SE=SD/$\sqrt{N}$, where SD is the standard deviation and $N$ is the sample size) of each combination of the ratios.
Non-detection cases are included in the median calculation; however, the other statistics are calculated without non-detection cases.
$F^{\rm p}$ indicates the peak flux density (in the unit of Jy) calculated by Gaussian spectrum line fitting  (Figure~\ref{fig:mom}). 
The units of $L_{\rm line}$,  $L'_{\rm line}$, $L_{{\rm 630\mu m}}$, and $S_{\rm cont}$, are $L_{\odot}$, ${\rm K~km~s^{-1}~pc^2}$, ${\rm erg~s^{-1}~Hz^{-1}}$, and Jy, respectively.
}\end{deluxetable}

\begin{figure}[!htbp]
\begin{center}
\includegraphics[angle=0,scale=0.2]{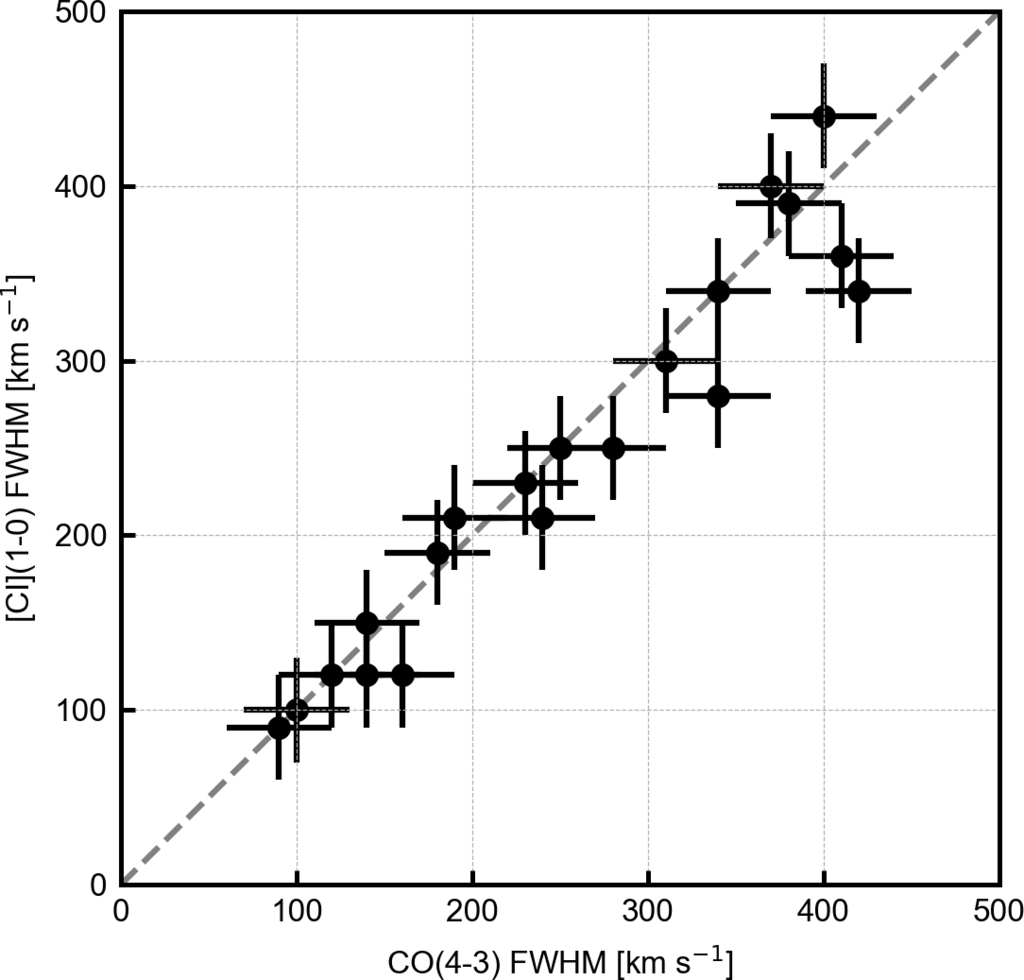}
\caption{The FWHM of [\ion{C}{1}]~(1--0) and CO~(4--3) line profiles. A constant 1$\sigma$ error of 30~km\,s$^{-1}$ is applied. The grey dashed line has a slope of unity. \label{fig:FWHM}}
\end{center}
\end{figure}

Figure~\ref{fig:FWHM} shows the FWHM of the  [\ion{C}{1}]~(1--0) and CO~(4--3) line profiles. 
In most case, the line shapes of [\ion{C}{1}]~(1--0) and CO~(4--3) lines and mom1 and mom2 maps appear similar (Figure \ref{fig:mom}), implying that the kinematical properties of the [\ion{C}{1}]~(1--0) and CO~(4--3) are the same for molecular gas distributions at the $\sim$kpc scale.
We manually selected galaxies used in Figure~\ref{fig:FWHM}, which can be well fitted by single Gaussian components (i.e., the used spectra are marked in Figure~\ref{fig:mom}.)
For example, we do not include NGC~232-2 (NGC~235) because the S/N of the spectrum is low.
We also avoid double peak emission, because a single Gaussian fitting may not work well (e.g., IC4280). However, this manual selection did not impact the results of this study. 

\subsection{[\ion{C}{1}] non-detection}\label{sec:ND}
ALMA/ACA recently identified a galaxy, NGC~6052 \citep{Michiyama_2020}, which may be a counterexample to the recent concept of the ubiquitous [\ion{C}{1}]/CO distribution during the cloud lifetime and the C$^0$-rich/CO-poor region due to the CO dissociation process, proposed by e.g., \citet{Papa_2004_theory, Papadopoulos_2018}.
During this survey, we further discovered the [CI] non-detection case of gas-rich type~2 AGN, NGC~7679. 
Figure~\ref{fig:Mdust_CO_SFR-CI} shows that the upper limits of [\ion{C}{1}] are small when we compare the global properties of galaxies, i.e., $M_{\rm dust}$, SFR, and $L'_{\rm CO(1-0)}$, demonstrating that NGC~6052 and NGC~7679 are candidates for the [\ion{C}{1}]-dark galaxies.
Two possible scenarios of weak [\ion{C}{1}]~(1--0) emission are C$^0$-poor/CO-rich environments or an environment with an extremely large [\ion{C}{1}]~(1--0) missing flux. A direct comparison of [\ion{C}{1}] and CO in the same uv-plane is necessary to distinguish between these two scenarios.

If the C$^0$-poor/CO-rich scenario is true, we may be able to use the [\ion{C}{1}]-deficit to identify young starbursts (suggested in NGC~6052; \citealt{Michiyama_2020}). 
Young starbursts are also observed in NGC~7679 as well.  For example, \citet{Gu_2001} detected clear higher-order Balmer absorption lines implying the presence of active star-forming regions in the central  $2\arcsec\times2\arcsec$ region.
By modeling the spectrum, they found that young ($<10$~Myr) and intermediate ($\sim100$~Myr) stellar age populations are dominant in the nuclear region of NGC~7679.
In addition, \citet{Gu_2001} compared the UV spectra of NGC~7679 and NGC~5135.
Among our samples, NGC~7679 showed the smallest  $L_{\rm [CI](1-0)}/L_{\rm CO(4-3)}$ of $<0.08$, and NGC~5135 has the largest $L_{\rm [CI](1-0)}/L_{\rm CO(4-3)}$ of $ 1.1\pm0.48$.
\citet{Gu_2001} demonstrated that the stellar wind absorption line, such as the P-Cygni profile of \ion{C}{4}$\lambda$1550, is more significant in NGC~7679 than that in NGC~5135;
this also suggests the younger stellar population in NGC~7679 than NGC~5135 while both galaxies host type~2 AGNs. The nuclear starburst activity in NGC~7679 may be related to the tidal interaction with NGC~7682 (with a separation of $\sim90$~kpc). The keywords ``galaxy interaction" and ``young starburst" are common among NGC~7679 and NGC~6052 \citep{Michiyama_2020}, whose [\ion{C}{1}]~(1--0) was not detected in our survey.
The ``[CI]-deficit" would be a hint to understand the timescale triggered by a galaxy interaction.

If an extremely large [\ion{C}{1}]~(1--0) missing flux scenario is true (i.e., $>90\%$ of
[\ion{C}{1}]~(1--0) flux is resolved out / outside the FoV).
This implies that [\ion{C}{1}] is not associated with the compact molecular gas, which is a direct material for star formation. In this case, [\ion{C}{1}] emission may be associated with atomic hydrogen (\ion{H}{1}) rather than with the molecular (H$_2$) phase. If the spatial structure is completely different between CO and [\ion{C}{1}], the comparison with dust distributions traced by the Band 8 continuum and HI high-resolution map -- which are accessible by cm telescopes (e.g., Very Large Telescope) -- are important for comprehensively understanding the different gas phases traced by CO/\ion{C}{1}/\ion{H}{1} and dust.\\

\begin{figure*}[!htbp]
\begin{center}
\includegraphics[angle=0,scale=0.7]{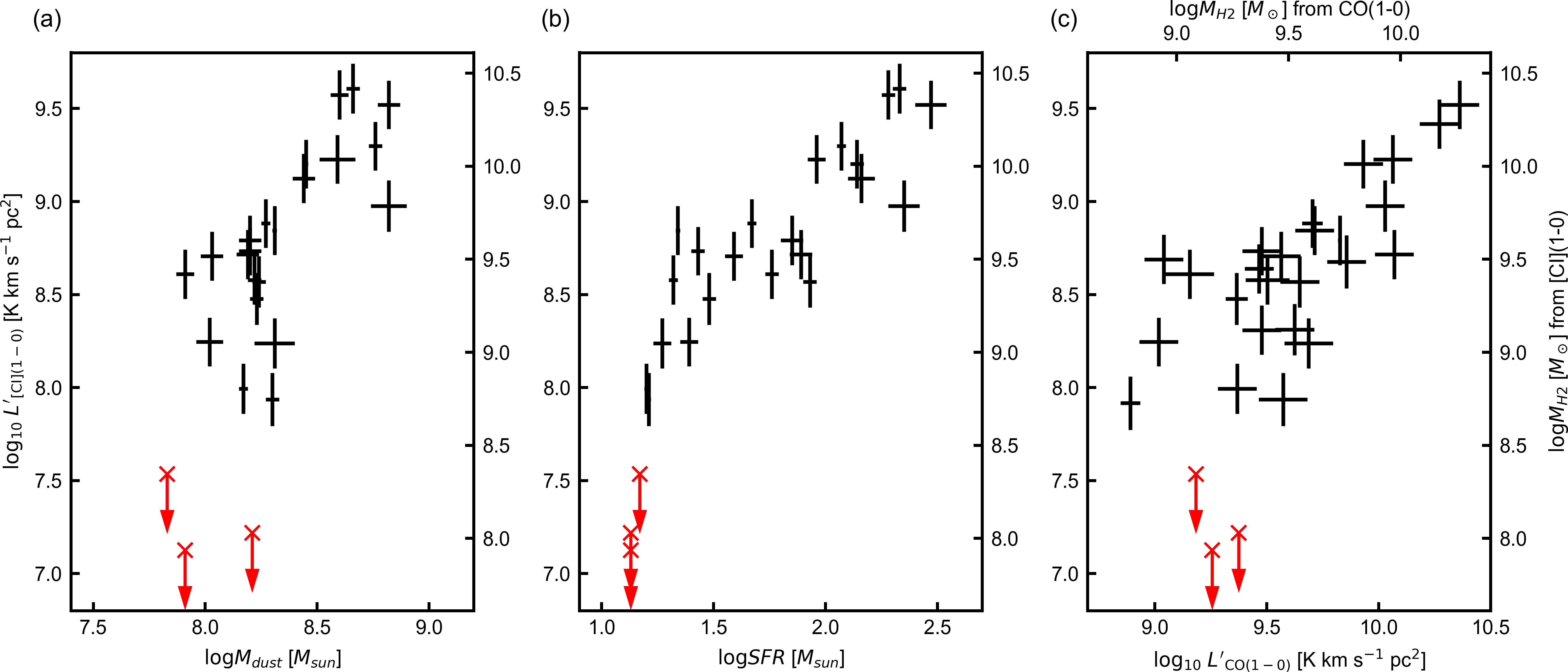}\label{fig:Mdust_CO_SFR-CI}
\caption{The relation between [\ion{C}{1}]~(1--0) luminosity and global properties of galaxies, i.e., (a) $M_{\rm dust}$, (b) SFR, and (c) CO~(1--0) luminosity.}
\label{fig:Mdust_CO_SFR-CI}
\end{center}
\end{figure*}

\subsection{What controls [CI] luminosities in galaxies?}\label{Sec:CI-ratio}
In this section, we attempt to determine the cause of the variation in $L'_{\rm [CI](1-0)}/L'_{\rm CO(4-3)}$ and  $L'_{\rm [CI](1-0)}/L_{\rm 630\mu m}$.
In particular, we investigate the effects of  galaxy merger activities, AGN, and dust temperature. 
In this section, we conclude that observed [\ion{C}{1}] luminosities are not correlated with the inferred mergers/AGNs and $T_{\rm dust}$.
The details are as follows.

\subsubsection{Galaxy merger}
\begin{figure*}[!htbp]
\begin{center}
\includegraphics[angle=0,scale=0.6]{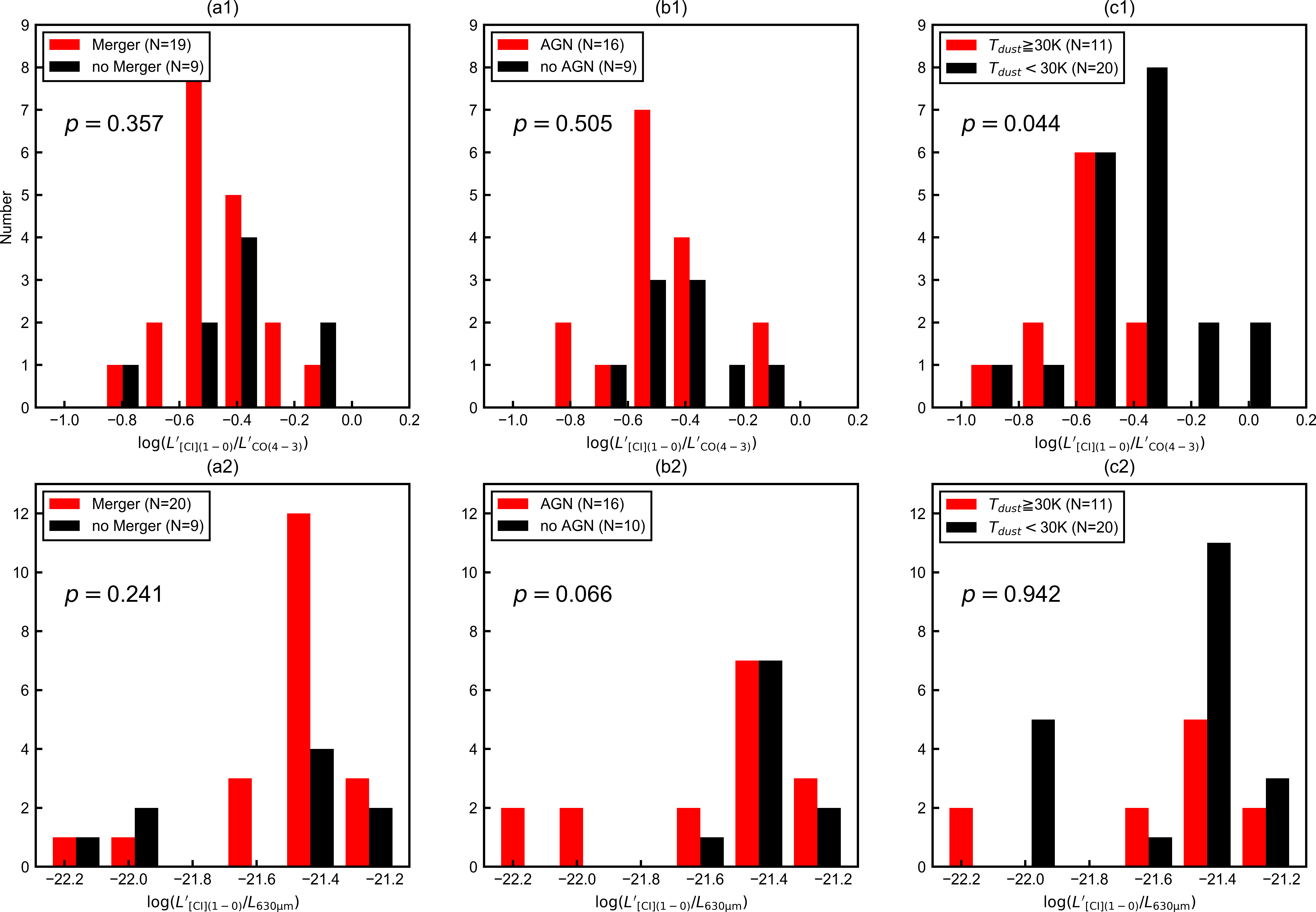}
\caption{(left) The distribution of (a1) $L'_{\rm [CI](1-0)}/L'_{\rm CO(4-3)}$ and (a2) $L'_{\rm [CI](1-0)}/L_{\rm 630 \mu m}$ for merger (red) and non-merger (black) galaxies. The p-value of the t-test between mergers and non-mergers is shown in each panel.  There was no significant difference between mergers and non-mergers in each plot. (middle) The distribution of (b1) $L'_{\rm [CI](1-0)}/L'_{\rm CO(4-3)}$ and (b2) $L'_{\rm [CI](1-0)}/L_{\rm 630 \mu m}$ for AGNs (red) and non-AGNs (black), respectively. There is no significant difference between AGNs and non-AGNs in each plot. (right) The distribution of (c1) $L'_{\rm [CI](1-0)}/L'_{\rm CO(4-3)}$ and (c2) $L'_{\rm [CI](1-0)}/L_{\rm 630 \mu m}$ for galaxies with $T_{\rm dust}\geqq30$~K (red) and $T_{\rm dust}<30$~K (black), respectively. The p-value of the t-test between high- and low-dust temperature is shown in each panel. There is possibly a negative relationship (i.e., lower $T_{\rm dust}$ may indicate higher $L'_{\rm [CI](1-0)}/L'_{\rm CO(4-3)}$, $p=0.044$) in panel (c1).
\label{fig:Merger_AGN_Tdust}}
\end{center}
\end{figure*}

Galaxy mergers are one possible event that affects [\ion{C}{1}]/CO ratio.
For example, \citet{Michiyama_2020} show that the small  [\ion{C}{1}]/CO line ratio in NGC~6052 is possibly due to the dense gas compressed at the collision front.
In addition, the other [\ion{C}{1}] non-detected galaxy, NGC~7679, also shows a young starburst, possibly triggered by a galaxy interaction (see details in Section~\ref{sec:ND}).
Therefore, a small  [\ion{C}{1}]/CO ratio may be common in merging galaxies.
We apply the selection criteria of merging galaxies used in \citet{Shangguan_2019} and used a t-test between the merger and non-merger groups (Figure~\ref{fig:Merger_AGN_Tdust}a). 
There was no significant difference in the mean values of $L'_{\rm [CI](1-0)}/L'_{\rm CO(4-3)}$  and $L'_{\rm [CI](1-0)}/L_{\rm 630\mu m}$  between merger and non-merger groups, suggesting that a merger process may not be a significant for controlling the [\ion{C}{1}]/CO ratio.
In the case of NGC~6052 and NGC~7679, we may see a very rare phase (e.g., very young starburst) during the merger process.\\

\subsubsection{AGN}
AGN is a possible event that affects the [\ion{C}{1}]/CO ratio.
For example, \citet{Meijerink_2005} and \citet{Meijerink_2007} demonstrated that the abundance of atomic carbon relative to CO will increase due to CO dissociation in X-ray dominated regions (XDRs), where strong X-rays from the central AGN control the chemical and physical properties of the ambient gas.
In addition, \citet{Izumi_2020} have investigated the [\ion{C}{1}] and CO distribution around the AGN of NGC~7469 with $\sim100$~pc resolution and found that the [\ion{C}{1}]~(1--0)/$^{12}$CO~(2--1) ratio was 20 times higher around the AGN than the starburst ring outside;
this indicates that [\ion{C}{1}] enhancement can be used to identify dust-obscured AGNs.
They believed that an elevated C$^0$/CO ratio (as expected in the XDR models) is required to explain the [\ion{C}{1}]-enhancement. 

We applied the AGN selection criteria used by \citet{Shangguan_2019} to identify AGNs among our observed nearby U/LIRGs.
We used a t-test to determine if there was a significant difference ($p<0.05$ or not) between the mean values  ($L'_{\rm [CI](1-0)}/L'_{\rm CO(4-3)}$  and $L'_{\rm [CI](1-0)}/L_{\rm 630\mu m}$) of the AGN and non-AGN groups (Figure~\ref{fig:Merger_AGN_Tdust}b).  
There is no significant difference between the AGNs and non-AGN for $L'_{\rm [CI](1-0)}/L'_{\rm CO(4-3)}$  and $L'_{\rm [CI](1-0)}/L_{\rm 630\mu m}$. 
We note that our ACA spatial resolution is $\sim1$~kpc. 
To understand the AGN properties, we need to achieve a higher resolution, as in \citet{Izumi_2020}.
For example,  $L_{\rm [CI](1-0)}/L_{\rm CO(4-3)}$ is $0.19\pm0.08$ for NGC~7469 in our survey, which is below the average value of observed galaxies.
This suggests that AGN enhances the [\ion{C}{1}]/CO ratio on a small scale ($\sim100$~pc), but AGN cannot control the ratio in star-forming regions outside ($>$kpc).
A higher spatial resolution is necessary to understand the [\ion{C}{1}] properties of AGNs.

\subsubsection{Dust temperature}
As explained in equation~(\ref{equ:MH2_cont}), the conversion from continuum flux density to $M_{\rm H2}$ depends on the assumption of the dust temperature (i.e., $\Gamma$  in equation~\ref{eq:RJ} depends on dust temperature).
For example, assuming $T_{\rm dust}=20$~K, the value of $M_{\rm H2}$ is approximately 1.1 times higher than that assuming $T_{\rm dust}=40$~K in equation~(\ref{equ:MH2_cont}).
Therefore, the line-to-continuum ratio may depend on the dust temperature. Figure~\ref{fig:Merger_AGN_Tdust}(c2) shows that  there is no evidence that the dust temperature controls the ratios of $L'_{\rm [CI](1-0)}/L_{\rm 630\mu m}$, suggesting that the dust temperature (i.e., conversion from 630~$\mu$ m to 850~$\mu$ m) is not the main systematic error in equation~(\ref{equ:MH2_cont}).

Conversely, Figure~\ref{fig:Merger_AGN_Tdust}(c1) suggests a negative relationship (i.e., lower $T_{\rm dust}$ may indicate higher $L'_{\rm [CI](1-0)}/L'_{\rm CO(4-3)}$, $p=0.04$).
This indicates that the excitation of [\ion{C}{1}] and/or CO may be related to dust temperature.
For example, a positive correlation is seen between $T_{\rm dust}$ and CO or [\ion{C}{1}] excitation conditions in spatially resolved studies in very nearby galaxies (i.e., $D_{\rm L}<20$~Mpc, \citealt{Koda_2020, Jiao_2019}), and a positive correlation was observed between $T_{\rm dust}$ and the [\ion{C}{1}] excitation temperatures that is estimated based on [\ion{C}{1}]~(2--1)/[\ion{C}{1}]~(1--0) ratio \citep{Jiao_2019}.
However, the range of $T_{\rm dust}$ and the number of samples are limited in our analysis and a further complete survey, including multi-[\ion{C}{1}] and CO transitions by ALMA, is necessary to understand the CO and [\ion{C}{1}] excitation condition.

\subsection{Comparison with high-z galaxies}\label{sec:highz}
The secondary aim of this study was to compare our nearby U/LIRGs with high-z galaxies.
In the previous sections, we investigated the uncertainty of the usage of Band~8 information, that is, [\ion{C}{1}]~(1--0), CO~(4--3), and the dust continuum as a tracer of bulk $M_{\rm H2}$ of the galaxies.
However, the most important reason for using Band~8 measurements is that we can directly compare nearby and high-z galaxies.

Using the high-z  [\ion{C}{1}]~(1--0) sample compiled by \citet{Valentino_2018, Valentino_2020}, we investigated the properties of high-z and nearby galaxies in this section.
We compared three groups: 29 nearby U/LIRGs from our survey, 22 galaxies on the main sequence at $z\sim1$ ($z\sim1$ MS), and 27 galaxies at $z=2-4$ who are categorized as SMGs  ($z=2-4$ SMGs) from \citet{Valentino_2018, Valentino_2020}.

We compared the nearby U/LIRGs, $z\sim1$ MS, and $z=2-4$ SMGs on the plane of $L_{\rm{[CI](1-0)}}/L_{\rm TIR}$ and $L_{\rm{[CI](1-0)}}/L_{\rm{CO(4-3)}}$.
The physical parameters of the PDR can be constrained in this plane \citep{Alaghband-Zadeh_2013,Bothwell_2017, Valentino_2018, Valentino_2020}.
We used the Photodissociation Region Toolbox
 (PDRT\footnote{\url{http://dustem.astro.umd.edu/pdrt/}}; \citealt{Kaufman_1999, Kaufman_2006, Pound_2008}).
This enabled us to investigate the density of H nuclei ($n_{\rm H}$ in units of cm$^{-3}$) and
incident far-ultraviolet (FUV) radiation field ($U_{\rm uv}$), assuming the plane-parallel PDR model.
The $U_{\rm uv}$ corresponds to photons at 6~eV$\leq h\nu <$ 13.6~eV in the unit of the average interstellar radiation field in the vicinity of the Sun ($G_0$).
The model provides the line intensities 
for each combination of $n_{\rm H}$ and $U_{\rm uv}$ 
by self-consistently solving for radiation transfer, thermal balance, and chemical processes.

\begin{figure}[!htbp]
\begin{center}
\includegraphics[angle=0,scale=0.2]{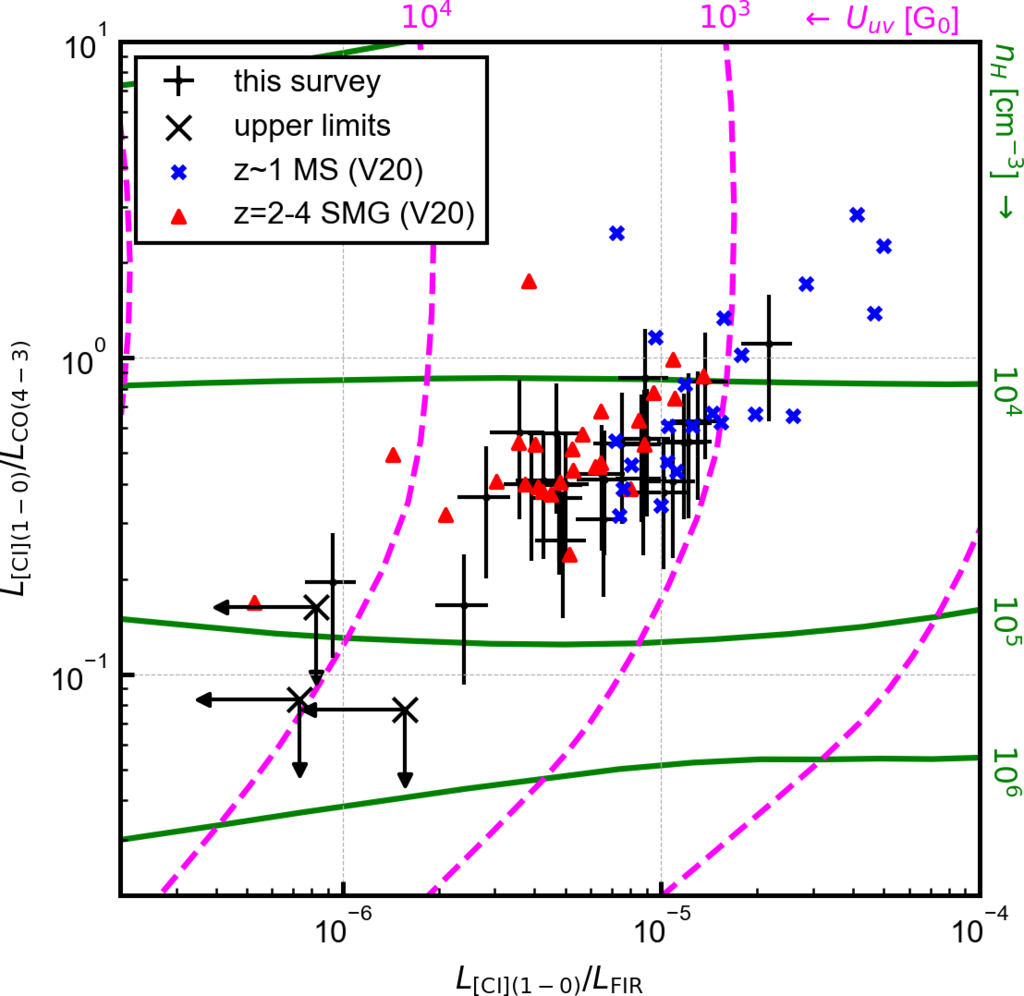}
\caption{$L_{\rm [CI](1-0)}/L_{\rm CO(4-3)}$ versus $L_{\rm [CI](1-0)}/L_{\rm TIR}$ for our U/LIRGs (black), z=2-4 SMGs (red), and z$\sim1$ MS (blue).
The arrows indicate the [\ion{C}{1}]~(1--0) upper limits.
The green solid and magenta dashed lines indicate the theoretical tracks (based on PDRT) for constant $n_{\rm H}$~[cm$^{-3}$] and $U_{\rm uv}$~[$G_0$], respectively ($G_0 =1.6\times10^{-3}$~ergs~cm$^{-2}$~s$^{-1}$; \citealt{Habing_1968}).
We assume that the typical [\ion{C}{1}]~(1--0) recovered flux of \input{RF_CI_new.tex}\%. 
The term ``V20" indicates galaxies compiled by \citet{Valentino_2020}. \label{fig:PDR}}
\end{center}
\end{figure}

Figure~\ref{fig:PDR} shows $L_{\rm [CI](1-0)}/L_{\rm TIR}$  versus $L_{\rm [CI](1-0)}/L_{\rm CO(4-3)}$ for nearby U/LIRGs, $z\sim1$ MS, and $z=2-4$ SMGs.
For nearby U/LIRGs, we use $L_{\rm TIR}$ from SED fitting (AGN contribution is subtracted) and assume that the typical ACA recovered [\ion{C}{1}] flux is $\sim 59~\%$ when we calculate $L_{\rm{[CI](1-0)}}/L_{\rm TIR}$.
The tracks for constant $n_{\rm H}$ and $U_{\rm uv}$ are indicated by the green solid and magenta dashed lines, respectively.
According to a plane-parallel PDR geometry,
the smaller [\ion{C}{1}]~(1--0)/CO~(4--3) ratio indicates the denser PDR because UV radiation cannot penetrate deeply into the molecular medium. 
For example, the small line ratio (e.g.,  $L_{\rm [CI](1-0)}/L_{\rm CO(4-3)}<0.1$) means high density ($>10^5$~cm$^{-3}$).
The small $L_{\rm [CI](1-0)}/L_{\rm TIR}$ is indicative of a high $U_{\rm uv}$.
We found that the [\ion{C}{1}]~(1--0) weak galaxies such as NGC~6052 and NGC~7679 (e.g., $L_{\rm [CI](1-0)}/L_{\rm CO(4-3)}<0.1$) are rare, including the high-z galaxies (see also \citealt{Israel_2002, Israel_2015}). 

\begin{figure}[!htbp]
\begin{center}
\includegraphics[angle=0,scale=0.8]{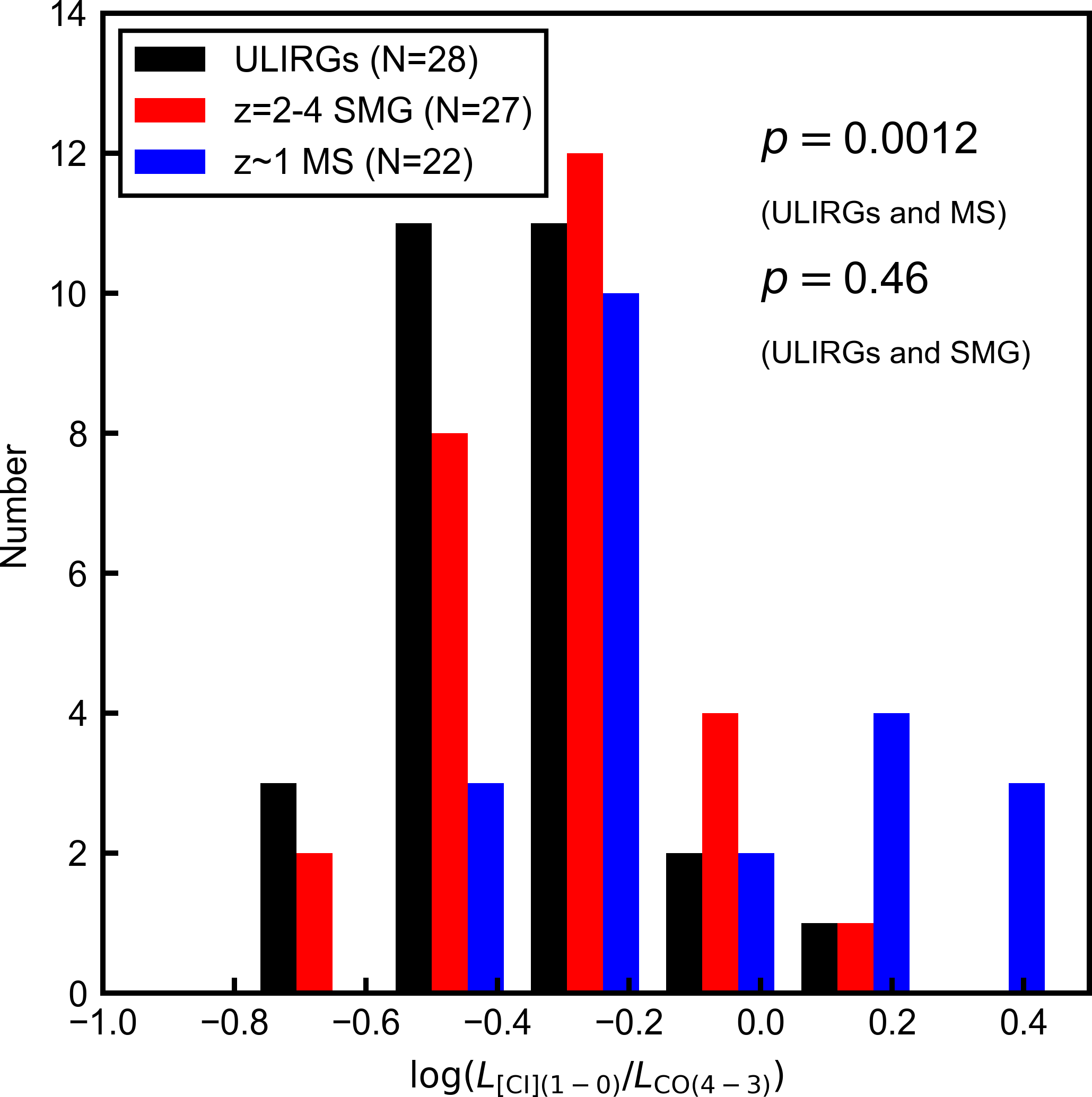}
\caption{The distribution of $L_{\rm [CI](1-0)}/L_{\rm CO(4-3)}$ for our nearby U/LIRGs (black), z=2-4 SMGs (red), and z$\sim$1 MS (blue). The t-test p-values between U/LIRGs and MS and U/LIRGs and SMGs are shown. There is a significant difference between U/LIRGs and MS but the distributions of U/LIRGs and SMGs are the same. The $N$ value means the number of galaxies used in this statistical test. \label{fig:Highz_KS}}
\end{center}
\end{figure}

We used  a t-test to determine if there is a significant difference among nearby U/LIRGs, $z\sim1$ MS, and $z=2-4$ SMGs  in terms of $L_{\rm [CI](1-0)}/L_{\rm CO(4-3)}$. Figure~\ref{fig:Highz_KS} demonstrates that the distribution of $L_{\rm [CI](1-0)}/L_{\rm CO(4-3)}$ in nearby U/LIRGs is the same as SMGs but is significantly different from MS.
For example, the median value of 
$L_{\rm [CI](1-0)}/L_{\rm CO(4-3)}$ is 
0.42, 0.47, and 0.66 
for nearby U/LIRGs, $z=2-4$ SMG, and $z\sim1$ MS, respectively. 
This means that the typical density is approximately 1.5 times larger in  U/LIRGs and  SMGs than in MS galaxies. 
We note that this suggestion, as shown in Figure~\ref{fig:Highz_KS}(a), does not strongly depend on the missing flux issue of our ACA survey.

\section{Summary}\label{sec:summary}
We present our recent ACA band 8 observations of [\ion{C}{1}]~(1--0),  CO~(4--3), and dust continuum emission in 36 nearby U/LIRGs.
The main results are as follows.
 \begin{itemize}
      \item Correlations were confirmed  among $L_{\rm [CI](1-0)}$, $L_{\rm CO(4-3)}$, and  $L_{\rm 630um}$. However, there was variation in the ratios among the three observables. For example, the range of [\ion{C}{1}]~(1--0)/CO~(4--3) luminosity ratio is (0.1-1.1), including the [\ion{C}{1}]~(1--0) non-detection case for a ratio of $<0.08$ (i.e., NGC~6052 and NGC~7679).
	\item The observed [\ion{C}{1}] luminosities did not correlate with inferred $T_{\rm dust}$, or AGN/merger morphology.
	\item As a case study for specific targets, we compare two AGN host galaxies; i.e., NGC~7679, which exhibits the smallest $L_{\rm [CI](1-0)}/L_{\rm CO(4-3)}<0.08$, and NGC~5135, which exhibits the largest $L_{\rm [CI](1-0)}/L_{\rm CO(4-3)}=1.1\pm0.48$. We found that the evidence of young starburst activity was more robust in NGC~7679 than in NGC~5135 and star formation activities in NGC~7679 may be related to the tidal interaction with the pair galaxy besides.
	\item We compare our nearby U/LIRGs and high-z galaxies, such as $z\sim1$ galaxies on MS and $z=2-4$ SMGs. The mean value of the [\ion{C}{1}]~(1--0)/CO(4--3) ratio of U/LIRGs is similar to that of SMGs but smaller than that of galaxies on MS.  This possibly indicates a higher hydrogen density in U/LIRGs and SMGs  than MS when we assume a simple photodissociation model.
 \end{itemize}

  \acknowledgments
This work was supported by the National Science Foundation of China (11721303, 11991052) and the National Key R\&D Program of China (2016YFA0400702).
This paper makes use of the following ALMA data: ADS/JAO.ALMA $\#$2018.1 00994.
ALMA is a partnership of ESO (representing its member states), NSF (USA) and NINS (Japan), together with NRC (Canada), MOST and ASIAA (Taiwan), and KASI (Republic of Korea), in cooperation with the Republic of Chile. The Joint ALMA Observatory is operated by ESO, AUI/NRAO and NAOJ.
DI is supported by JSPS KAKENHI Grant Number JP18H03725.
The authors appreciate Prof. Masami Ouchi (The University of Tokyo), Prof. Takuya Hashimoto (Tsukuba University), and  Prof. Yoshiyuki Inoue (Osaka University) who provided a comfortable and fruitful research environment during the COVID-19 pandemic time.

\bibliographystyle{aasjournal}
\bibliography{Michiyama_2021_arXiv}

\appendix
\setcounter{section}{0}
\renewcommand{\thesection}{\Alph{section}}
\setcounter{figure}{0}
\renewcommand{\thefigure}{\Alph{section}.\arabic{figure}}
\setcounter{table}{0}
\renewcommand{\thetable}{\Alph{section}.\arabic{table}}
\maxdeadcycles=1000

\section{Appendix}

\clearpage
\subsection{Figures}

\begin{figure}[!htbp]\label{fig:SED}
\figurenum{A1.1}
\begin{center}
\includegraphics[angle=0,scale=0.5]{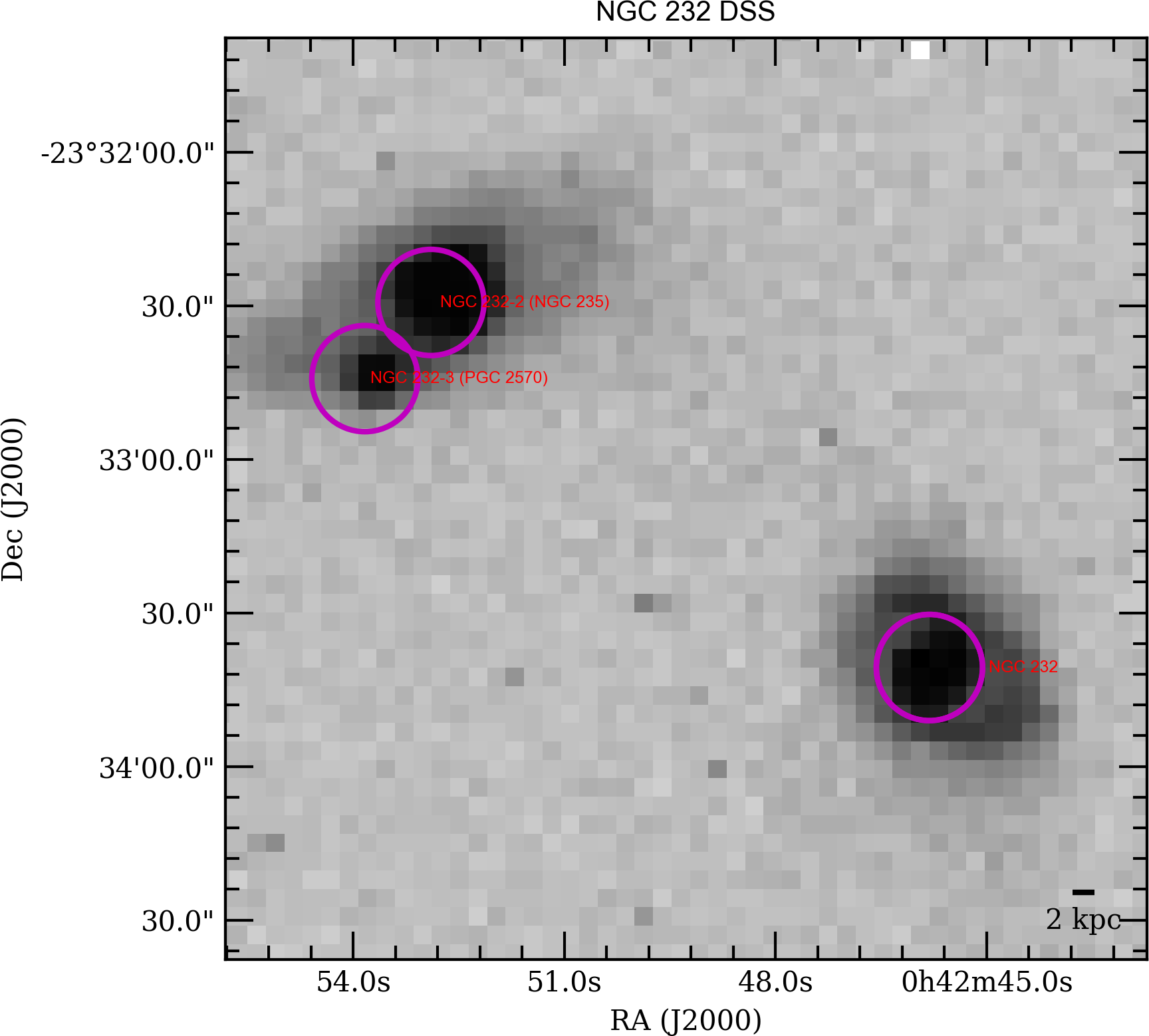}
\caption{The optical image obtained by Digitized Sky Survey (DSS), the fits images are obtained by astroquery.skyview modules in Python. The magenta circle indicates the [C I](1–0) FoV.}
\end{center}
\end{figure}

\begin{figure}[!htbp]
\figurenum{A1.2}
\begin{center}
\includegraphics[angle=0,scale=0.5]{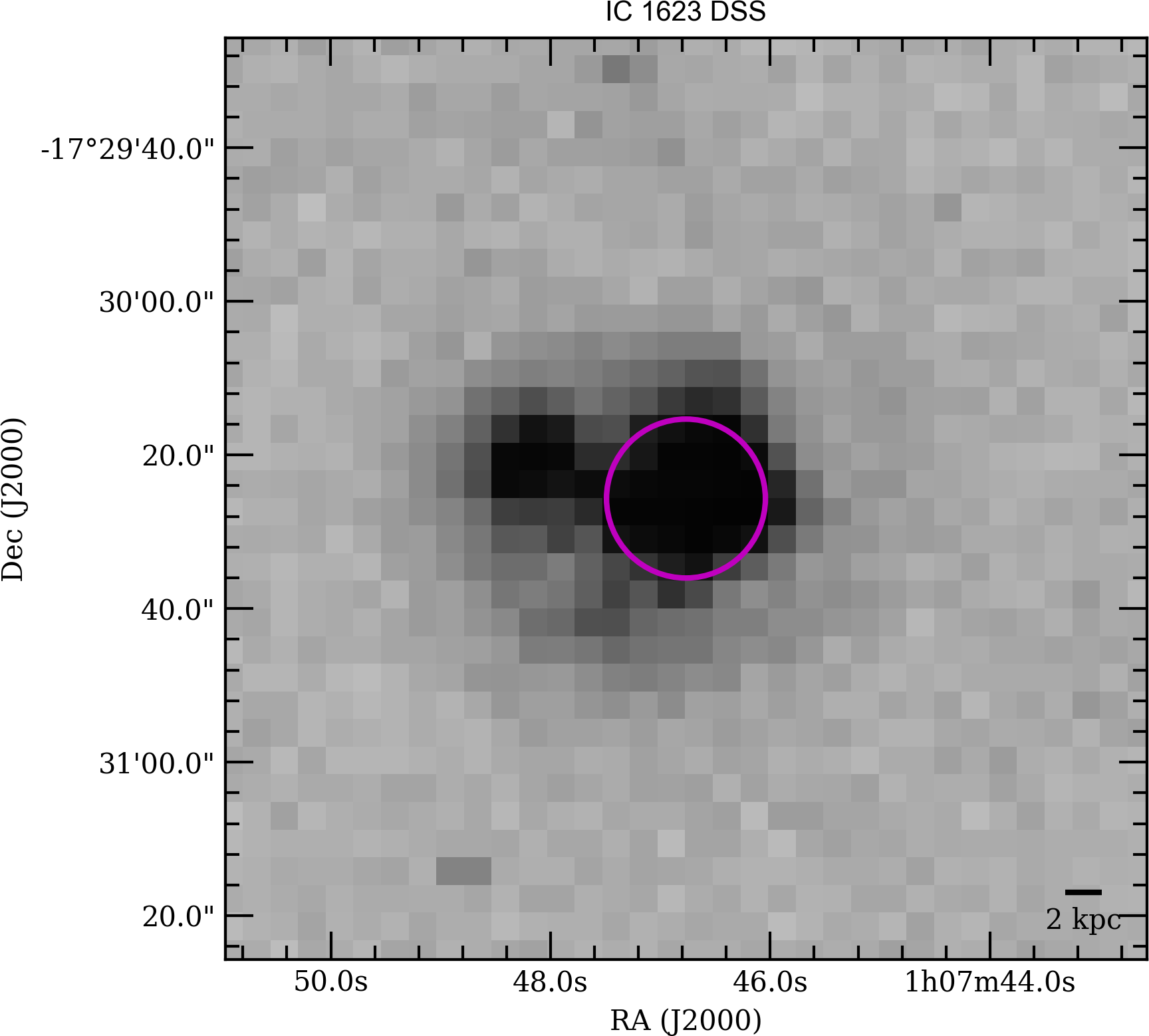}
\caption{Same as Figure A1.1.}
\end{center}
\end{figure}

\begin{figure}[!htbp]
\figurenum{A1.3}
\begin{center}
\includegraphics[angle=0,scale=0.5]{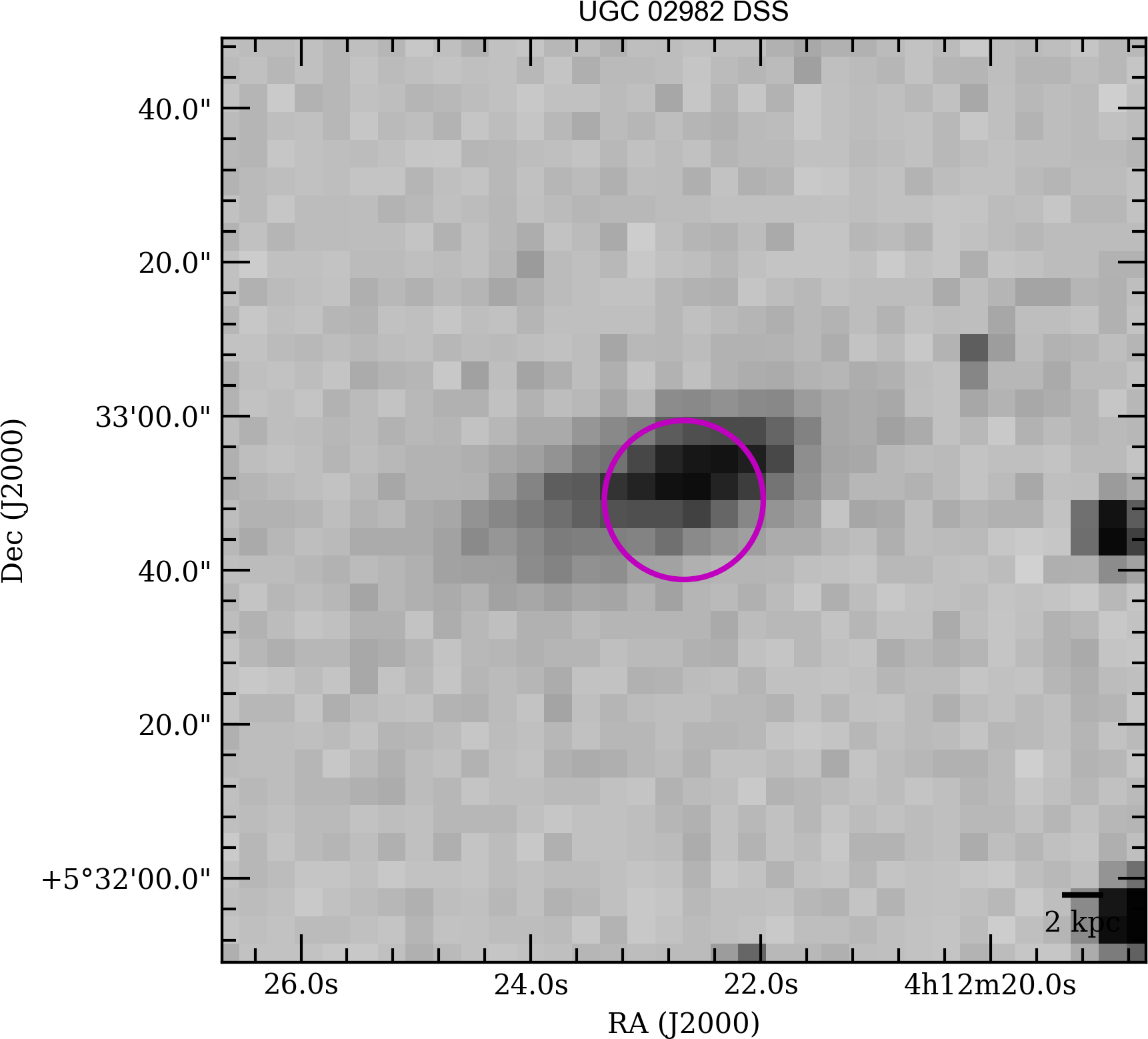}
\caption{Same as Figure A1.1.}
\end{center}
\end{figure}

\begin{figure}[!htbp]
\figurenum{A1.4}
\begin{center}
\includegraphics[angle=0,scale=0.5]{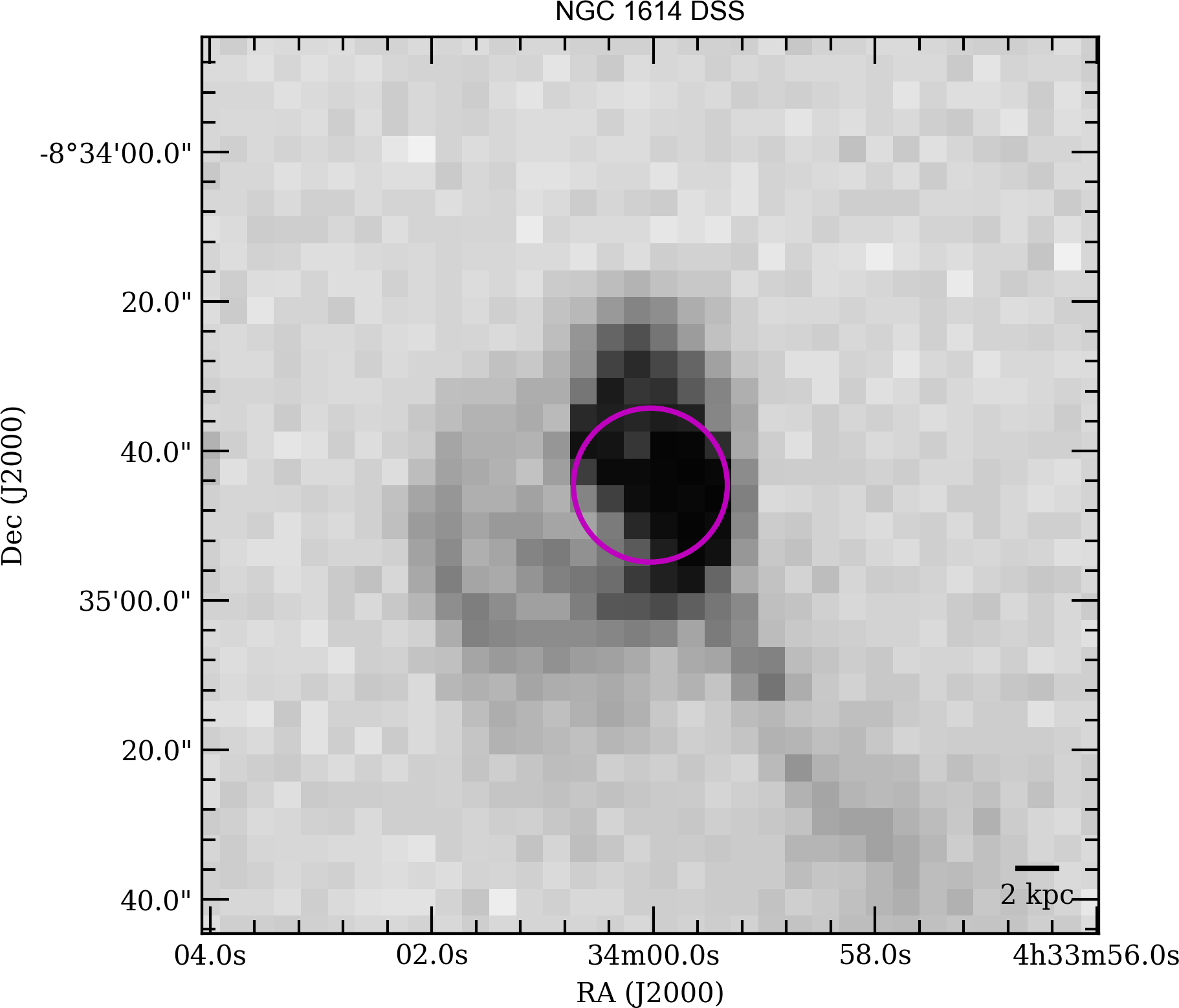}
\caption{Same as Figure A1.1.}
\end{center}
\end{figure}

\begin{figure}[!htbp]
\figurenum{A1.5}
\begin{center}
\includegraphics[angle=0,scale=0.5]{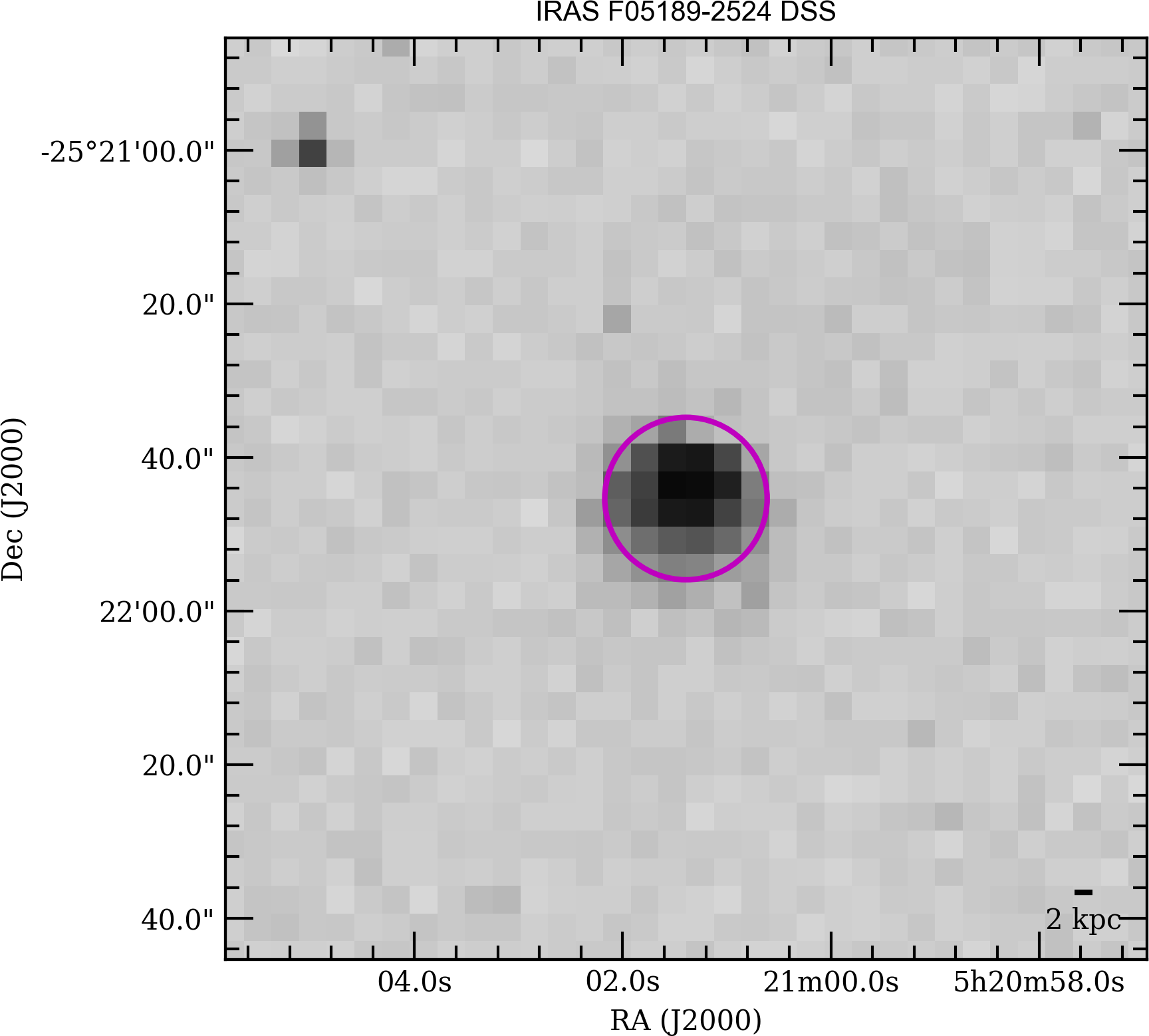}
\caption{Same as Figure A1.1.}
\end{center}
\end{figure}

\begin{figure}[!htbp]
\figurenum{A1.6}
\begin{center}
\includegraphics[angle=0,scale=0.5]{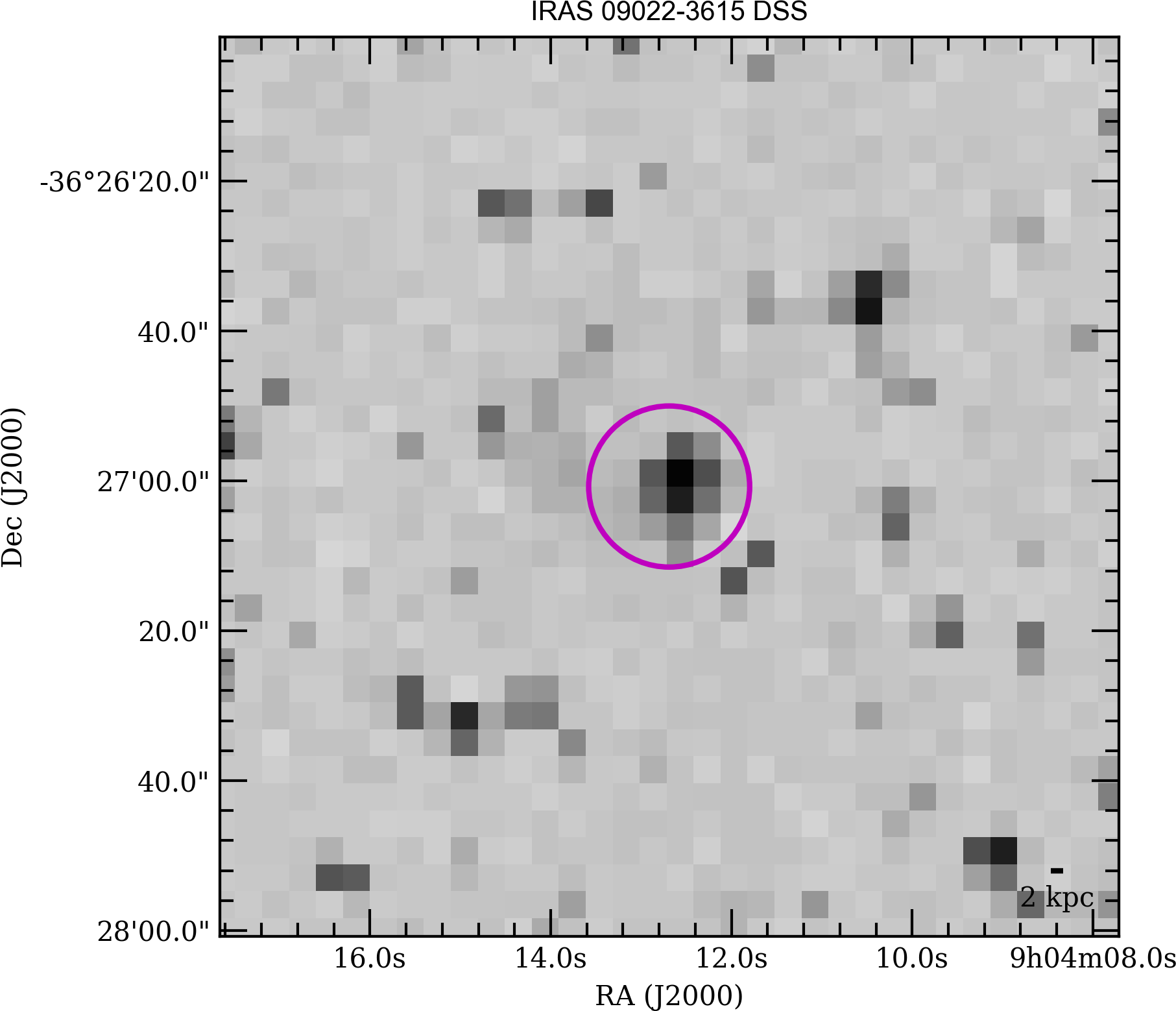}
\caption{Same as Figure A1.1.}
\end{center}
\end{figure}

\begin{figure}[!htbp]
\figurenum{A1.7}
\begin{center}
\includegraphics[angle=0,scale=0.5]{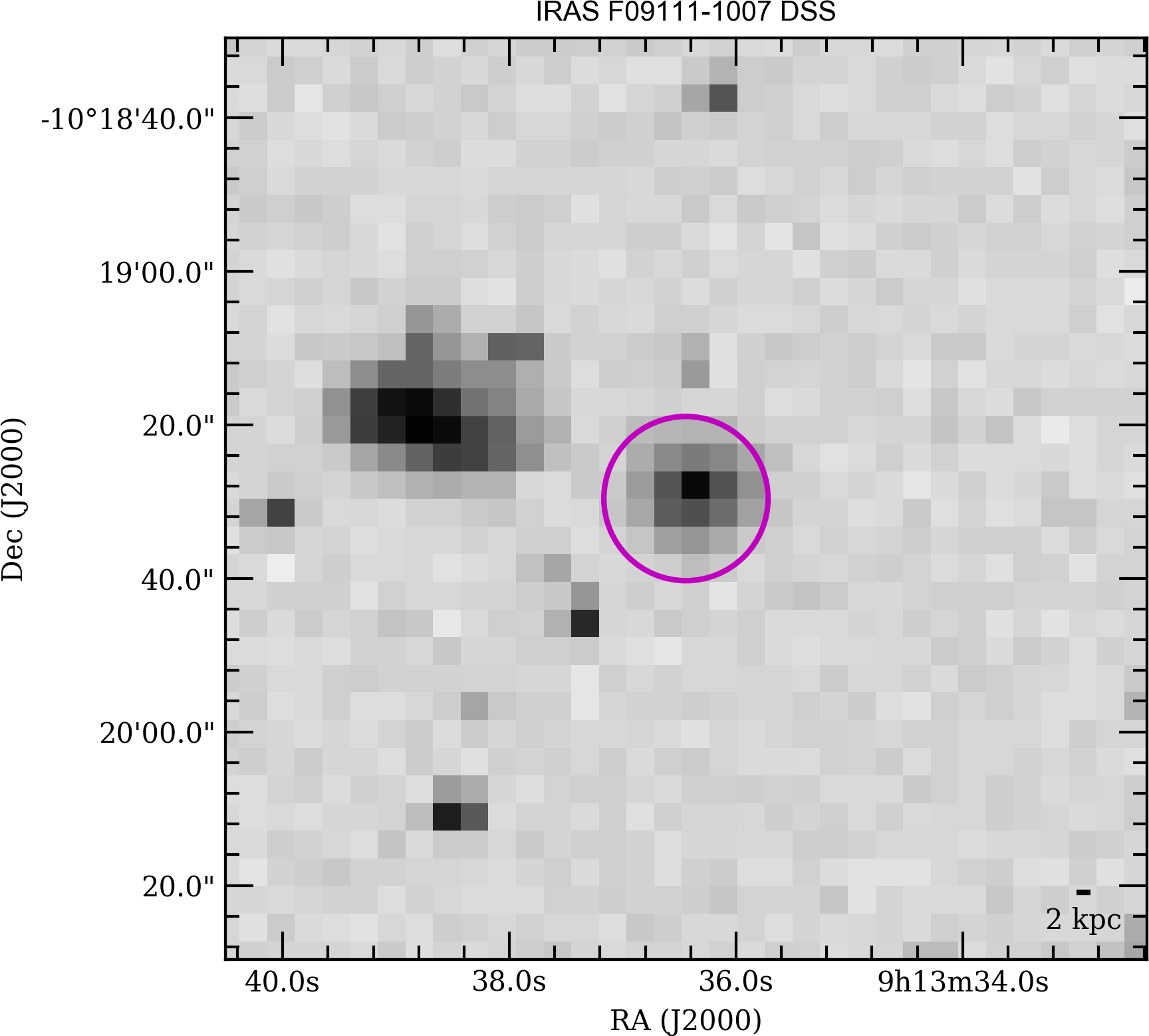}
\caption{Same as Figure A1.1.}
\end{center}
\end{figure}

\begin{figure}[!htbp]
\figurenum{A1.8}
\begin{center}
\includegraphics[angle=0,scale=0.5]{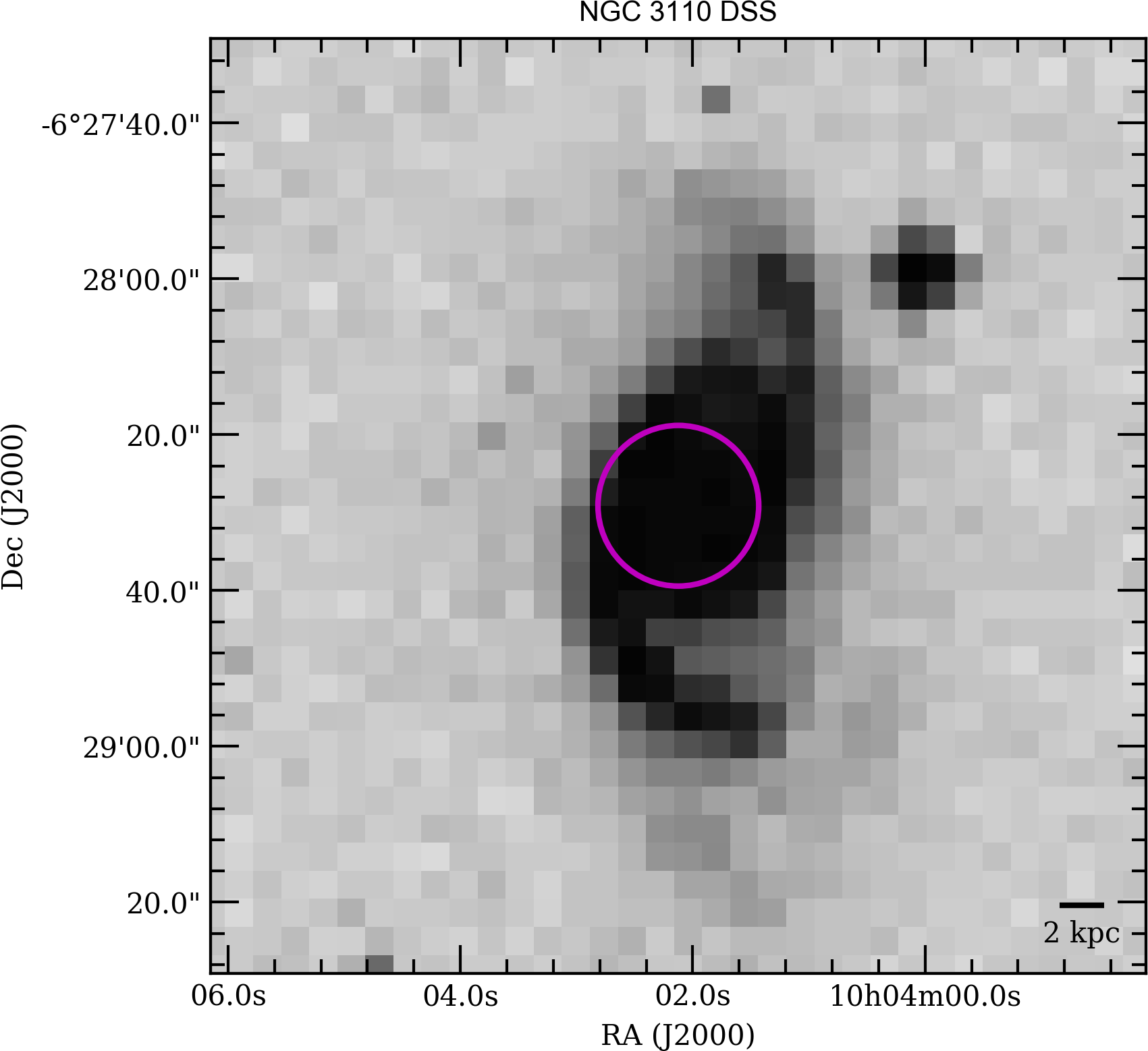}
\caption{Same as Figure A1.1.}
\end{center}
\end{figure}

\begin{figure}[!htbp]
\figurenum{A1.9}
\begin{center}
\includegraphics[angle=0,scale=0.5]{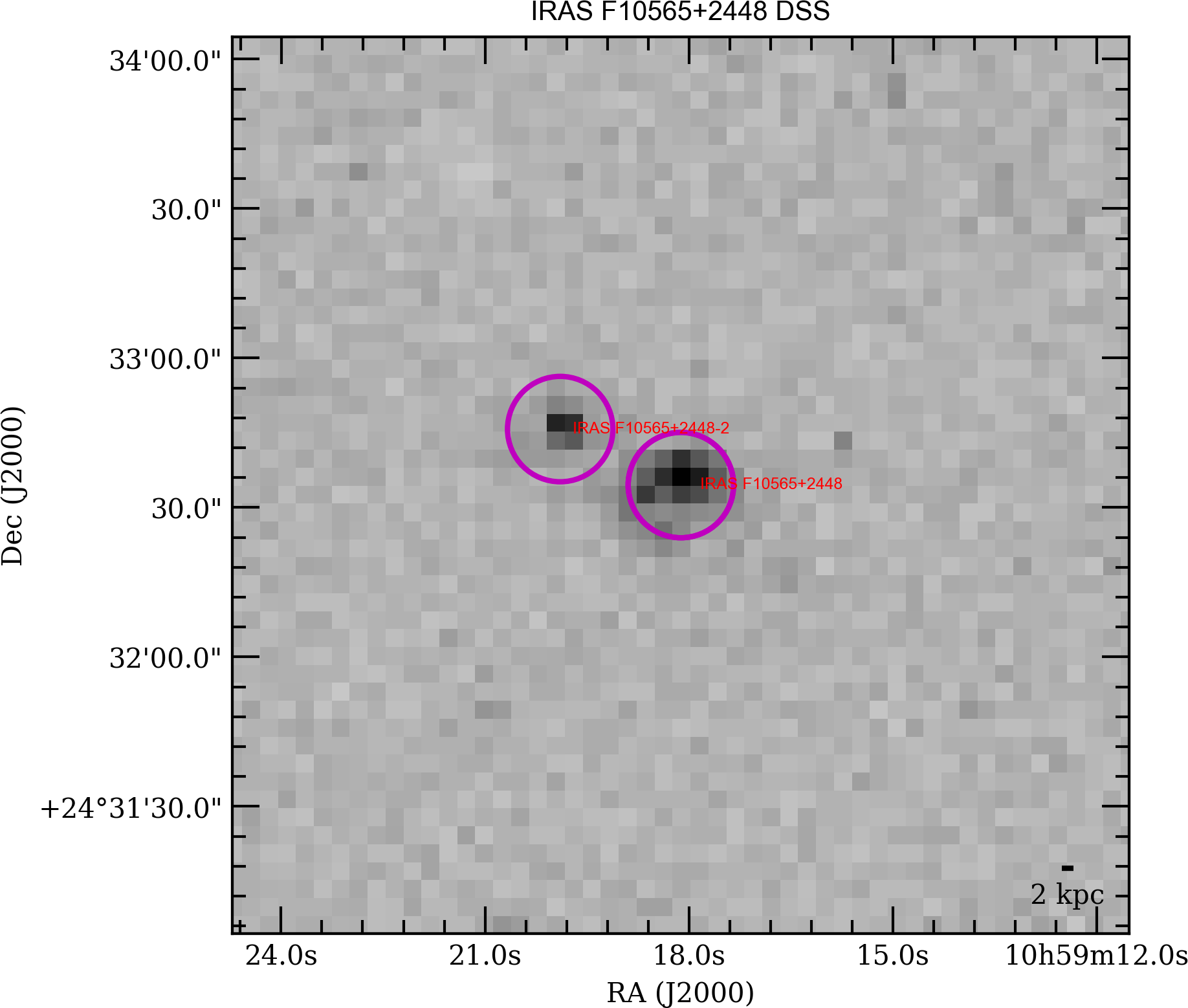}
\caption{Same as Figure A1.1.}
\end{center}
\end{figure}

\begin{figure}[!htbp]
\figurenum{A1.10}
\begin{center}
\includegraphics[angle=0,scale=0.5]{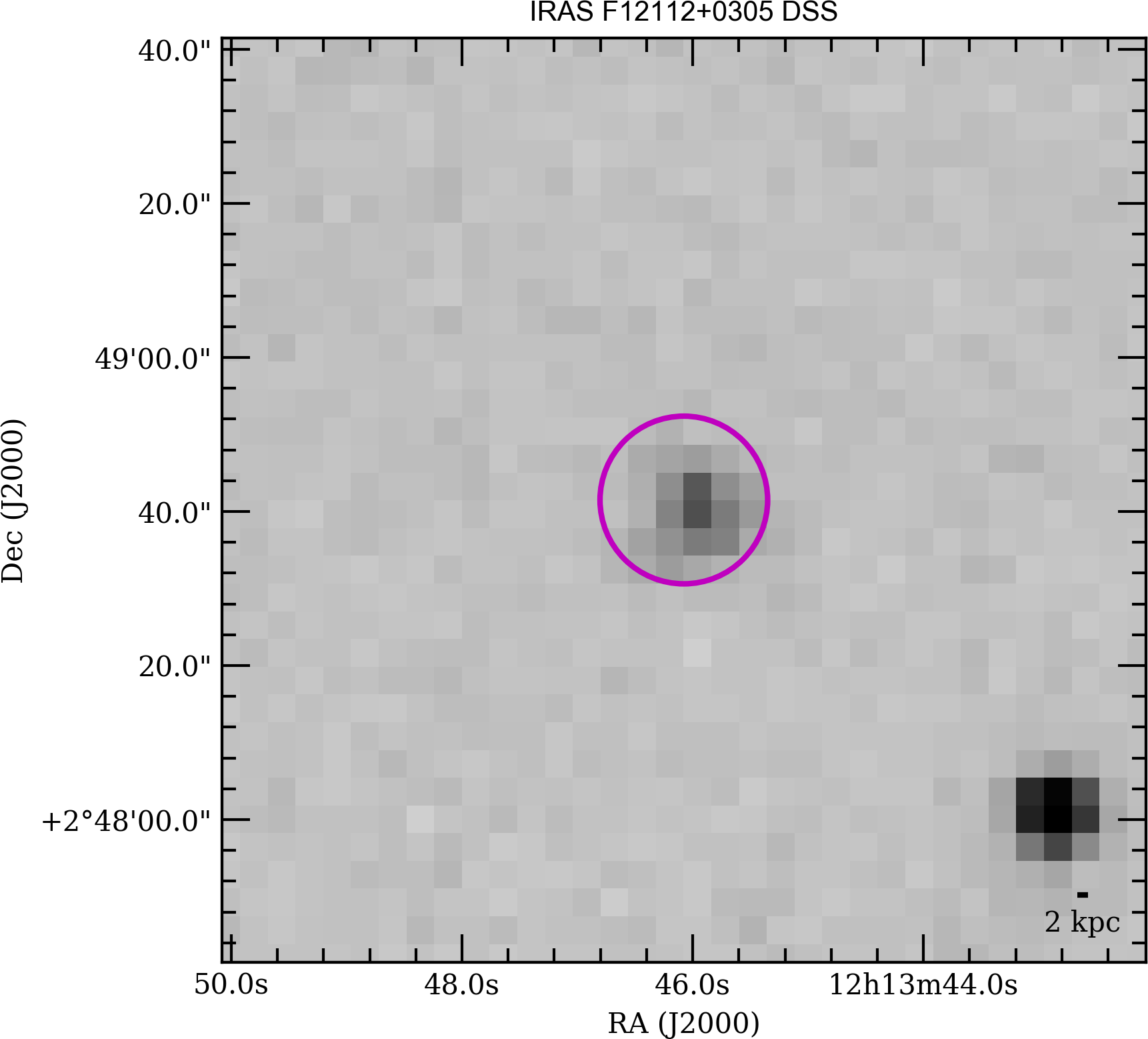}
\caption{Same as Figure A1.1.}
\end{center}
\end{figure}

\begin{figure}[!htbp]
\figurenum{A1.11}
\begin{center}
\includegraphics[angle=0,scale=0.5]{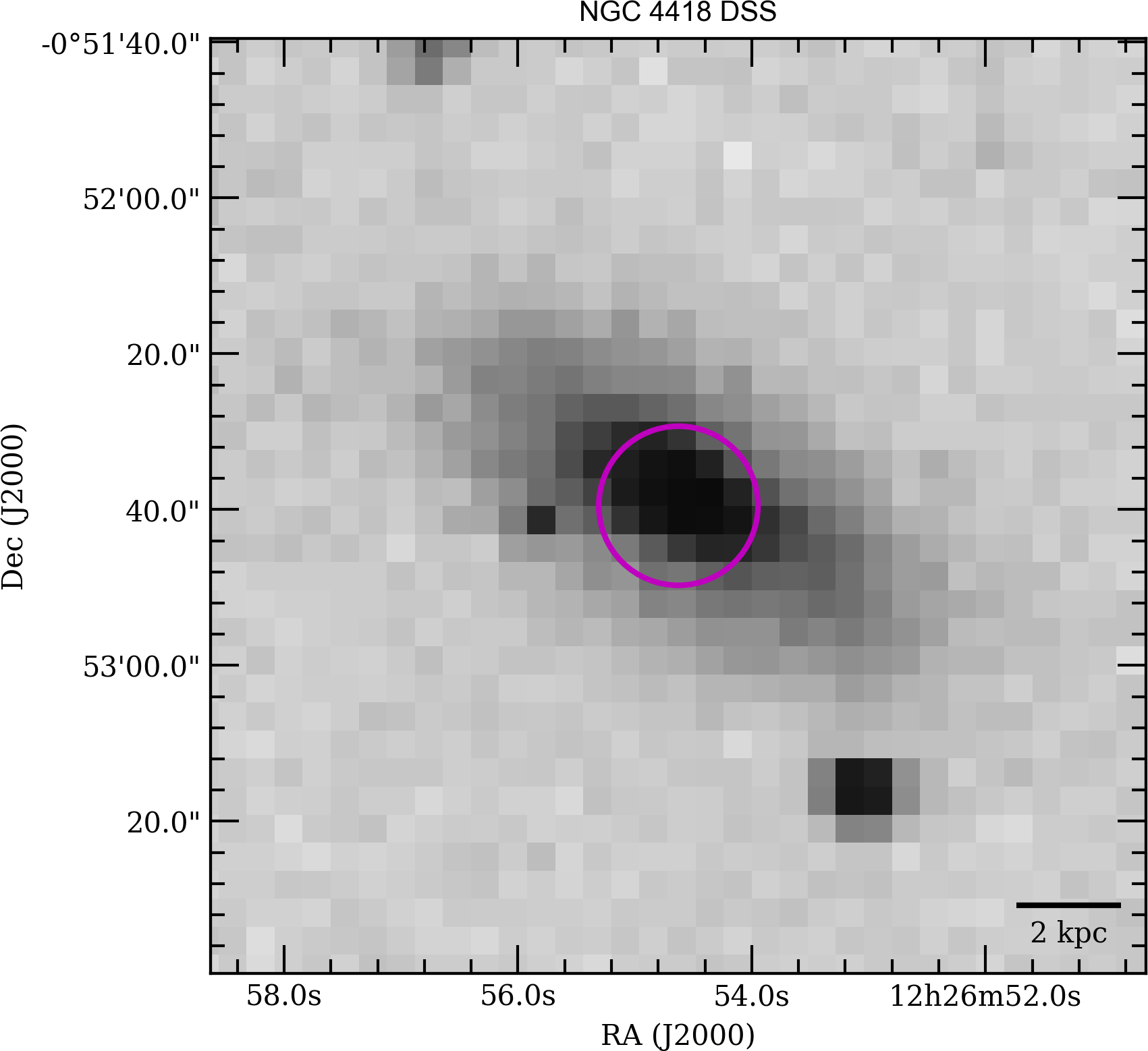}
\caption{Same as Figure A1.1.}
\end{center}
\end{figure}

\begin{figure}[!htbp]
\figurenum{A1.12}
\begin{center}
\includegraphics[angle=0,scale=0.5]{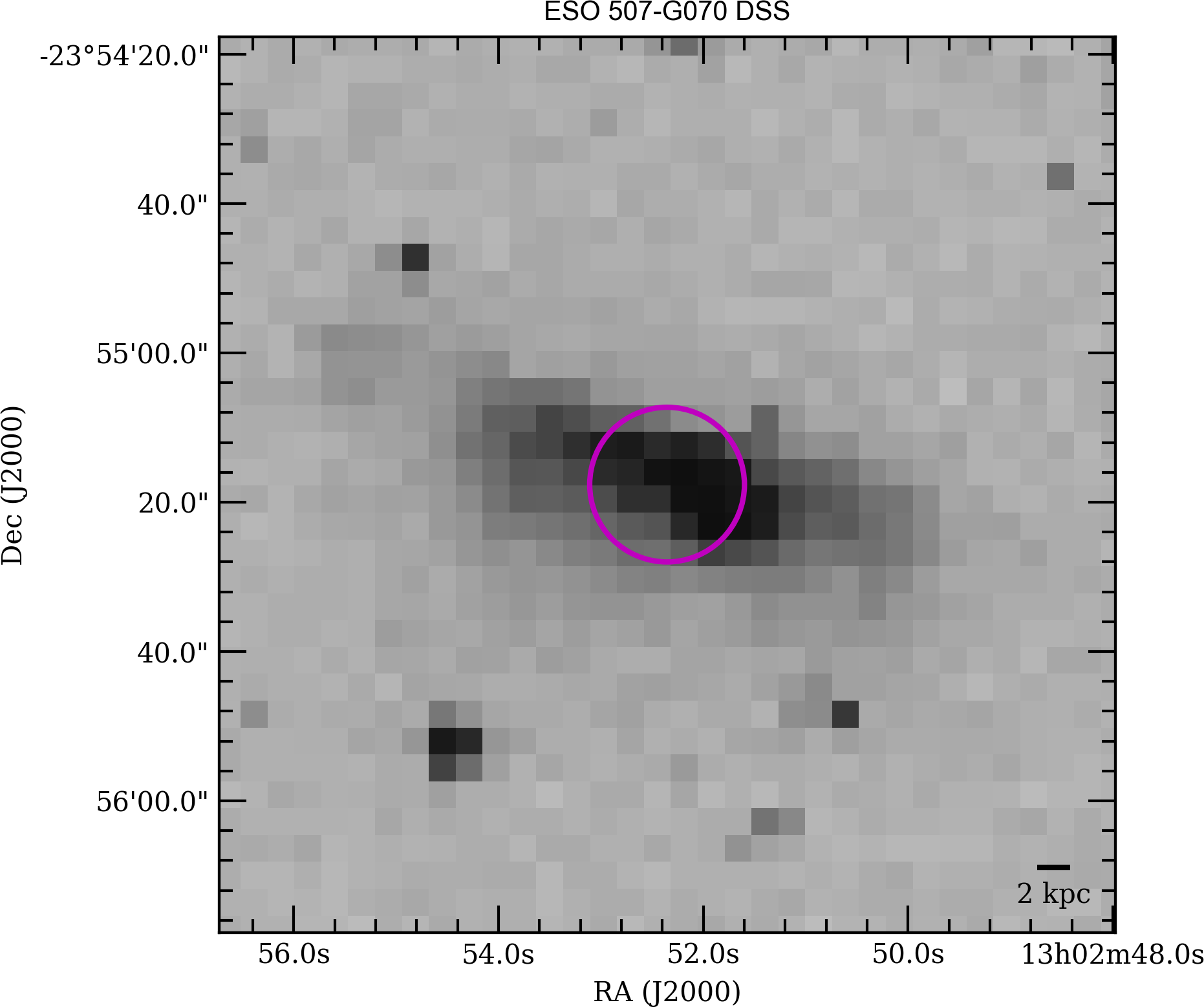}
\caption{Same as Figure A1.1.}
\end{center}
\end{figure}

\begin{figure}[!htbp]
\figurenum{A1.13}
\begin{center}
\includegraphics[angle=0,scale=0.5]{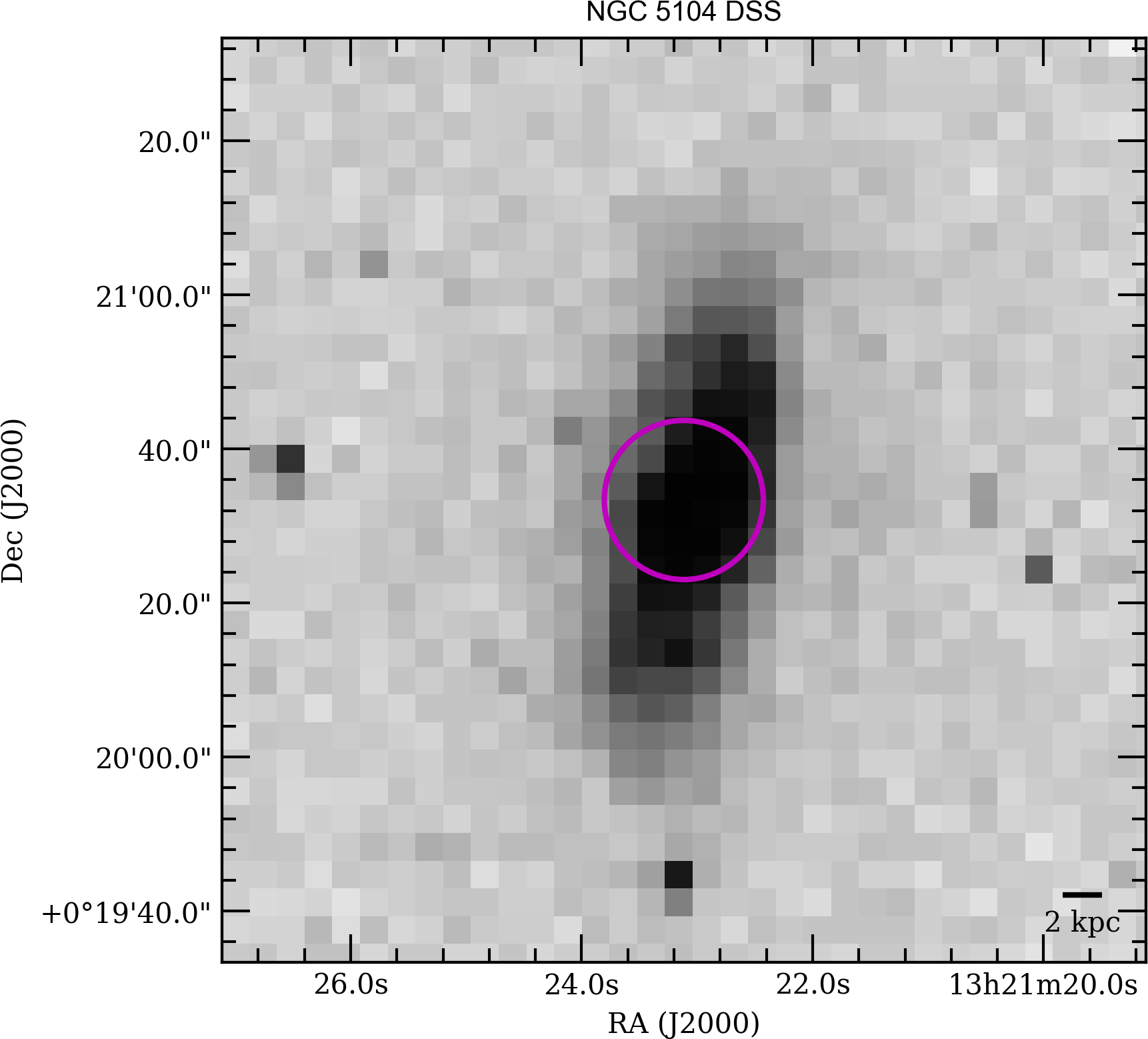}
\caption{Same as Figure A1.1.}
\end{center}
\end{figure}

\begin{figure}[!htbp]
\figurenum{A1.14}
\begin{center}
\includegraphics[angle=0,scale=0.5]{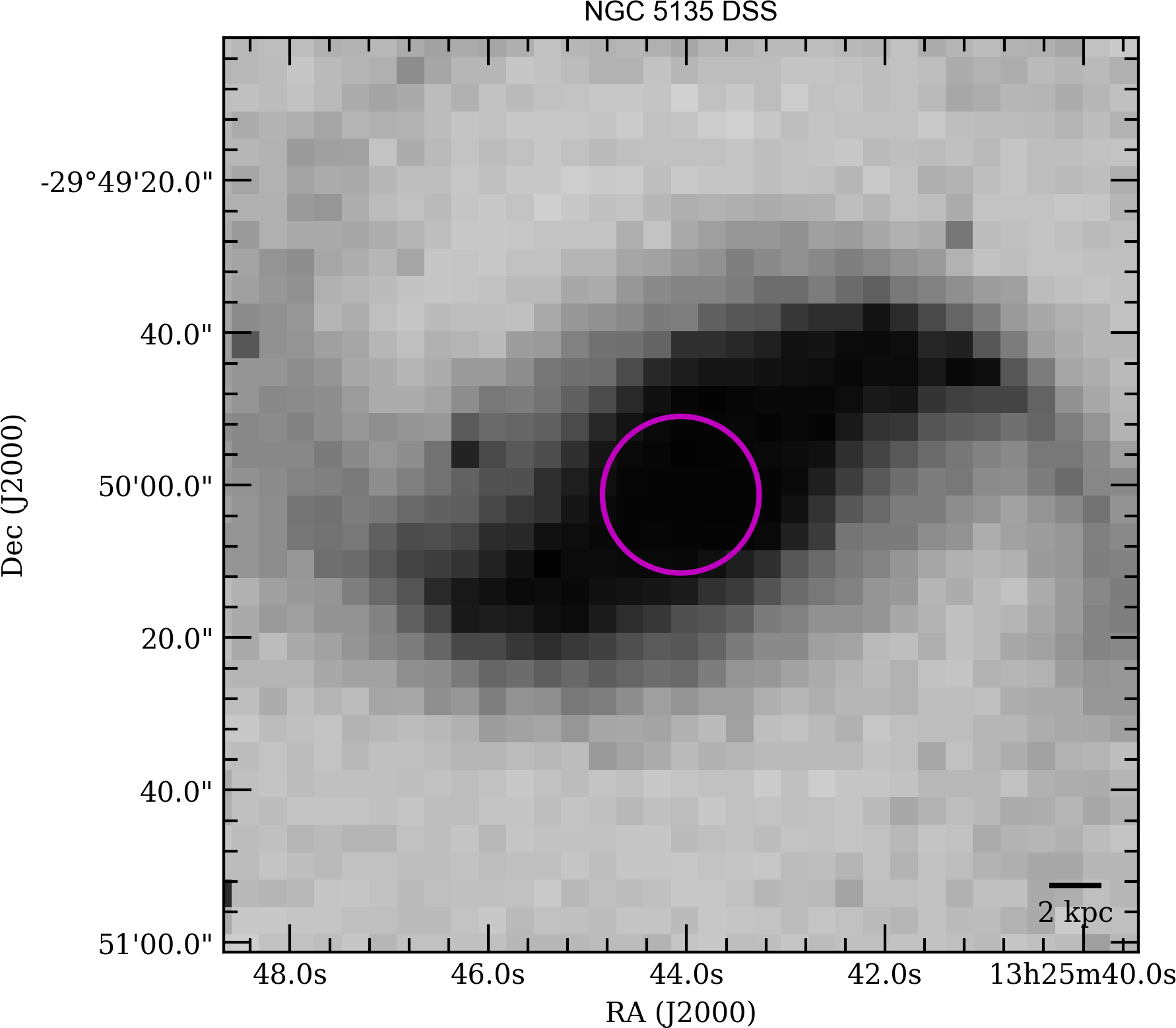}
\caption{Same as Figure A1.1.}
\end{center}
\end{figure}

\begin{figure}[!htbp]
\figurenum{A1.15}
\begin{center}
\includegraphics[angle=0,scale=0.5]{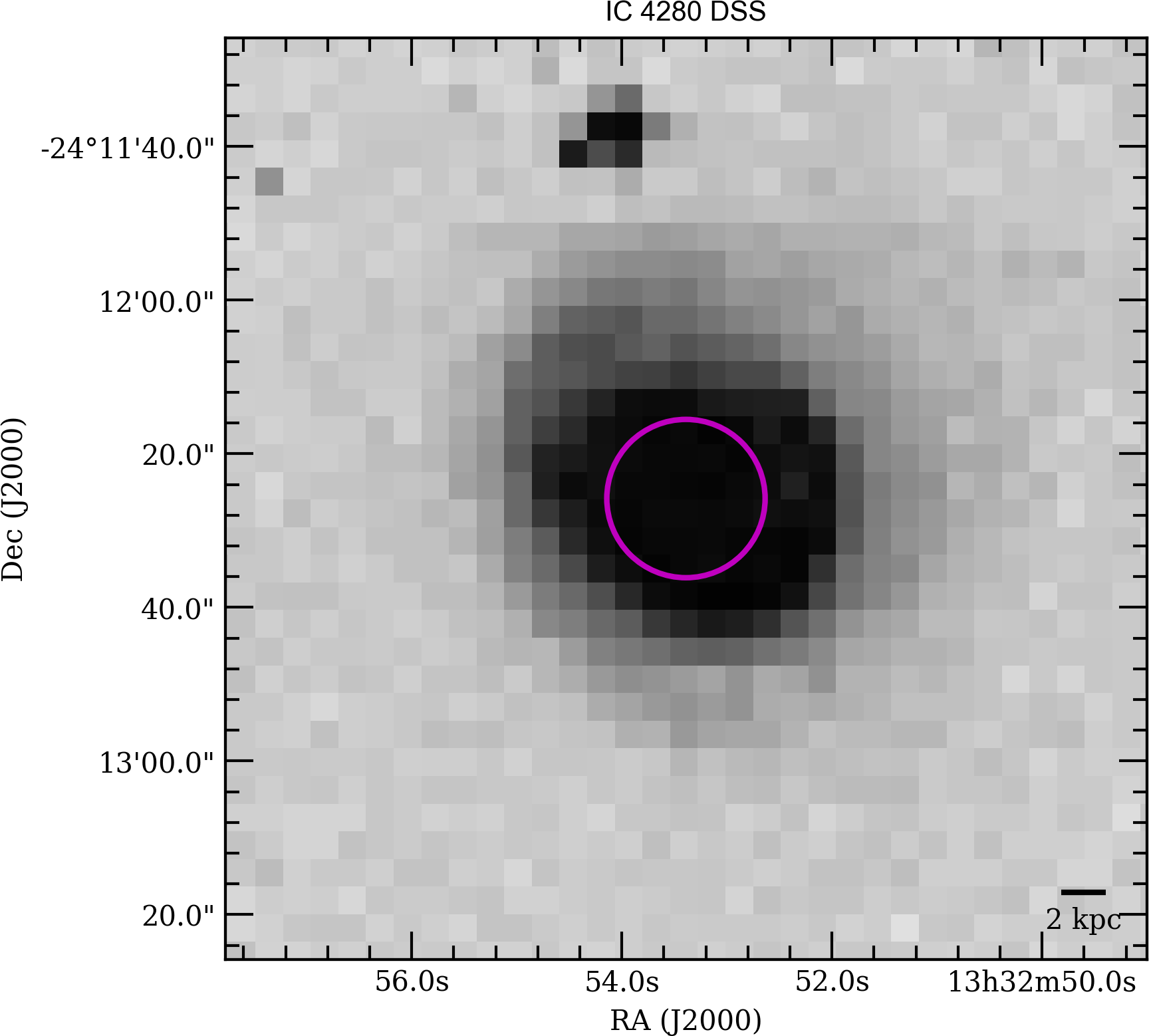}
\caption{Same as Figure A1.1.}
\end{center}
\end{figure}

\clearpage
\begin{figure*}[!htbp]\label{fig:mom}
\figurenum{A2.1}
\begin{center}
\includegraphics[angle=0,scale=0.4]{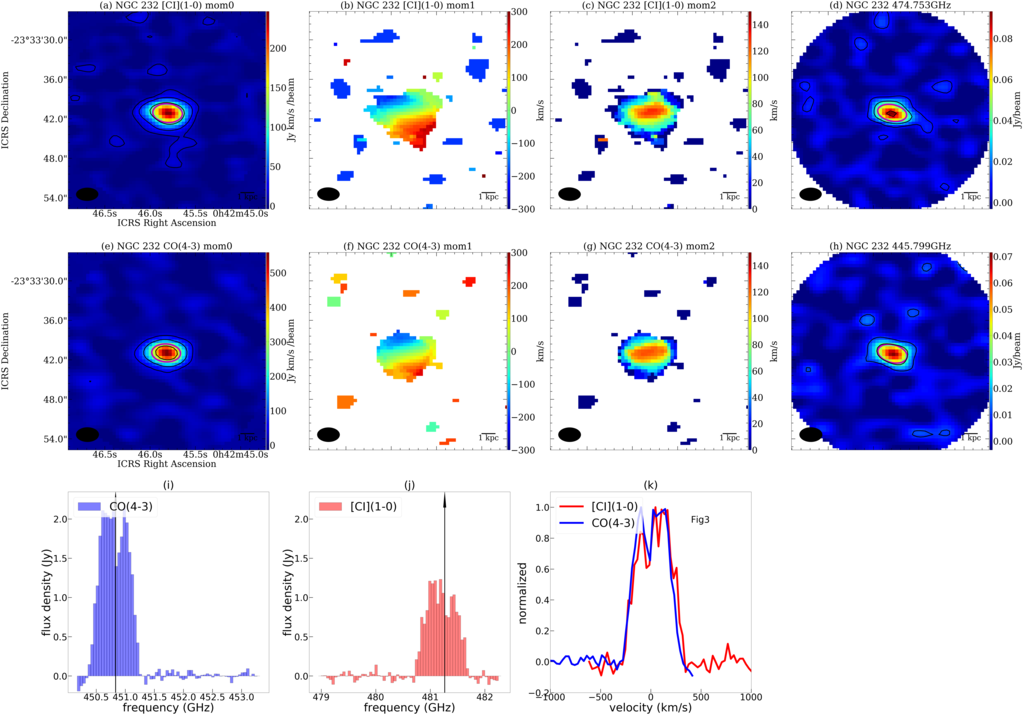}
\caption{(a) The velocity integrated intensity map (mom0) of [\ion{C}{1}]~(1--0) emission line. The $n$th contours are at $3n\sigma$ where $\sigma$ is the rms revel of the map shown in Table~\ref{tab:obs_res}. The magenta circle indicates the ACA FoV calculated by equation~\ref{eq:FoV}.
(b) The velocity field map (mom1) of [\ion{C}{1}]~(1--0) emission line.
(c) The velocity dispersion map (mom2) of [\ion{C}{1}]~(1--0) emission line.
The mom1 and mom2 maps are made by clipping 3$\sigma$ emission in data cube.
(d) The continuum emission map associated with [\ion{C}{1}]~(1--0) observation. The $n$th contours are at $3n\sigma$ where $\sigma$ is the rms revel of the map shown in Table~\ref{tab:obs_res}. 
(e)-(h) The same map with (a)-(d) but for CO~(4--3) emission.  
(i) and (j) The CO~(4--3) and [\ion{C}{1}]~(1--0) spectrum with the aperture of FoV.
The spectrum were derived from ${\tt specflux}$ task in ${\tt CASA}$.
(k) The normalized CO~(4--3) and [\ion{C}{1}]~(1--0) spectrum. The velocity is calculated as the offset from the systematic velocity shown in Table~\ref{tab:target}.
In the case of CO~(4--3) detection and [\ion{C}{1}]~(1--0) non-detection,  [\ion{C}{1}]~(1--0) moment maps are made by assuming same velocity range for integration.
If both CO~(4--3) and [\ion{C}{1}]~(1--0) are non-detection, we integrated $\pm200$~km~s$^{-1}$ from systematic velocities.}
\end{center}
\end{figure*}

\begin{figure*}[!htbp]
\figurenum{A2.2}
\begin{center}
\includegraphics[angle=0,scale=0.4]{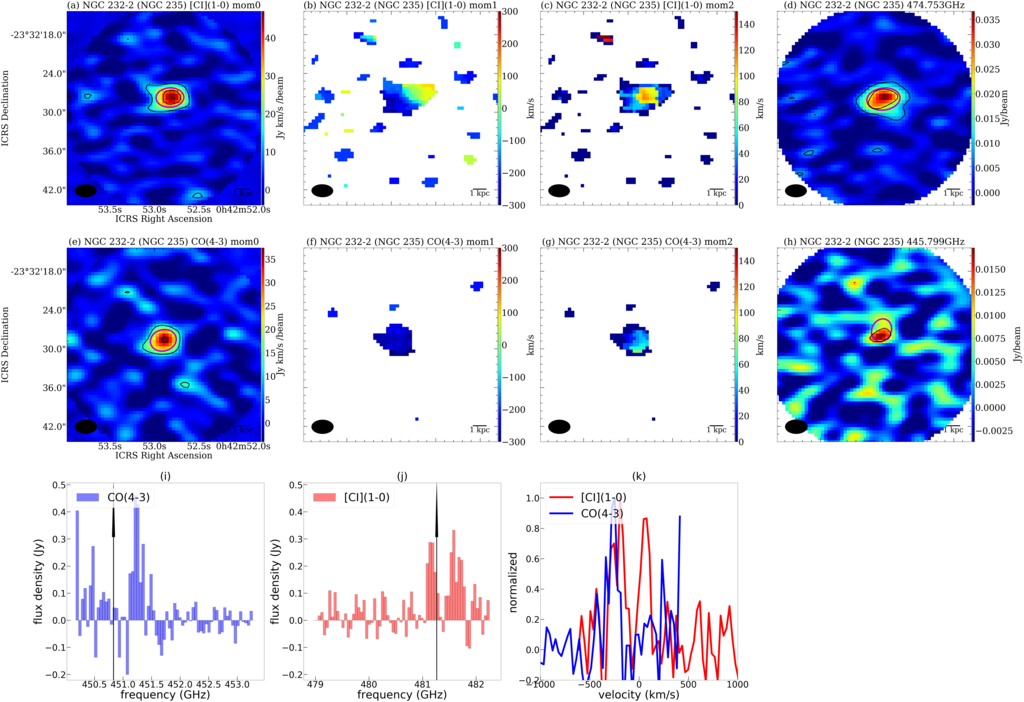}
\caption{Same as Figure A2.1.}
\end{center}
\end{figure*}

\begin{figure*}[!htbp]
\figurenum{A2.3}
\begin{center}
\includegraphics[angle=0,scale=0.4]{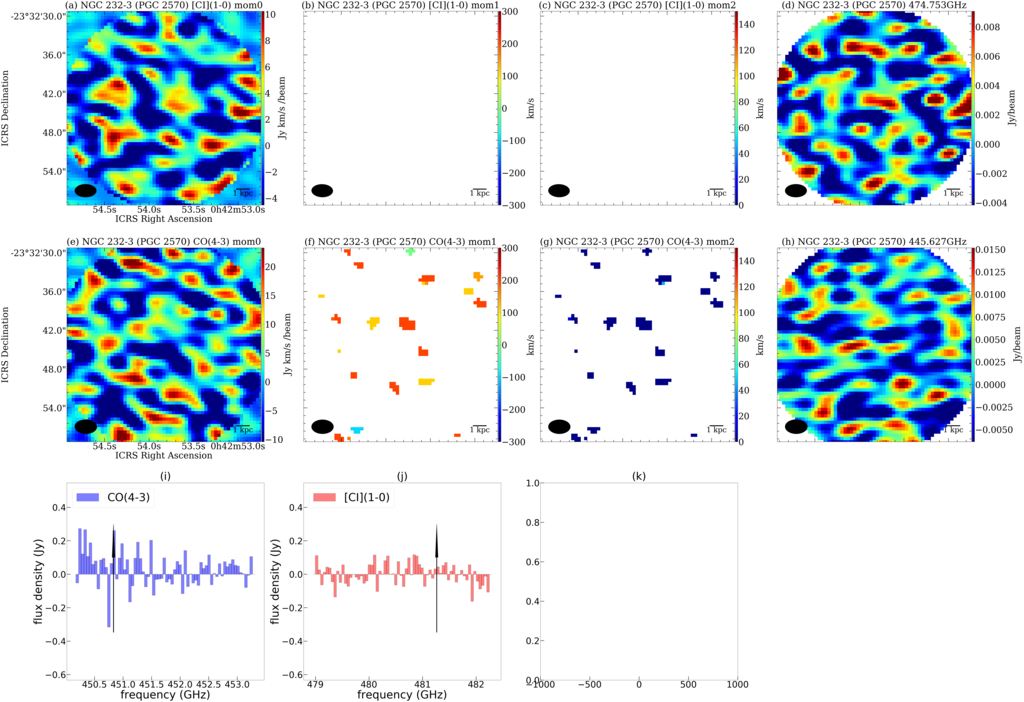}
\caption{Same as Figure A2.1.}
\end{center}
\end{figure*}

\begin{figure*}[!htbp]
\figurenum{A2.4}
\begin{center}
\includegraphics[angle=0,scale=0.4]{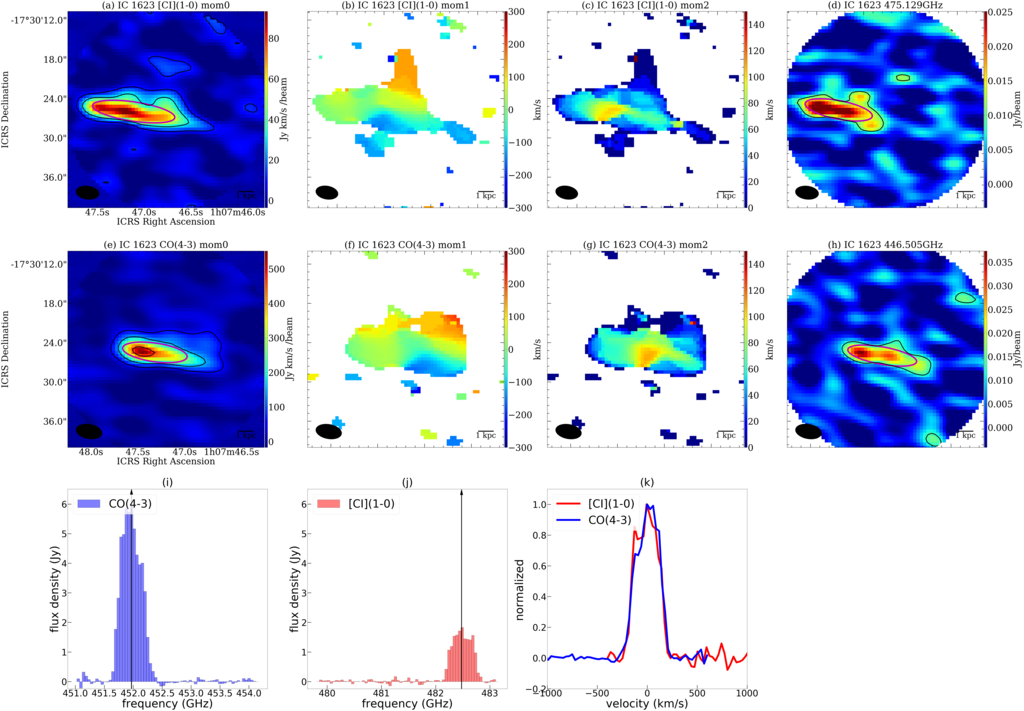}
\caption{Same as Figure A2.1.}
\end{center}
\end{figure*}

\begin{figure*}[!htbp]
\figurenum{A2.5}
\begin{center}
\includegraphics[angle=0,scale=0.4]{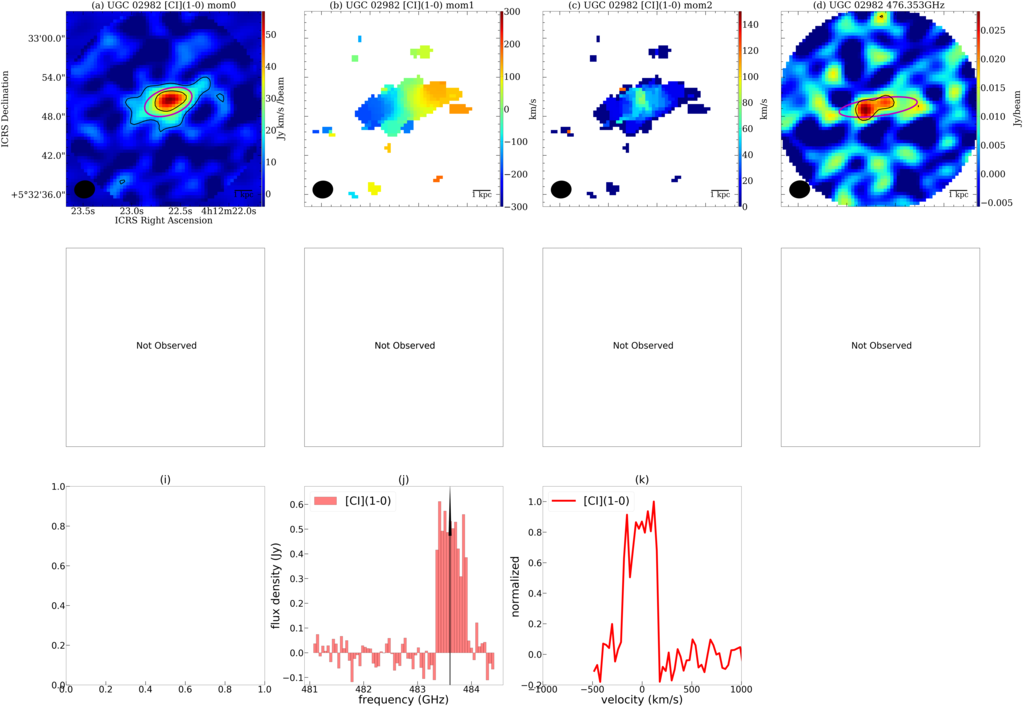}
\caption{Same as Figure A2.1.}
\end{center}
\end{figure*}

\begin{figure*}[!htbp]
\figurenum{A2.6}
\begin{center}
\includegraphics[angle=0,scale=0.4]{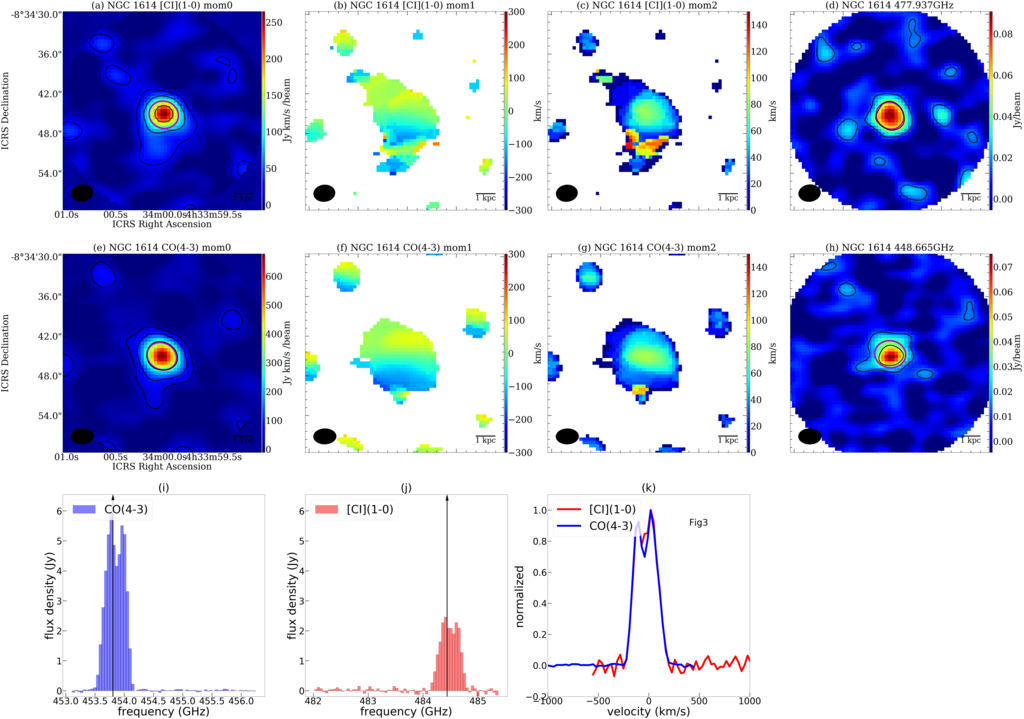}
\caption{Same as Figure A2.1.}
\end{center}
\end{figure*}

\begin{figure*}[!htbp]
\figurenum{A2.7}
\begin{center}
\includegraphics[angle=0,scale=0.4]{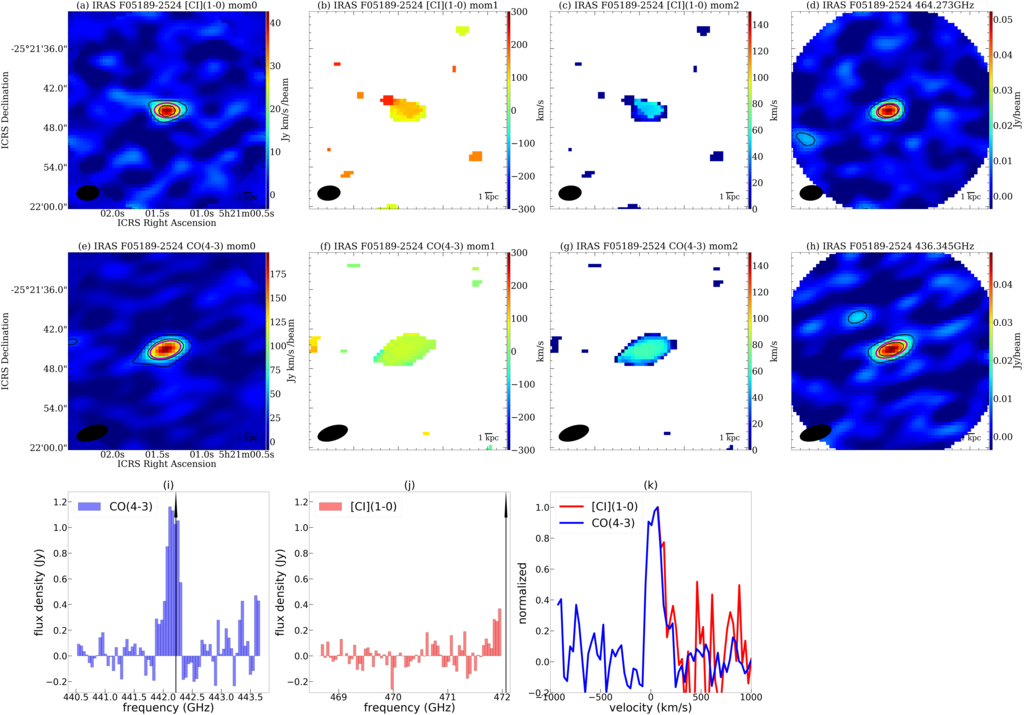}
\caption{Same as Figure A2.1.}
\end{center}
\end{figure*}

\begin{figure*}[!htbp]
\figurenum{A2.8}
\begin{center}
\includegraphics[angle=0,scale=0.4]{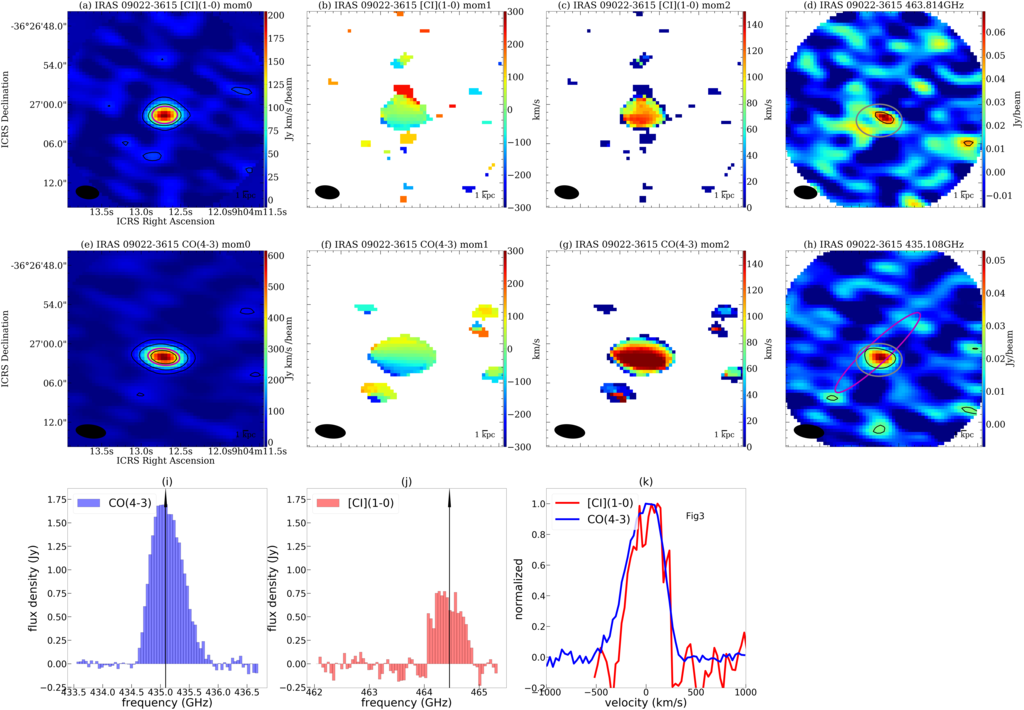}
\caption{Same as Figure A2.1.}
\end{center}
\end{figure*}

\begin{figure*}[!htbp]
\figurenum{A2.9}
\begin{center}
\includegraphics[angle=0,scale=0.4]{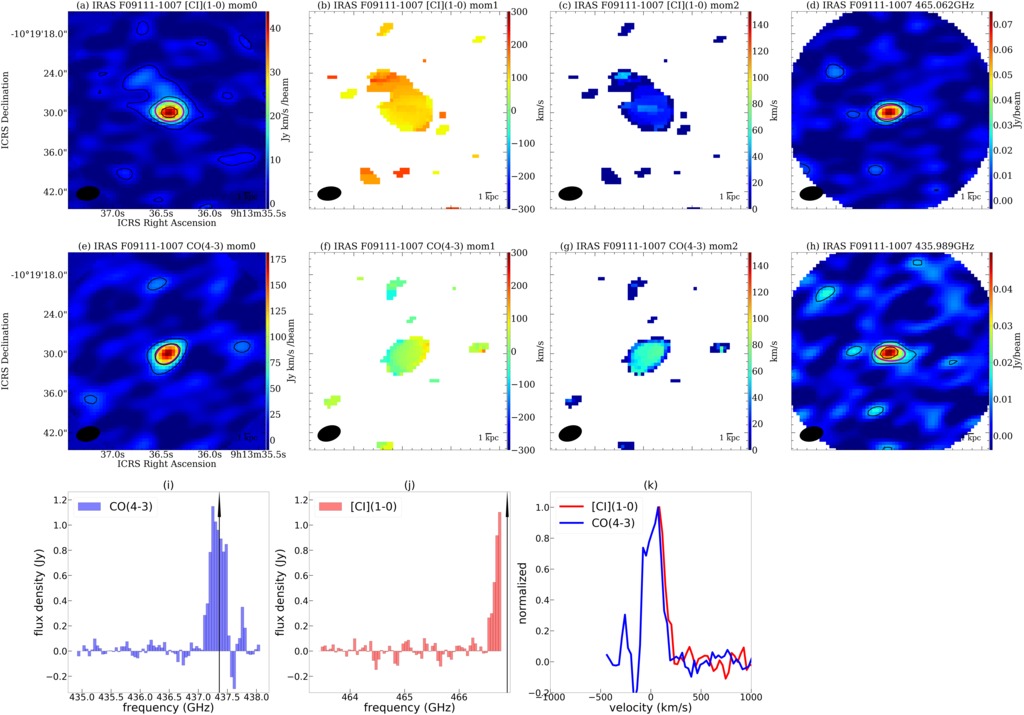}
\caption{Same as Figure A2.1.}
\end{center}
\end{figure*}

\begin{figure*}[!htbp]
\figurenum{A2.10}
\begin{center}
\includegraphics[angle=0,scale=0.4]{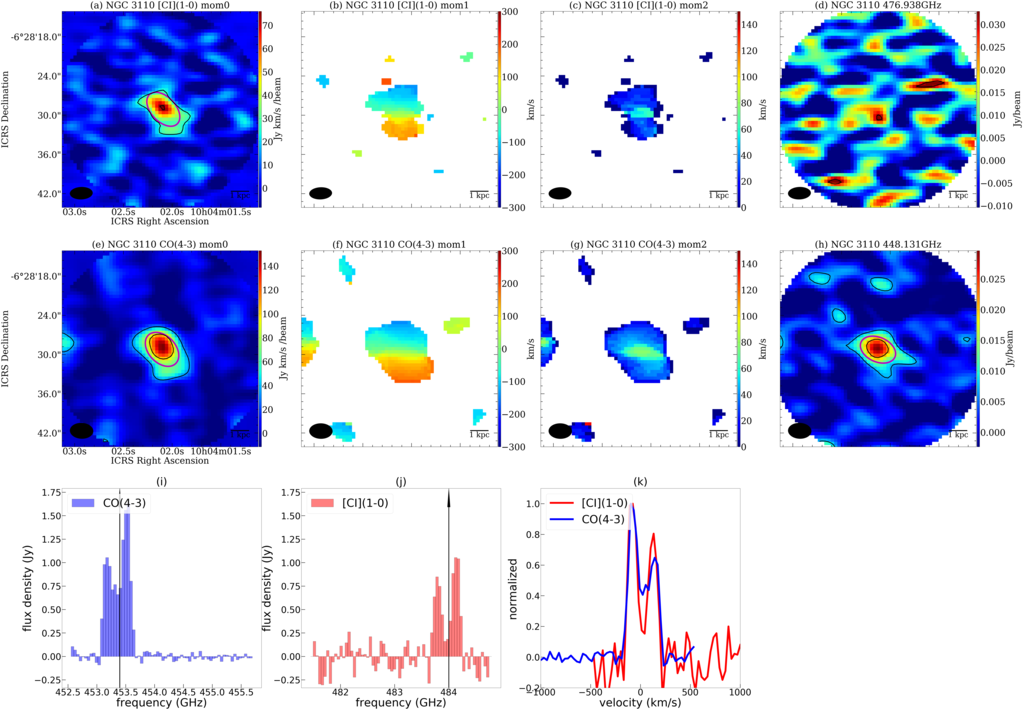}
\caption{Same as Figure A2.1.}
\end{center}
\end{figure*}

\begin{figure*}[!htbp]
\figurenum{A2.11}
\begin{center}
\includegraphics[angle=0,scale=0.4]{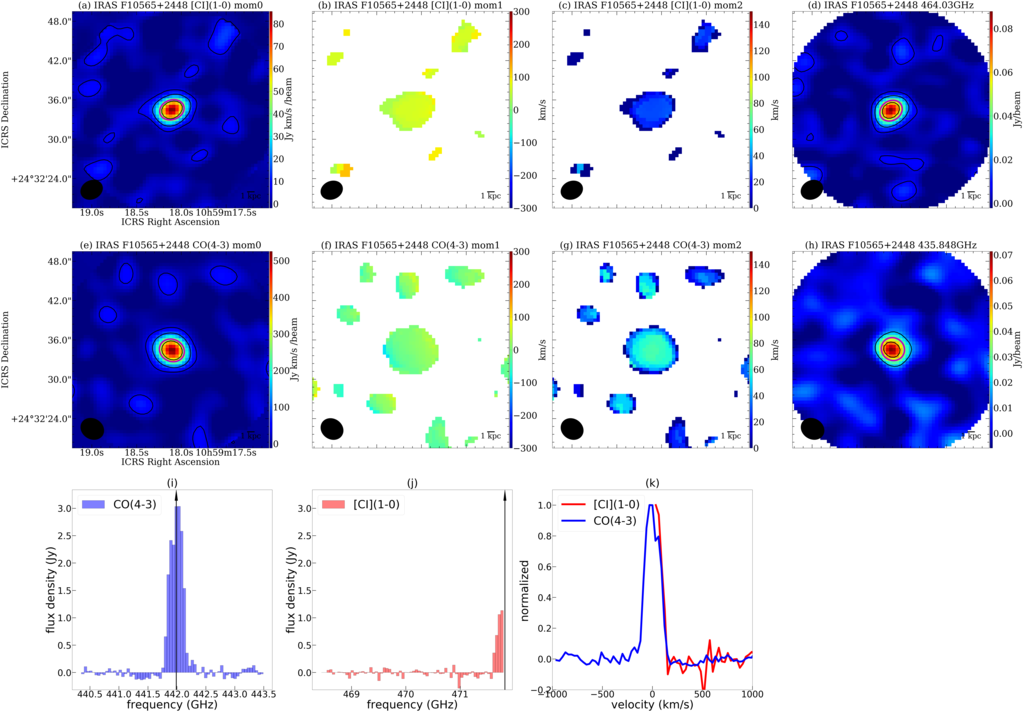}
\caption{Same as Figure A2.1.}
\end{center}
\end{figure*}

\begin{figure*}[!htbp]
\figurenum{A2.12}
\begin{center}
\includegraphics[angle=0,scale=0.4]{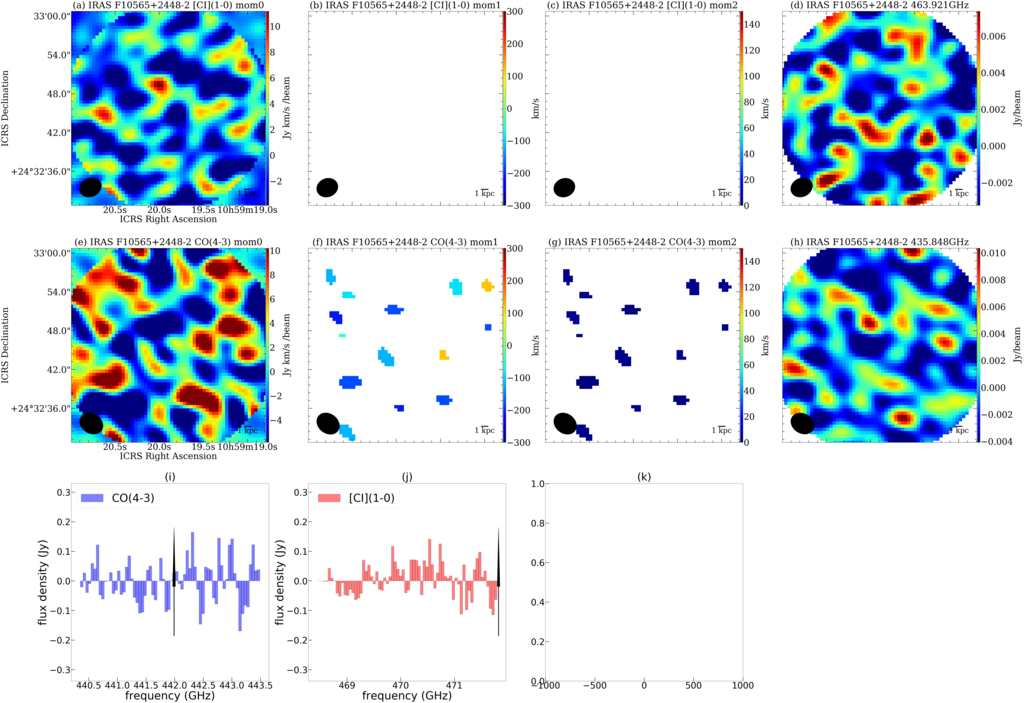}
\caption{Same as Figure A2.1.}
\end{center}
\end{figure*}

\begin{figure*}[!htbp]
\figurenum{A2.13}
\begin{center}
\includegraphics[angle=0,scale=0.4]{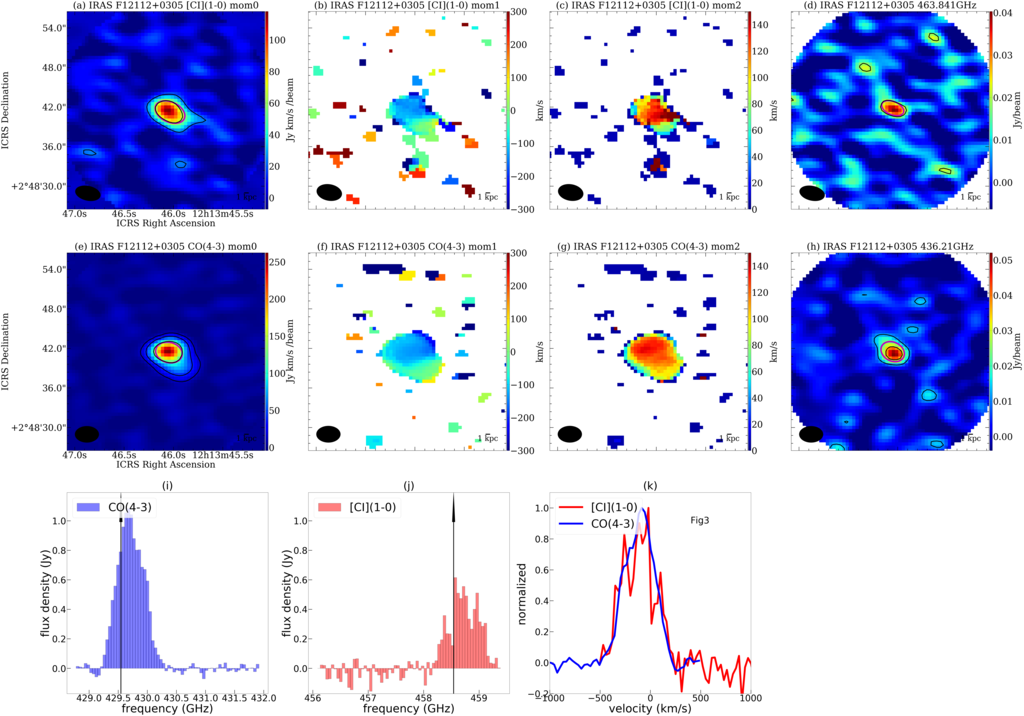}
\caption{Same as Figure A2.1.}
\end{center}
\end{figure*}

\begin{figure*}[!htbp]
\figurenum{A2.14}
\begin{center}
\includegraphics[angle=0,scale=0.4]{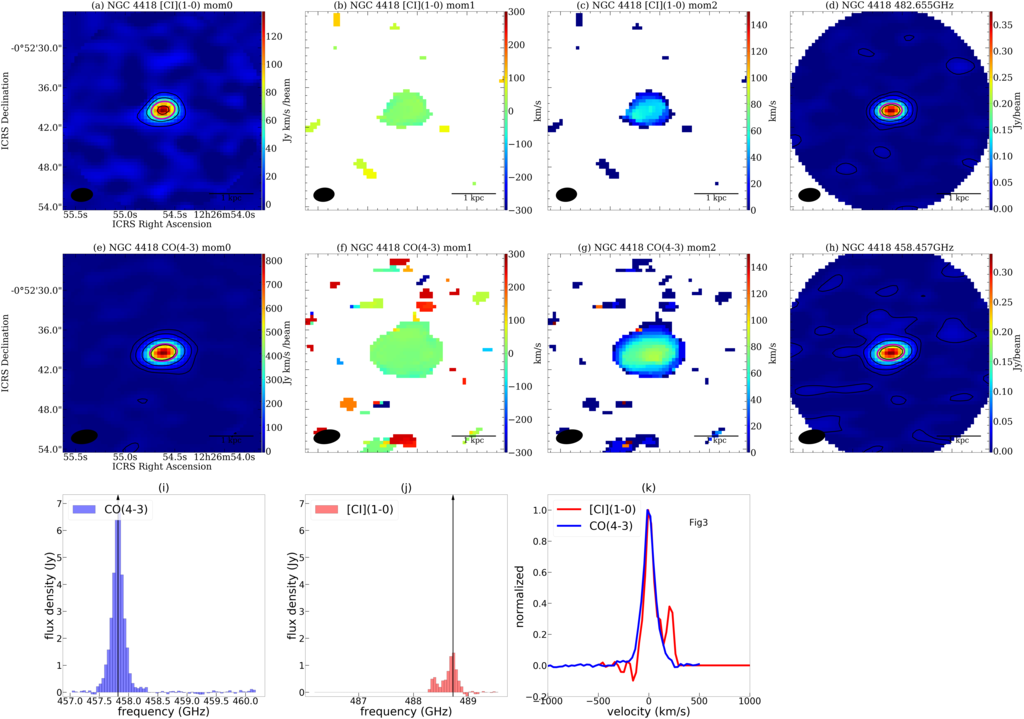}
\caption{Same as Figure A2.1.}
\end{center}
\end{figure*}

\begin{figure*}[!htbp]
\figurenum{A2.15}
\begin{center}
\includegraphics[angle=0,scale=0.4]{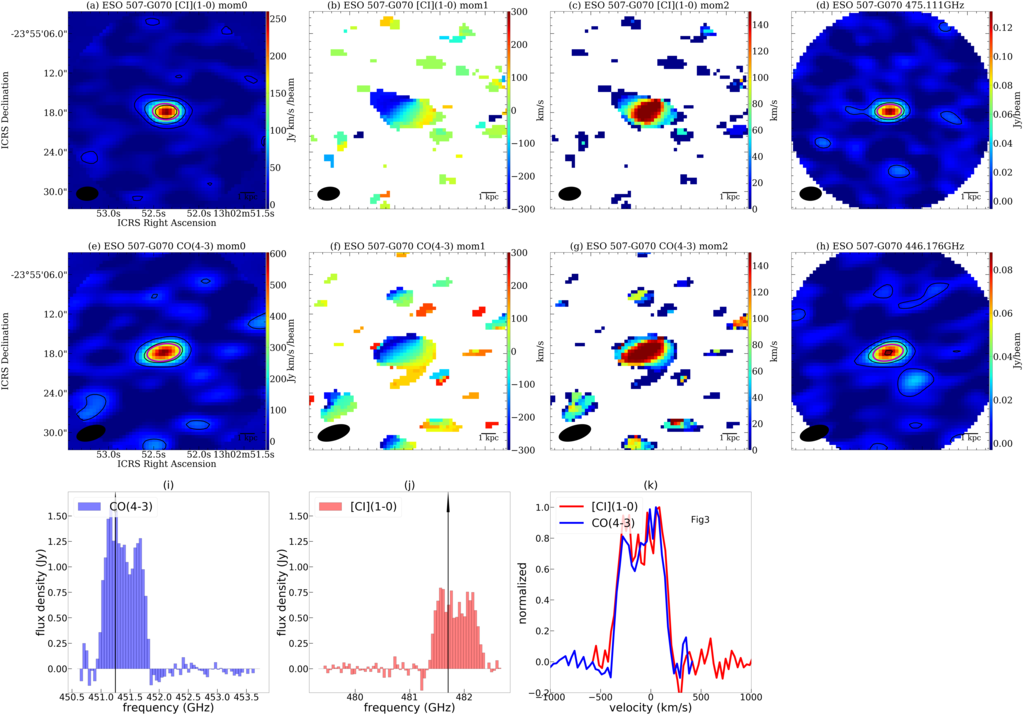}
\caption{Same as Figure A2.1.}
\end{center}
\end{figure*}

\begin{figure*}[!htbp]
\figurenum{A2.16}
\begin{center}
\includegraphics[angle=0,scale=0.4]{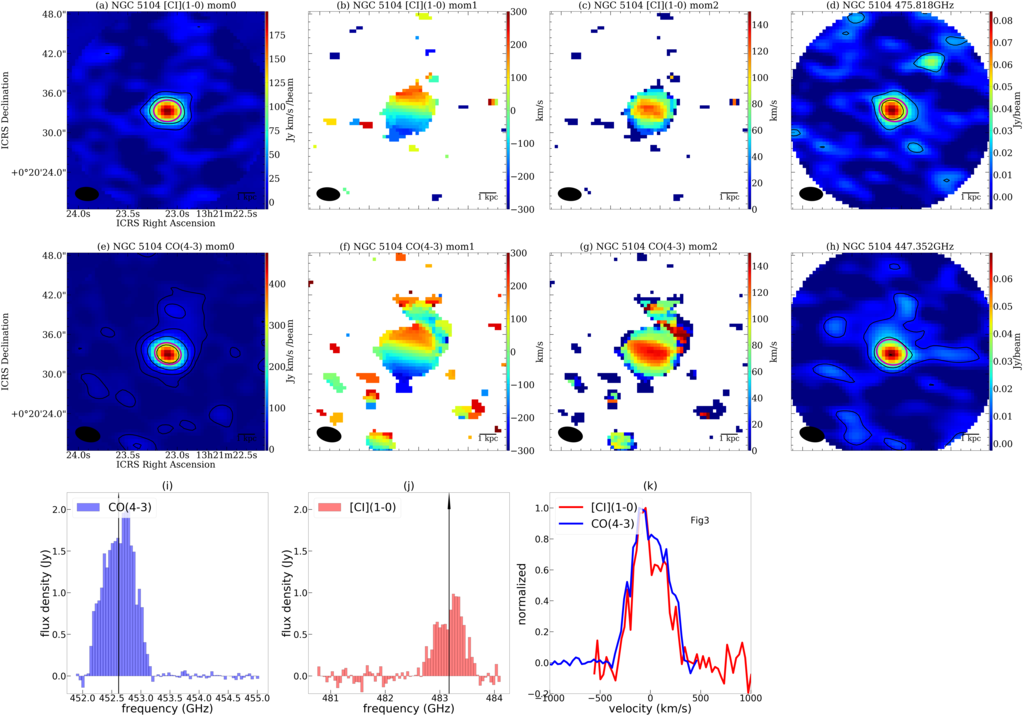}
\caption{Same as Figure A2.1.}
\end{center}
\end{figure*}

\begin{figure*}[!htbp]
\figurenum{A2.17}
\begin{center}
\includegraphics[angle=0,scale=0.4]{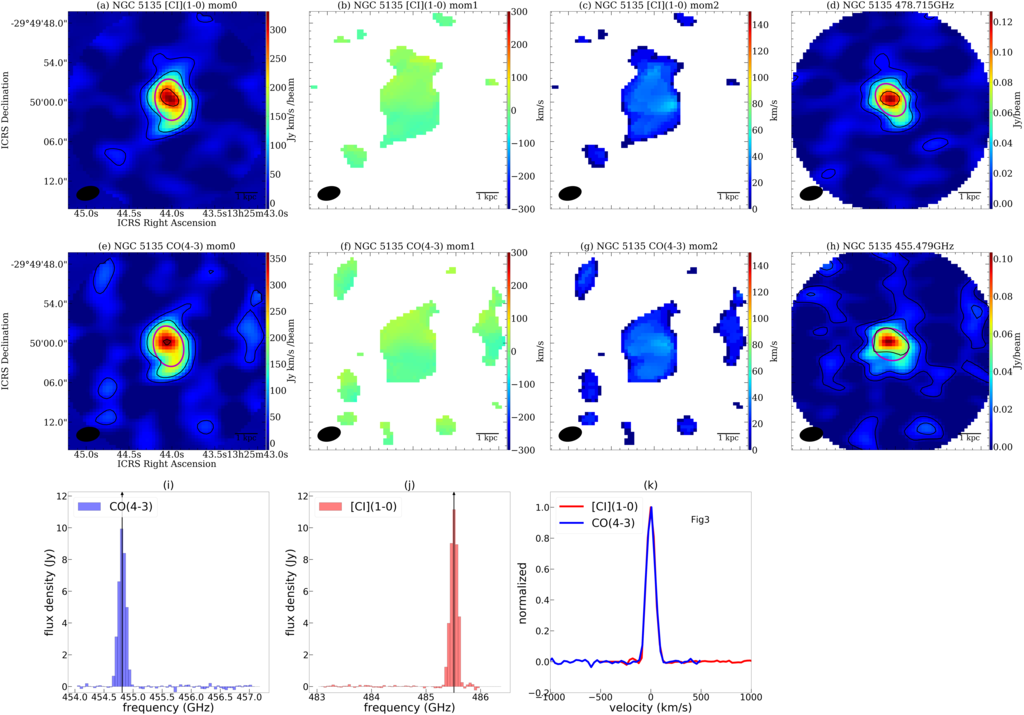}
\caption{Same as Figure A2.1.}
\end{center}
\end{figure*}

\begin{figure*}[!htbp]
\figurenum{A2.18}
\begin{center}
\includegraphics[angle=0,scale=0.4]{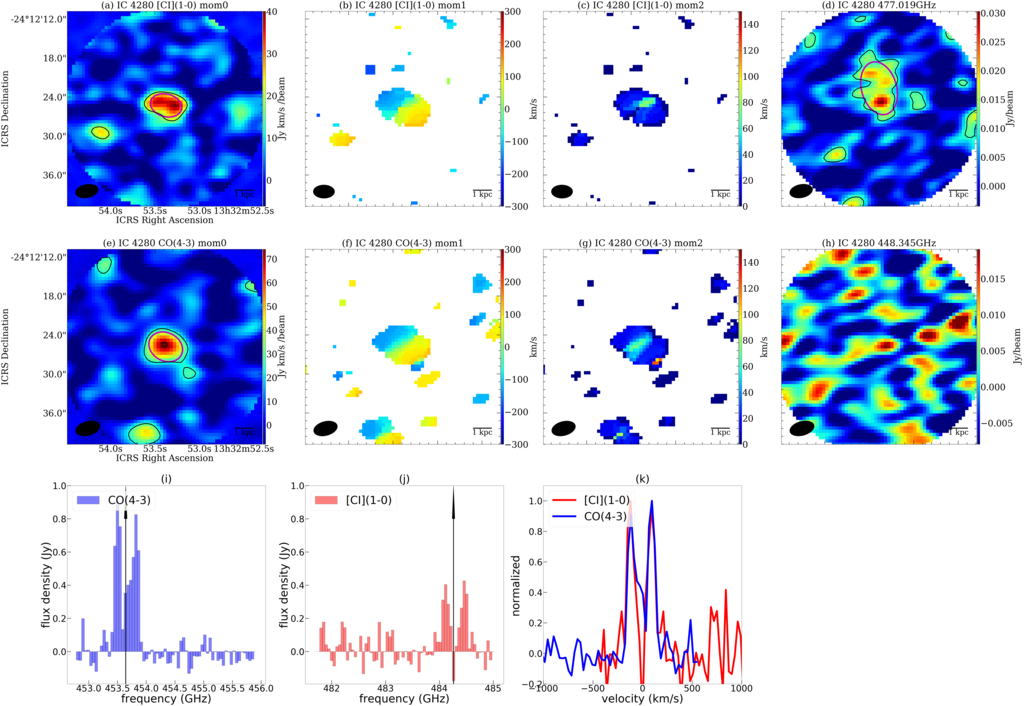}
\caption{Same as Figure A2.1.}
\end{center}
\end{figure*}

\begin{figure*}[!htbp]
\figurenum{A2.19}
\begin{center}
\includegraphics[angle=0,scale=0.4]{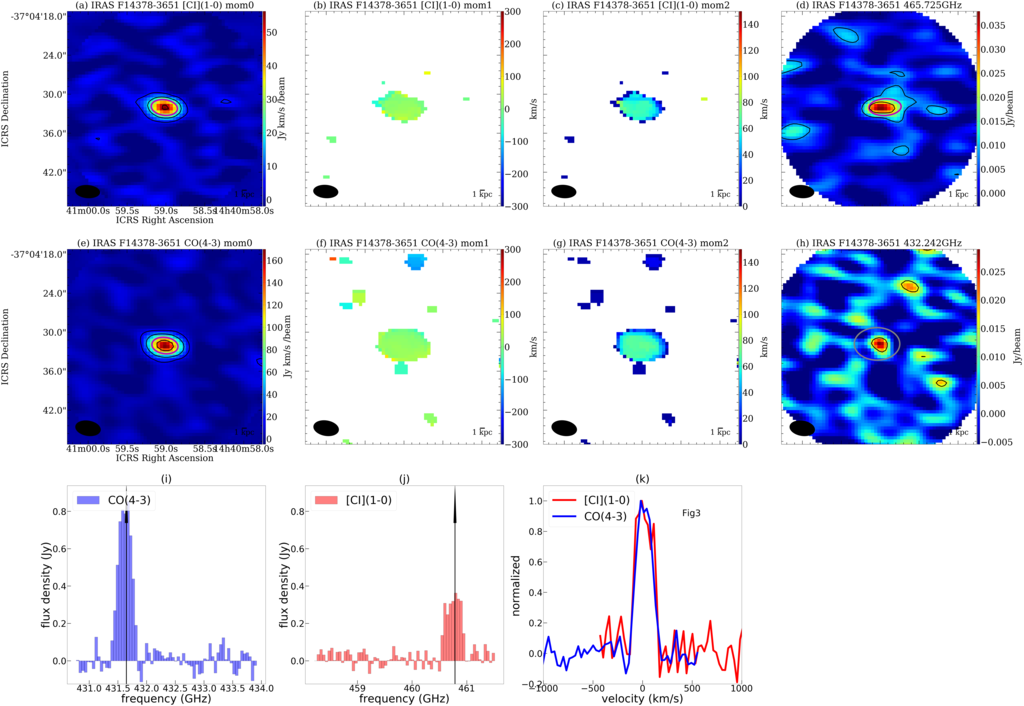}
\caption{Same as Figure A2.1.}
\end{center}
\end{figure*}

\begin{figure*}[!htbp]
\figurenum{A2.20}
\begin{center}
\includegraphics[angle=0,scale=0.4]{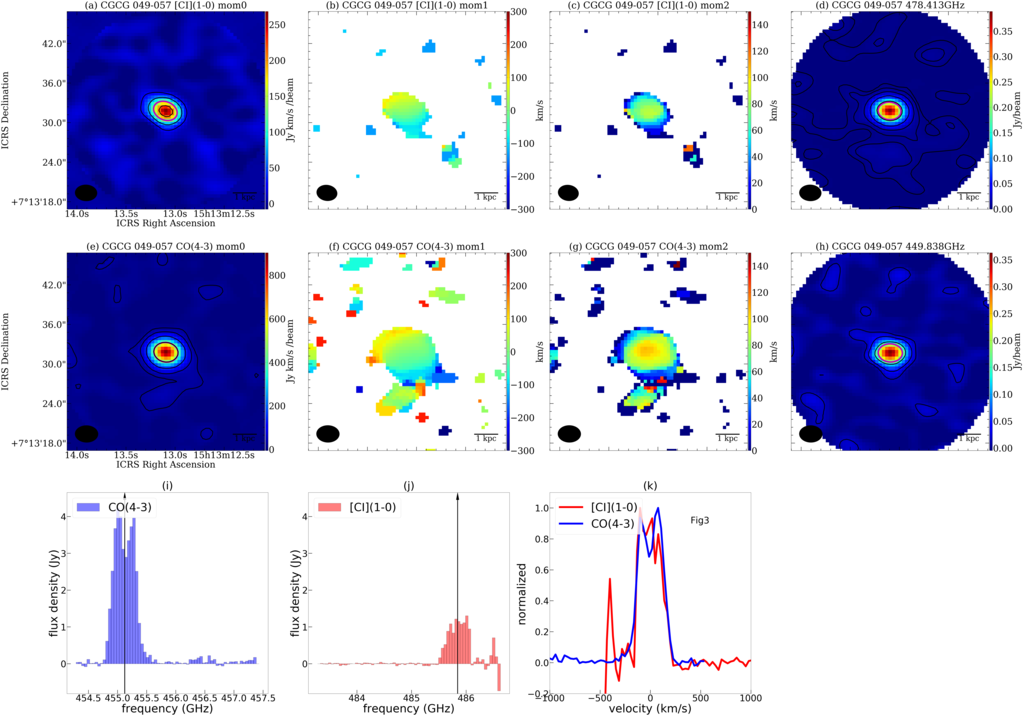}
\caption{Same as Figure A2.1.}
\end{center}
\end{figure*}

\begin{figure*}[!htbp]
\figurenum{A2.21}
\begin{center}
\includegraphics[angle=0,scale=0.4]{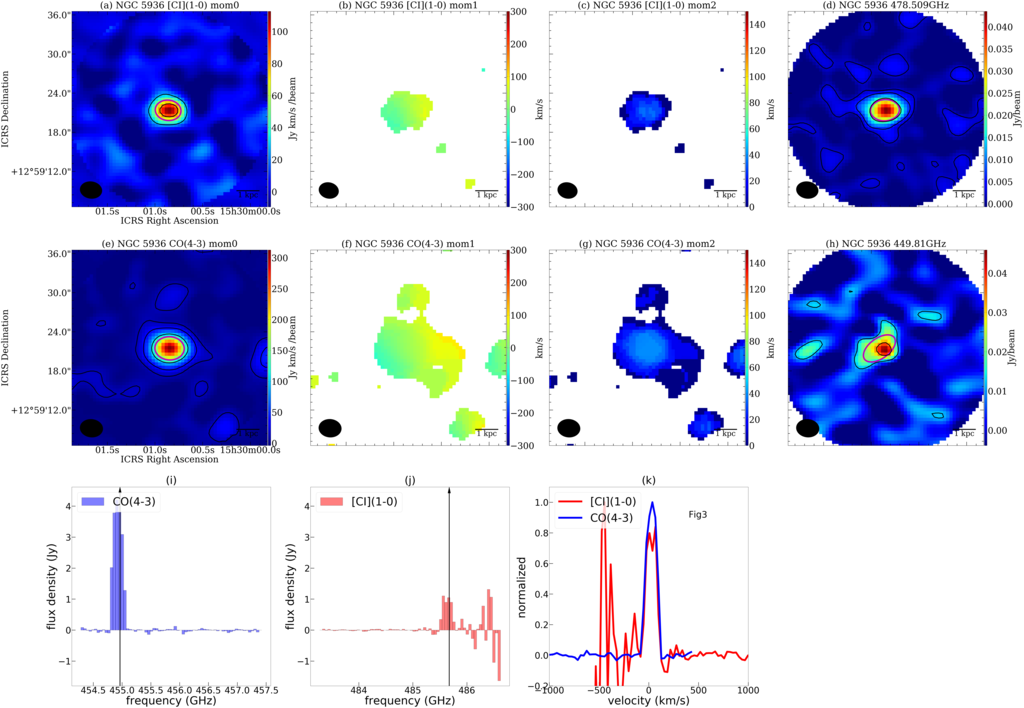}
\caption{Same as Figure A2.1.}
\end{center}
\end{figure*}

\begin{figure*}[!htbp]
\figurenum{A2.22}
\begin{center}
\includegraphics[angle=0,scale=0.4]{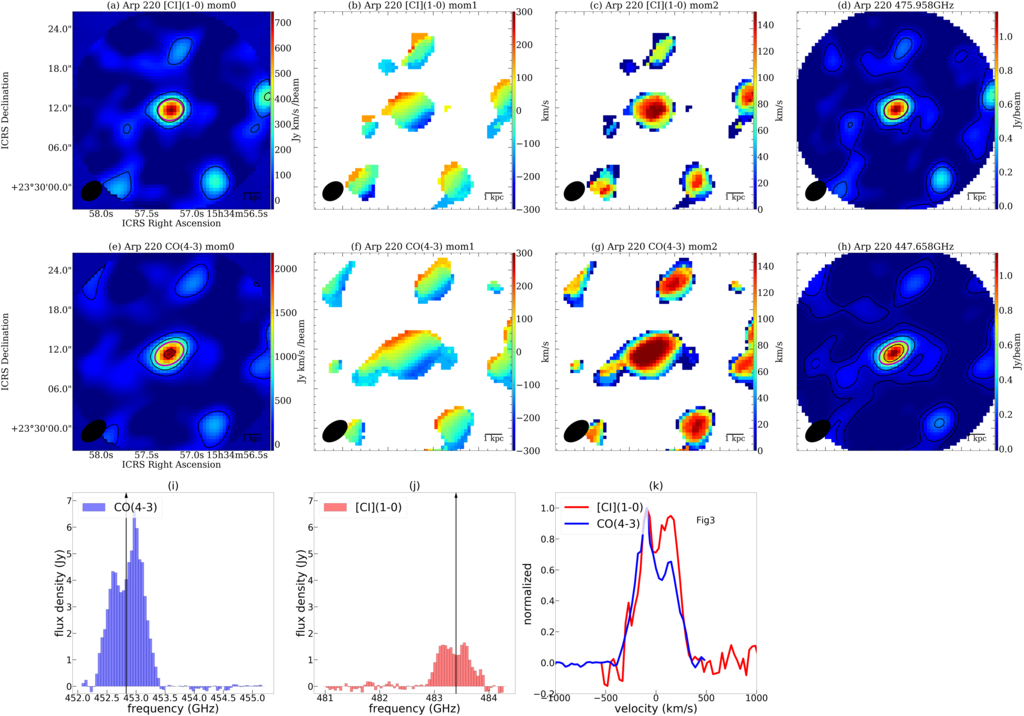}
\caption{Same as Figure A2.1.}
\end{center}
\end{figure*}

\begin{figure*}[!htbp]
\figurenum{A2.23}
\begin{center}
\includegraphics[angle=0,scale=0.4]{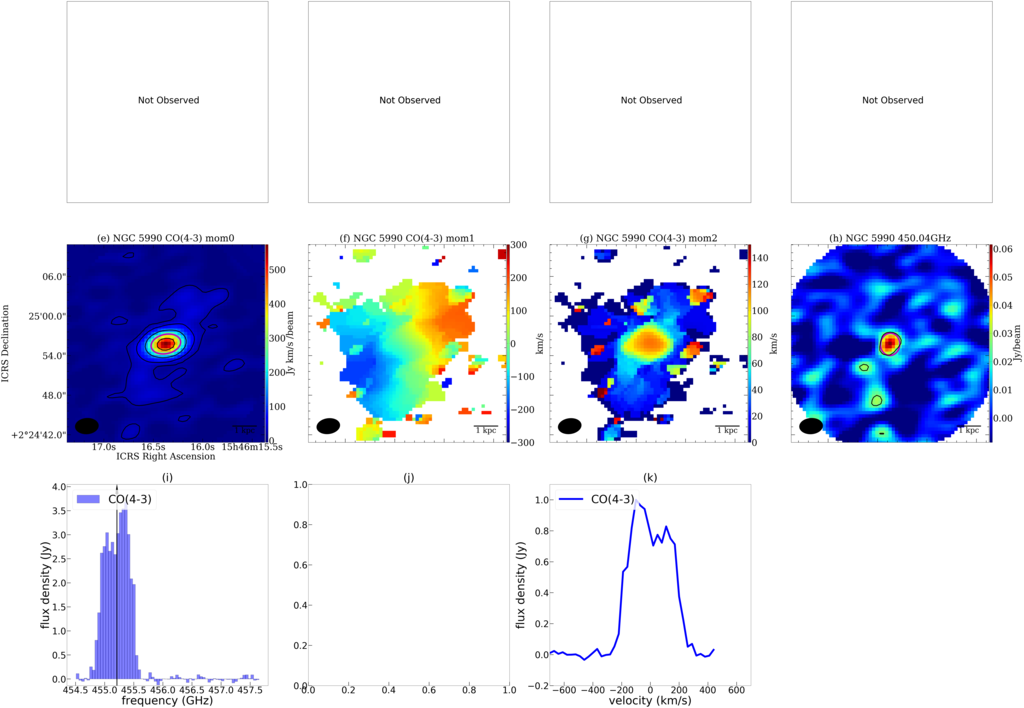}
\caption{Same as Figure A2.1.}
\end{center}
\end{figure*}

\begin{figure*}[!htbp]
\figurenum{A2.24}
\begin{center}
\includegraphics[angle=0,scale=0.4]{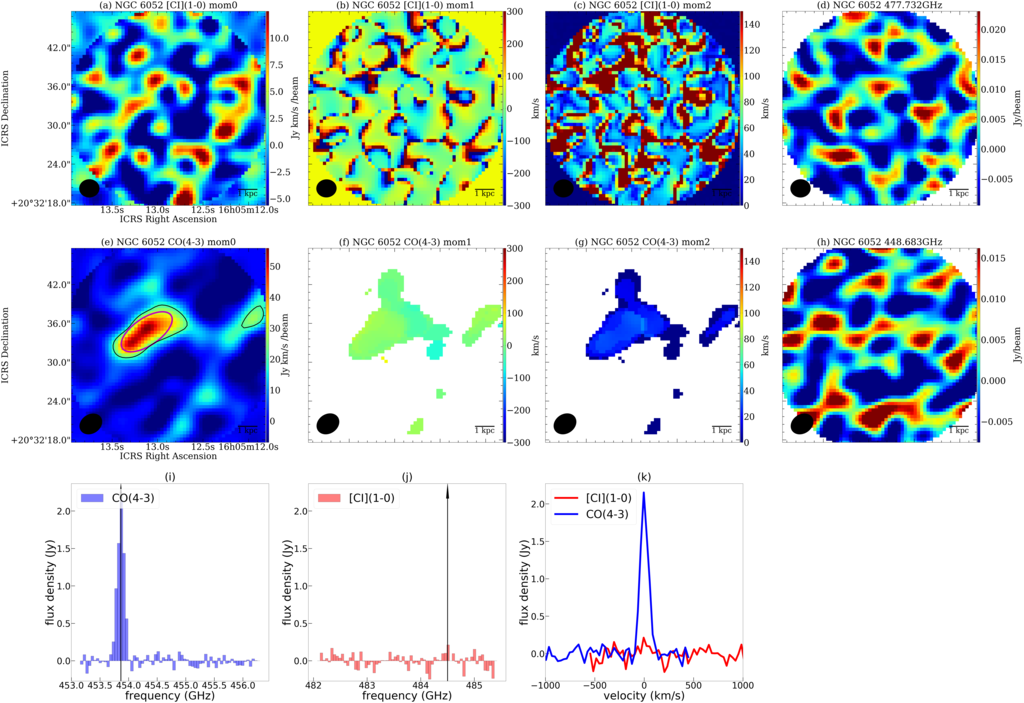}
\caption{Same as Figure A2.1.}
\end{center}
\end{figure*}

\begin{figure*}[!htbp]
\figurenum{A2.25}
\begin{center}
\includegraphics[angle=0,scale=0.4]{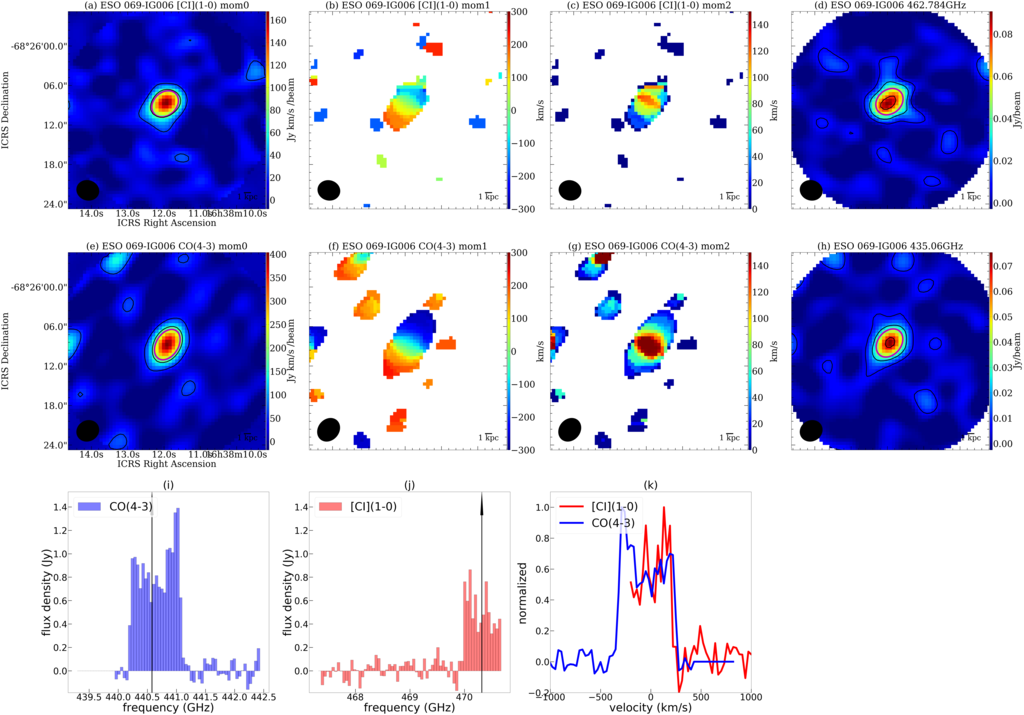}
\caption{Same as Figure A2.1.}
\end{center}
\end{figure*}

\begin{figure*}[!htbp]
\figurenum{A2.26}
\begin{center}
\includegraphics[angle=0,scale=0.4]{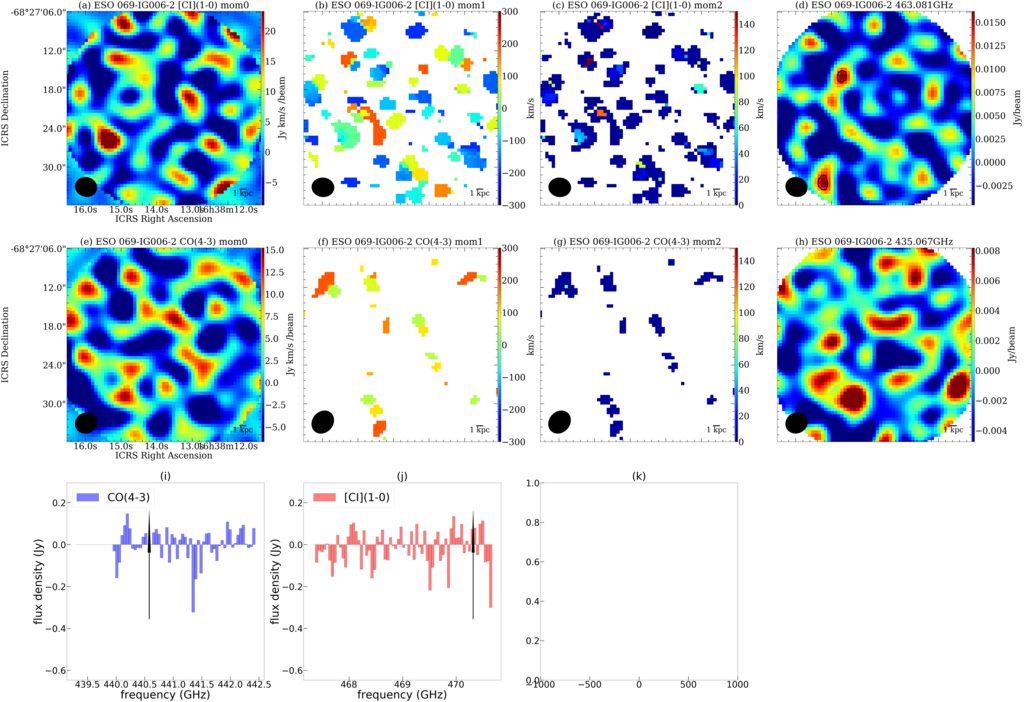}
\caption{Same as Figure A2.1.}
\end{center}
\end{figure*}

\begin{figure*}[!htbp]
\figurenum{A2.27}
\begin{center}
\includegraphics[angle=0,scale=0.4]{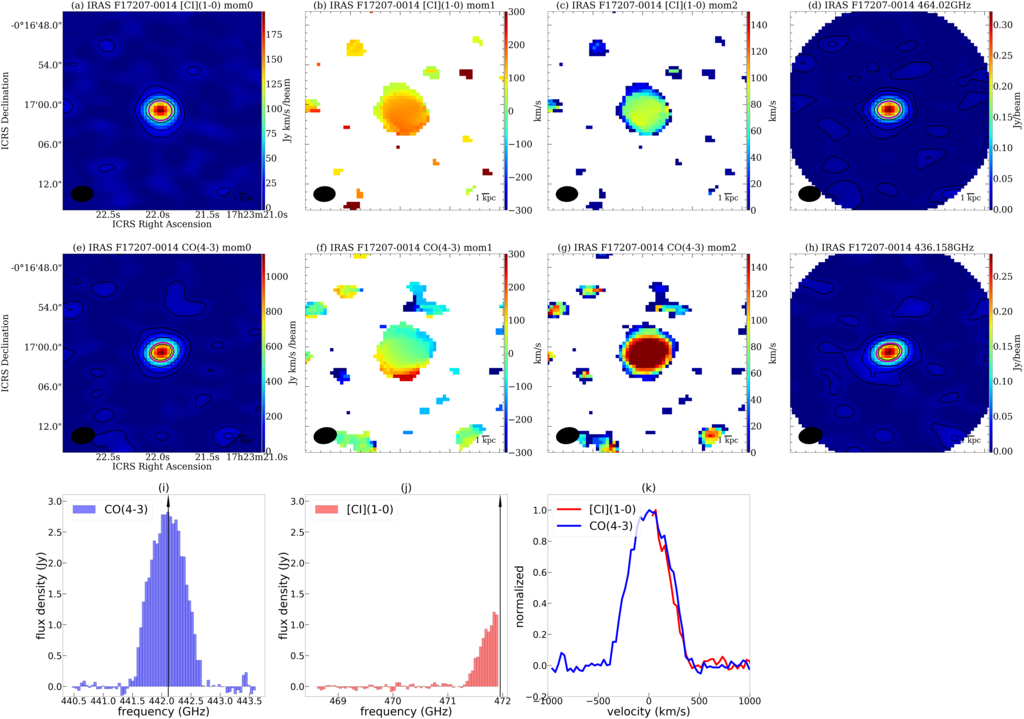}
\caption{Same as Figure A2.1.}
\end{center}
\end{figure*}

\begin{figure*}[!htbp]
\figurenum{A2.28}
\begin{center}
\includegraphics[angle=0,scale=0.4]{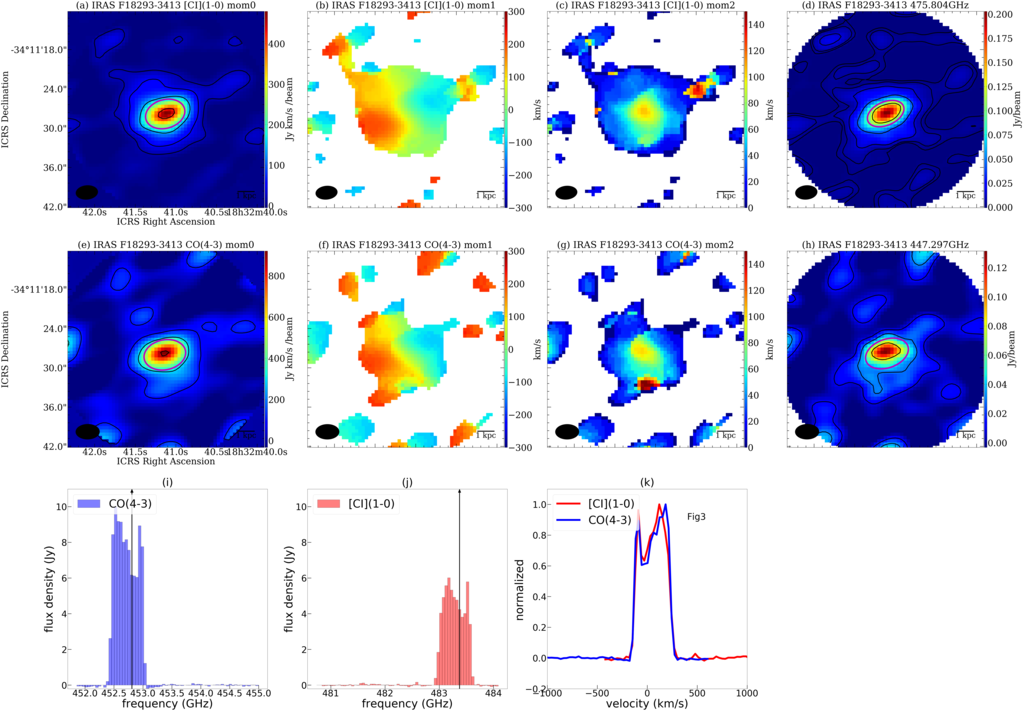}
\caption{Same as Figure A2.1.}
\end{center}
\end{figure*}

\begin{figure*}[!htbp]
\figurenum{A2.29}
\begin{center}
\includegraphics[angle=0,scale=0.4]{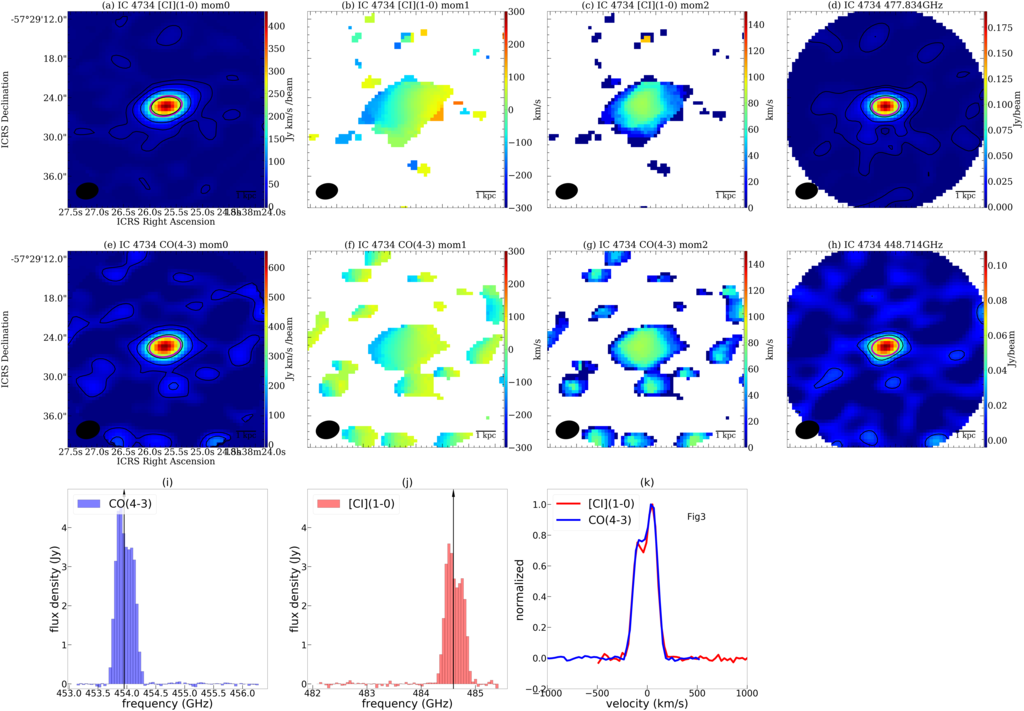}
\caption{Same as Figure A2.1.}
\end{center}
\end{figure*}

\begin{figure*}[!htbp]
\figurenum{A2.30}
\begin{center}
\includegraphics[angle=0,scale=0.4]{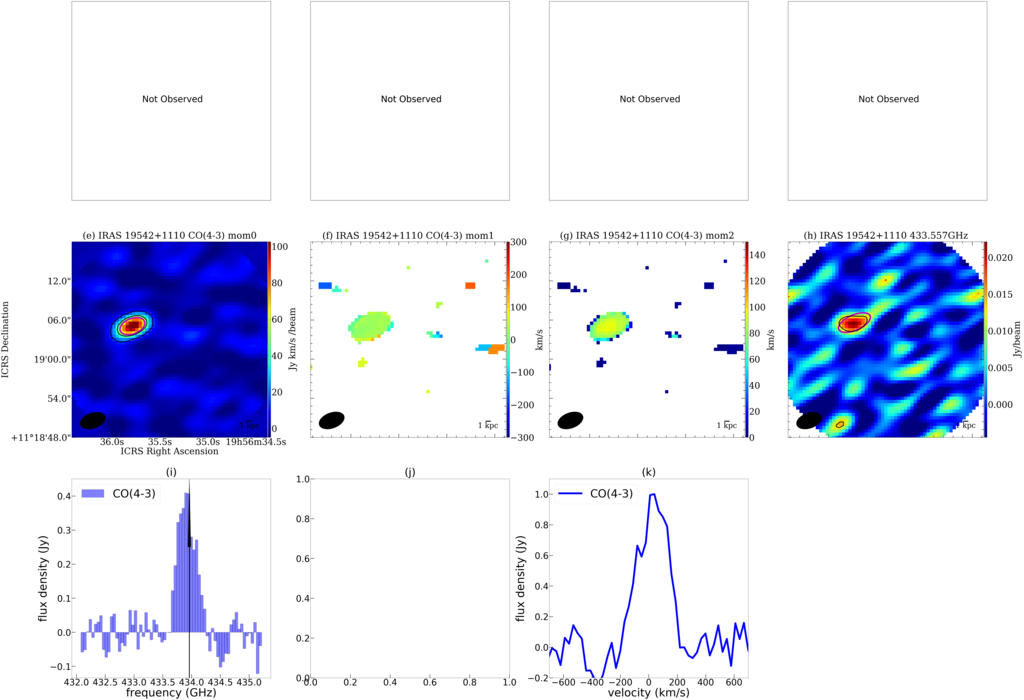}
\caption{Same as Figure A2.1.}
\end{center}
\end{figure*}

\begin{figure*}[!htbp]
\figurenum{A2.31}
\begin{center}
\includegraphics[angle=0,scale=0.4]{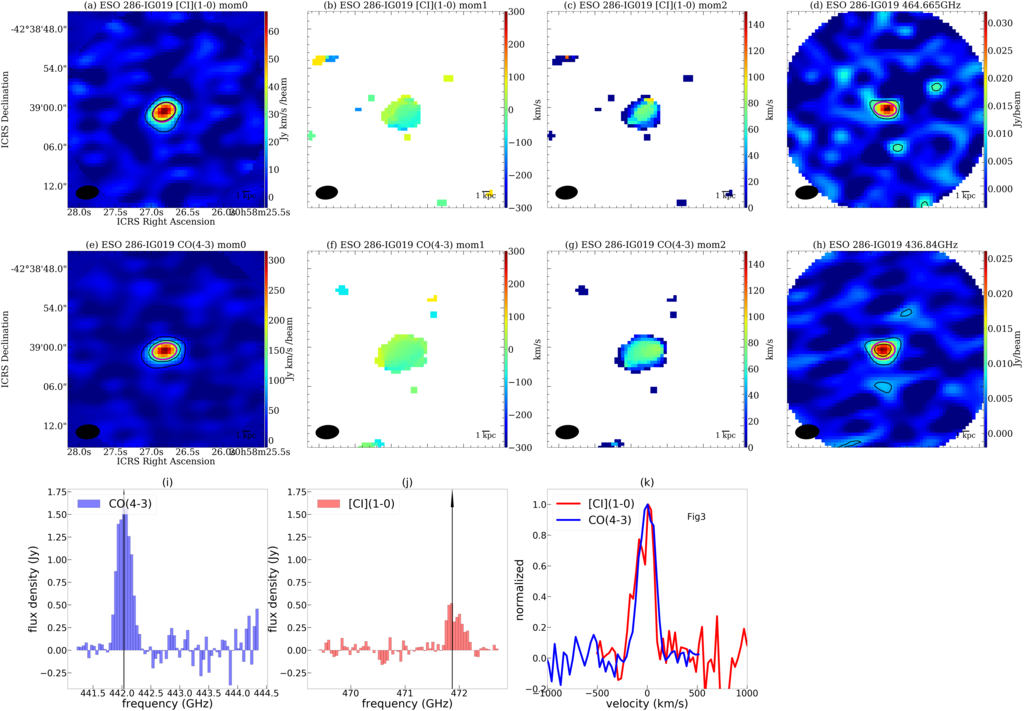}
\caption{Same as Figure A2.1.}
\end{center}
\end{figure*}

\begin{figure*}[!htbp]
\figurenum{A2.32}
\begin{center}
\includegraphics[angle=0,scale=0.4]{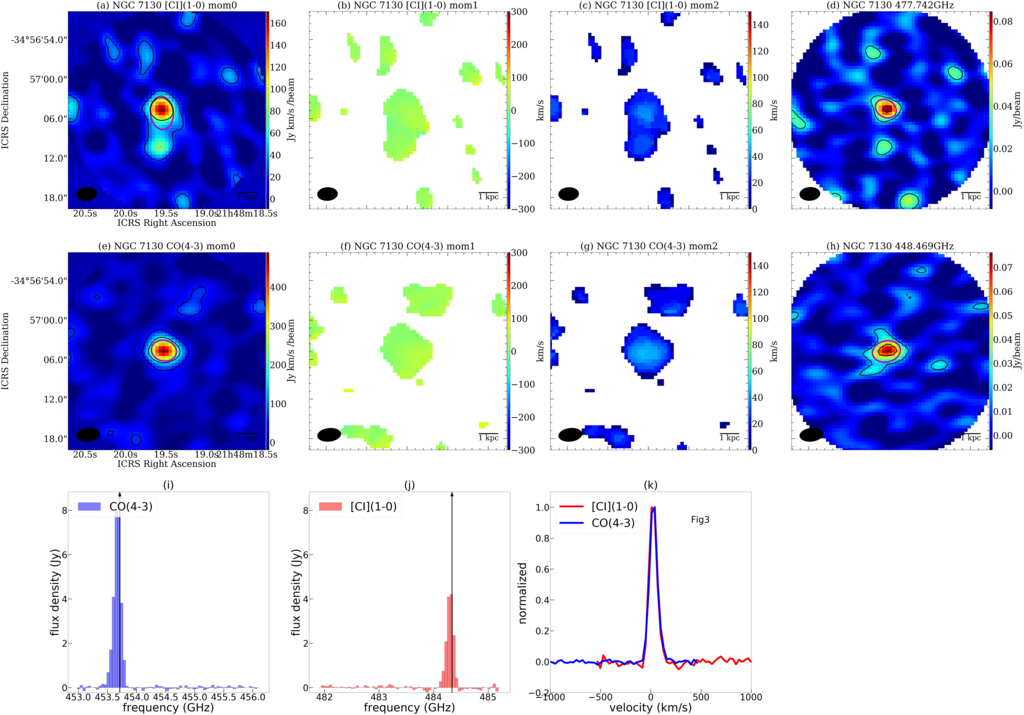}
\caption{Same as Figure A2.1.}
\end{center}
\end{figure*}

\begin{figure*}[!htbp]
\figurenum{A2.33}
\begin{center}
\includegraphics[angle=0,scale=0.4]{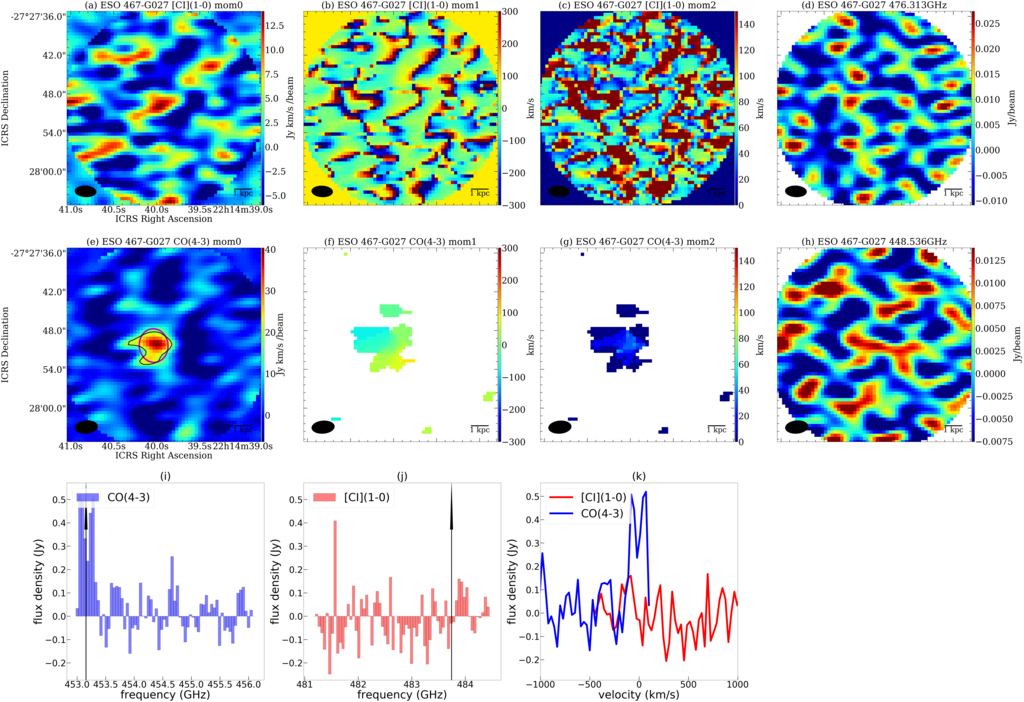}
\caption{Same as Figure A2.1.}
\end{center}
\end{figure*}

\begin{figure*}[!htbp]
\figurenum{A2.34}
\begin{center}
\includegraphics[angle=0,scale=0.4]{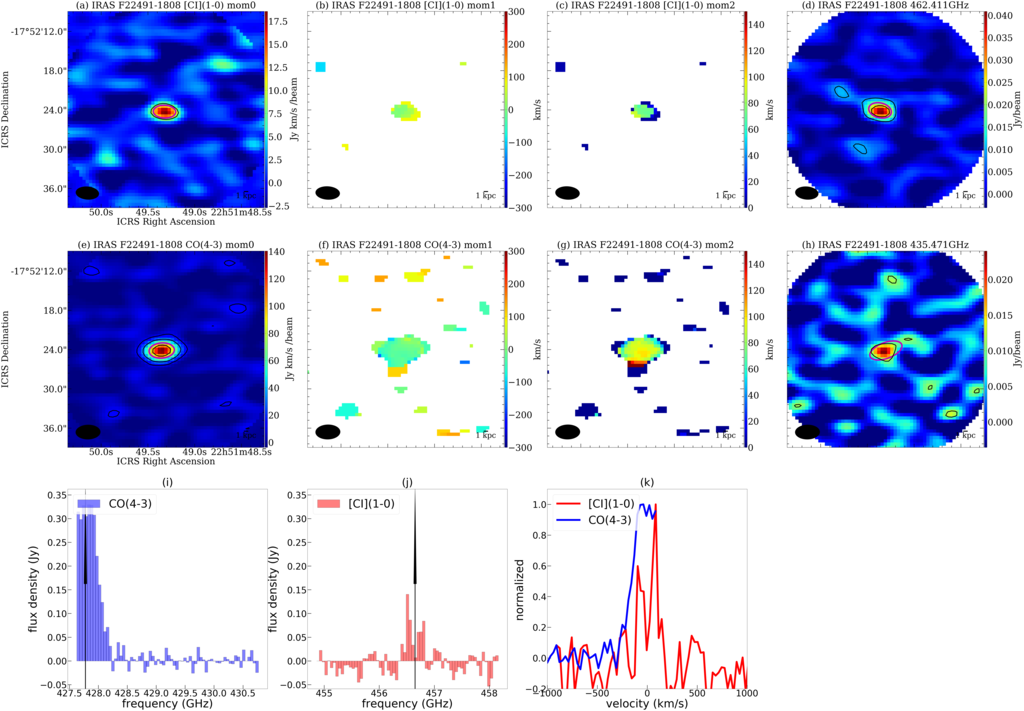}
\caption{Same as Figure A2.1.}
\end{center}
\end{figure*}

\begin{figure*}[!htbp]
\figurenum{A2.35}
\begin{center}
\includegraphics[angle=0,scale=0.4]{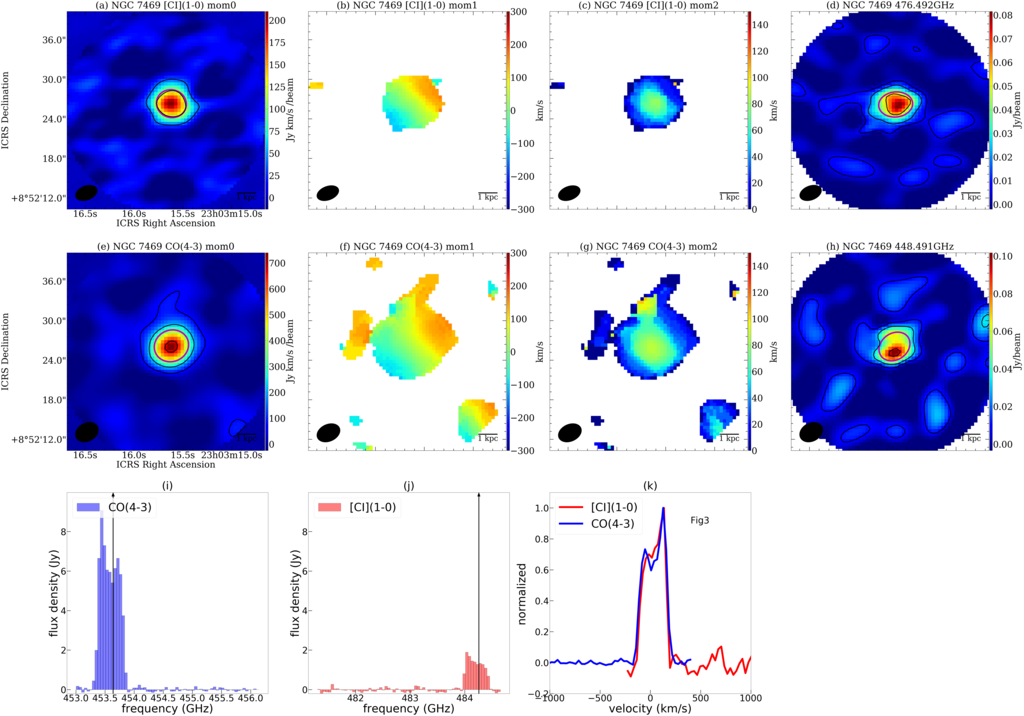}
\caption{Same as Figure A2.1.}
\end{center}
\end{figure*}

\begin{figure*}[!htbp]
\figurenum{A2.36}
\begin{center}
\includegraphics[angle=0,scale=0.4]{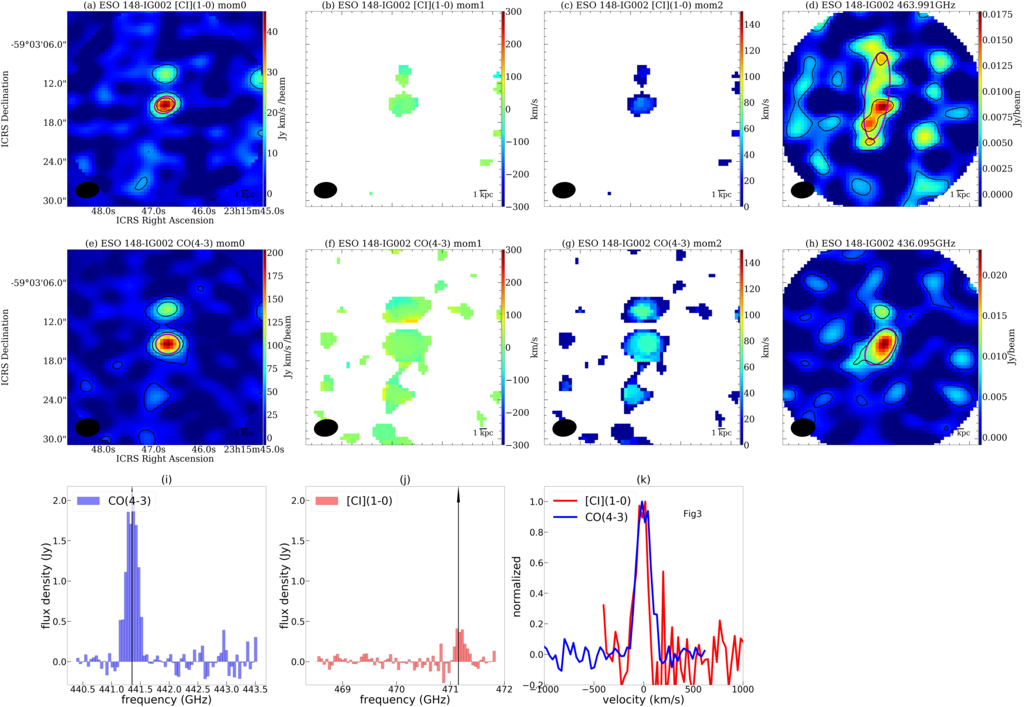}
\caption{Same as Figure A2.1.}
\end{center}
\end{figure*}

\begin{figure*}[!htbp]
\figurenum{A2.37}
\begin{center}
\includegraphics[angle=0,scale=0.4]{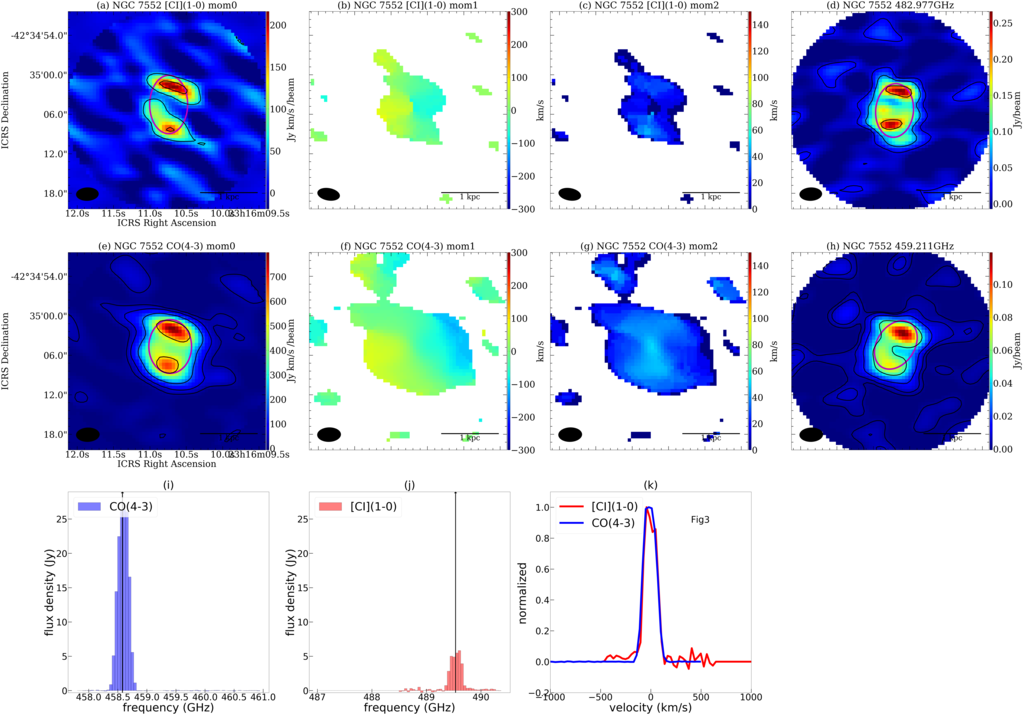}
\caption{Same as Figure A2.1.}
\end{center}
\end{figure*}

\begin{figure*}[!htbp]
\figurenum{A2.38}
\begin{center}
\includegraphics[angle=0,scale=0.4]{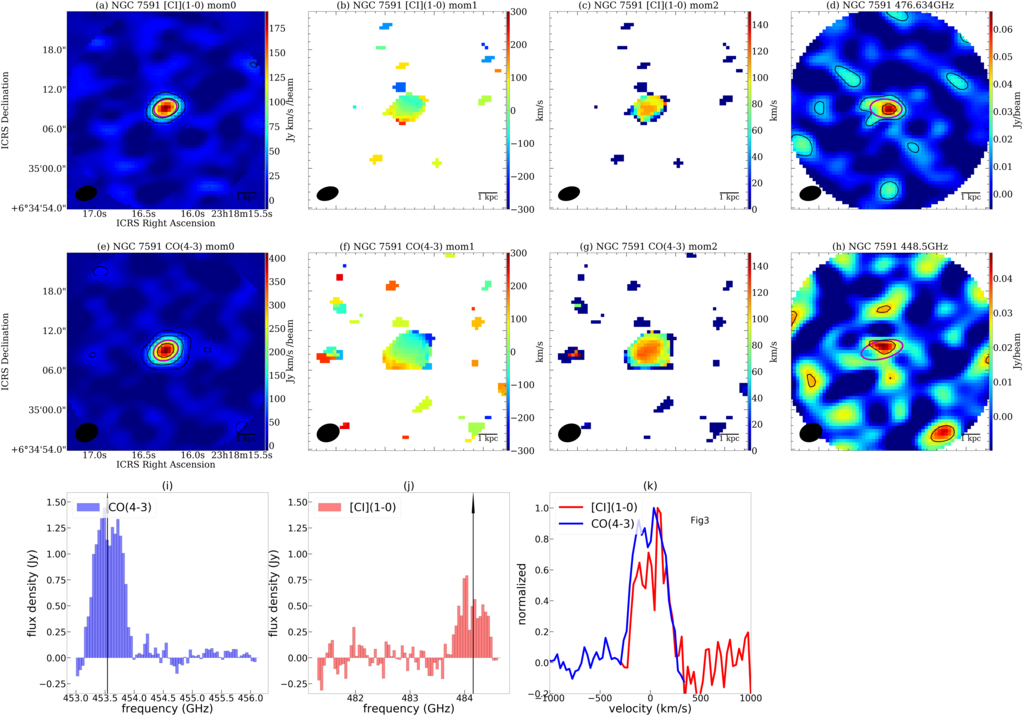}
\caption{Same as Figure A2.1.}
\end{center}
\end{figure*}

\begin{figure*}[!htbp]
\figurenum{A2.39}
\begin{center}
\includegraphics[angle=0,scale=0.4]{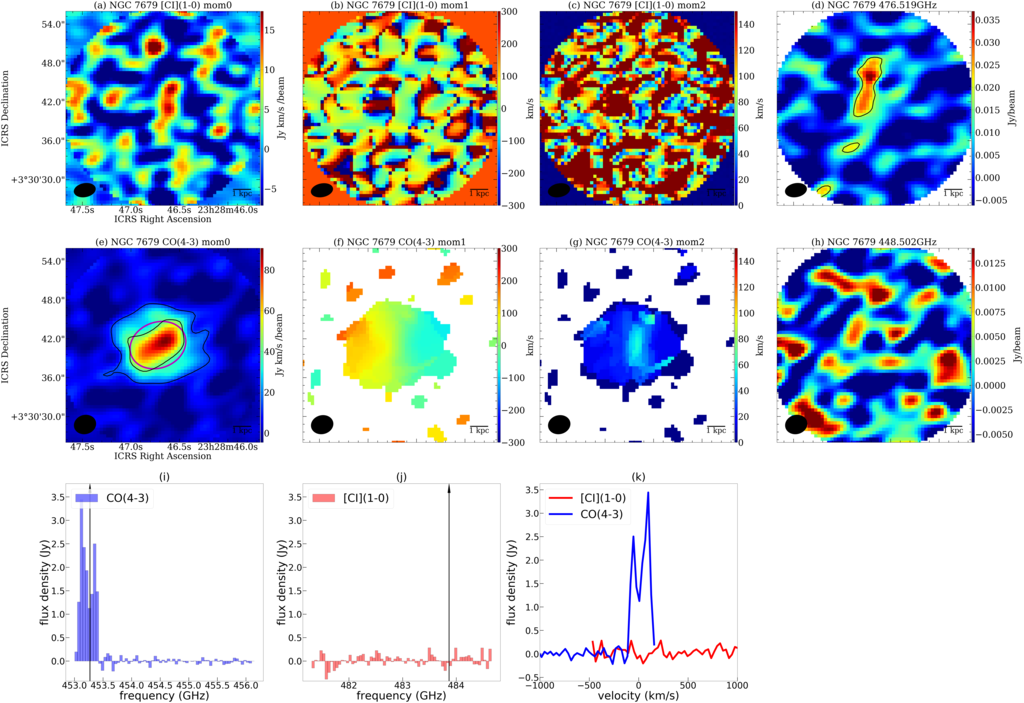}
\caption{Same as Figure A2.1.}
\end{center}
\end{figure*}

\begin{figure*}[!htbp]
\figurenum{A2.40}
\begin{center}
\includegraphics[angle=0,scale=0.4]{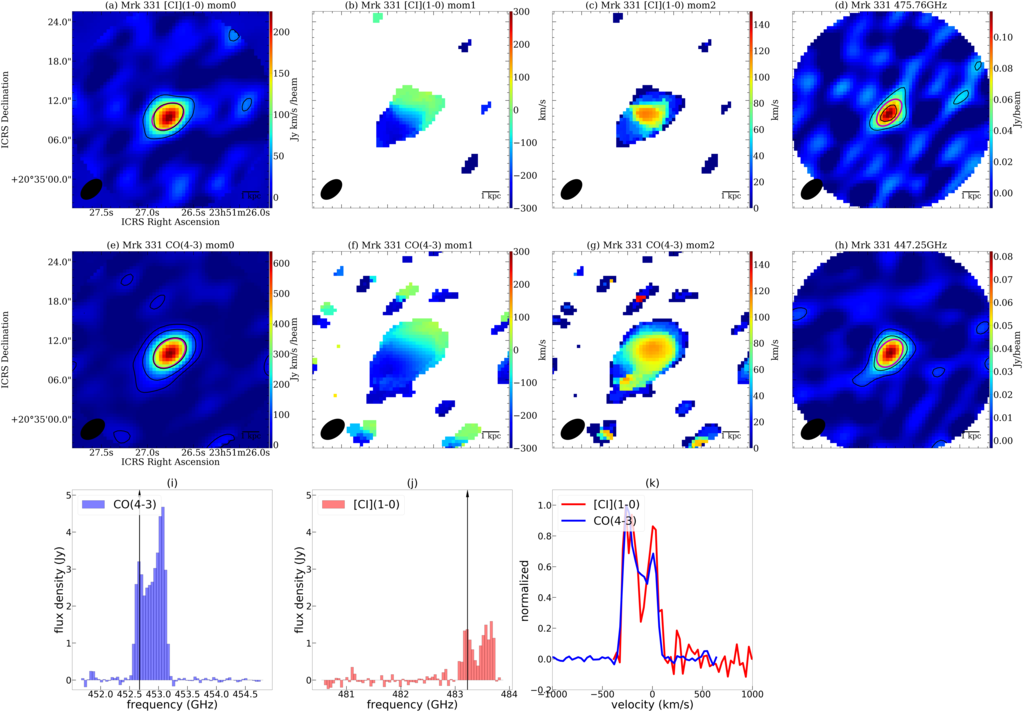}
\caption{Same as Figure A2.1.}
\end{center}
\end{figure*}

\clearpage
\subsection{Tables}
\startlongtable
\tabletypesize{\scriptsize}
\tablewidth{1000pt}
\begin{deluxetable*}{llcccccc}
\tablecaption{Target list \label{tab:target}}
\tablehead{
\colhead{Target Name} &
\colhead{coodinates (ICRS)} &
\colhead{redshift} &
\colhead{$D_{\rm L}$} &
\colhead{$1\farcs$} &
\colhead{$V_{\rm sys}$} &
\colhead{CO~(4--3) sky} &
\colhead{[CI]~(1--0) sky} \\
\colhead{} &
\colhead{hmsdms} &
\colhead{} &
\colhead{Mpc} &
\colhead{kpc} &
\colhead{km~s$^{-1}$} &
\colhead{GHz} &
\colhead{GHz} 
}
\decimalcolnumbers
\startdata 
              NGC 232 &  00h42m45.8s -23d33m40.7s &  0.0226 &   102 &  0.47 &      6637 &        450.8 &      481.3 \\
  NGC 232-2 (NGC 235) &  00h42m52.9s -23d32m29.4s &     $-$ &   $-$ &   $-$ &       $-$ &          $-$ &        $-$ \\
 NGC 232-3 (PGC 2570) &  00h42m53.8s -23d32m44.3s &     $-$ &   $-$ &   $-$ &       $-$ &          $-$ &        $-$ \\
              IC 1623 &  01h07m46.8s -17d30m25.7s &  0.0201 &    90 &  0.42 &      5898 &        452.0 &      482.5 \\
            UGC 02982 &  04h12m22.7s +05d32m49.1s &  0.0177 &    79 &  0.37 &      5213 &        453.0 &      483.6 \\
             NGC 1614 &  04h34m00.0s -08d34m44.6s &  0.0159 &    71 &  0.34 &      4703 &        453.8 &      484.4 \\
     IRAS F05189-2524 &  05h21m01.4s -25d21m45.4s &  0.0426 &   194 &  0.87 &     12239 &        442.2 &      472.1 \\
      IRAS 09022-3615 &  09h04m12.7s -36d27m00.8s &  0.0596 &   276 &  1.19 &     16874 &        435.1 &      464.5 \\
\enddata

\tablecomments{
(1) Target names. If the galaxy is the pair phase of a merger process, the companion's names are described as ``-2" and ``-3." 
(2) Pointing center of the ACA observation.
(3) The redshift is based on NED. In the case of IRAS~F14378-3651, NGC~4418, and IRAS~19542+1110, the redshifts are based on \citet{Rose_2018}, \citet{Gonzalez_2012}, and \citet{Veilleux_2013}, respectively.
(4) Luminosity distance.
(5) Physical scale corresponding to $1\farcs$.
(6) The systematic velocity calculated by $V_{\rm sys}=cz/(1+z)$, where $c$ is the speed of light). In the case of a pair of galaxies, we apply the same systematic velocity to identify the line emission.
(7), (8) Observed sky frequencies of [\ion{C}{1}]~(1--0) and CO~(4--3).}
\end{deluxetable*}

\startlongtable
\tabletypesize{\scriptsize}
\begin{deluxetable*}{lccccccc}
\tablecaption{Supplemental data \label{tab:SED}}
\tablehead{
\colhead{Target Name} &
\colhead{$\log$ SFR} &
\colhead{$\log M_{\rm dust}$} &
\colhead{$T_{\rm dust}$} &
\colhead{$\log L_{\rm TIR}$} &
\colhead{AGN} &
\colhead{Merger} &
\colhead{$L'_{\rm CO(1-0)}$} \\
 &
\colhead{$M_{\odot}$~yr$^{-1}$} &
\colhead{$M_{\odot}$} &
\colhead{K} &
\colhead{$L_{\odot}$} 
 &
 &
 &
\colhead{$10^{6}$~${\rm K~km~s^{-1}~pc^2}$} \\}
\decimalcolnumbers
\startdata
              NGC 232 &             -- &             -- &            -- &              -- &       Y &              Y &            -- \\
  NGC 232-2 (NGC 235) &             -- &             -- &            -- &              -- &      -- &             -- &            -- \\
 NGC 232-3 (PGC 2570) &             -- &             -- &            -- &              -- &      -- &             -- &            -- \\
              IC 1623 &  1.67$\pm$0.02 &  8.27$\pm$0.02 &  27.5$\pm$0.7 &   11.6$\pm$0.02 &       N &              Y &    505$\pm$53 \\
            UGC 02982 &  1.27$\pm$0.04 &  8.31$\pm$0.09 &  23.6$\pm$0.6 &  11.21$\pm$0.05 &       N &              Y &   486$\pm$121 \\
             NGC 1614 &  1.76$\pm$0.03 &  7.91$\pm$0.04 &  32.1$\pm$1.0 &   11.7$\pm$0.03 &       N &              Y &    143$\pm$35 \\
     IRAS F05189-2524 &  1.89$\pm$0.05 &  8.19$\pm$0.05 &  34.4$\pm$1.1 &  11.82$\pm$0.05 &       Y &              Y &  1173$\pm$234 \\
      IRAS 09022-3615 &  2.28$\pm$0.03 &   8.6$\pm$0.04 &  34.4$\pm$1.0 &  12.22$\pm$0.04 &       Y &              Y &            -- \\
\enddata

\tablecomments{
(1) Target names. (2)-(5) SFR, $M_{\rm dust}$, $T_{\rm dust}$, and $L_{\rm TIR}$ estimated in \citet{Shangguan_2019}.
$L_{\rm TIR}$ is for  8 - 1000 $\mu$m and the AGN contribution has been subtracted.
(6) Y and N represent AGN and non-AGN, respectively.
(7) Y and N represent mergers and non-mergers, respectively.
The classification of AGNs and mergers was from \citet{Shangguan_2019}.
(8) CO~(1--0) luminosity measured by \citet{Kame_2016}.
}
\end{deluxetable*}

\startlongtable
\begin{deluxetable*}{llcccc}
\tablecaption{Observation log \label{tab:obs}}
\tablehead{
\colhead{Target Name} &
\colhead{date} &
\colhead{line} &
\colhead{$T_{\rm sys}$(ave.)} &
\colhead{$N_{\rm ant}$} &
\colhead{MRS (representative freq.)}\\
\colhead{} &
\colhead{y/m/d} &
\colhead{} &
\colhead{K} &
\colhead{} &
\colhead{$\arcsec$ (GHz)}
}
\decimalcolnumbers
\startdata 
                       Arp220, NGC 6052 &  2018/11/15 &    [\ion{C}{1}]~(1--0) &          793 &    10 &  13.81 (484.53) \\
                               NGC 5104 &   2018/12/3 &  CO~(4--3) &          429 &    12 &   14.6 (452.66) \\
                       IRAS F14378-3651 &   2018/12/5 &    [\ion{C}{1}]~(1--0) &          367 &    11 &  14.35 (460.62) \\
                       IRAS F22491-1808 &   2018/12/5 &  CO~(4--3) &         1243 &    12 &  15.42 (428.44) \\
                        IRAS 09022-3615 &   2019/1/12 &    [\ion{C}{1}]~(1--0) &          949 &    11 &   14.23 (464.4) \\
                               NGC 5990 &   2019/1/13 &  CO~(4--3) &         1454 &    11 &  14.51 (455.25) \\
                        IRAS 09022-3615 &   2019/1/13 &  CO~(4--3) &           -- &    10 &   15.4 (434.32) \\
                               NGC 3110 &   2019/3/16 &    [\ion{C}{1}]~(1--0) &          995 &    11 &  13.66 (483.84) \\
\enddata

\tablecomments{
(1) Target names. 
(2) Date of observation. 
(3) Observed line. 
(4) Typical systematic temperature. If the systematic temperature is not shown in QA0 report, the value is shown as ``--."
(5) Number of antennas.
(6) MRS scale shown in the QA0 report.}
\end{deluxetable*}

\startlongtable
\tabletypesize{\scriptsize}
\begin{deluxetable*}{llcccccccc}
\tablecaption{Imaging results \label{tab:obs_res}}
\tablehead{
\colhead{Target Name} &
\colhead{Beam} &
\colhead{Beam} &
\colhead{rms (cube)} &
\colhead{rms (cube)} &
\colhead{rms} &
\colhead{rms} &
\colhead{rms} &
\colhead{v-range} &
\colhead{v-range} \\
\colhead{} &
\colhead{CO~(4--3)} &
\colhead{[\ion{C}{1}]~(1--0)} &
\colhead{CO~(4--3)} &
\colhead{[\ion{C}{1}]~(1--0)} &
\colhead{CO~(4--3)} &
\colhead{[\ion{C}{1}]~(1--0)} &
\colhead{cont} &
\colhead{CO~(4--3)} &
\colhead{[\ion{C}{1}]~(1--0)} \\
\colhead{} &
\colhead{( $\arcsec$ , $\arcsec$ )} &
\colhead{( $\arcsec$ , $\arcsec$ )} &
\colhead{*1} &
\colhead{*1} &
\colhead{*2} &
\colhead{*2} &
\colhead{*1} &
\colhead{km~s$^{-1}$} &
\colhead{km~s$^{-1}$}
}
\decimalcolnumbers
\startdata 
              NGC 232 &  3.4$\times$2.2 &  3.4$\times$2.0 &                63 &              29 &                14 &               4 &      3.4,3.2 &   [-220,290] &  [-190,350] \\
  NGC 232-2 (NGC 235) &  3.4$\times$2.2 &  3.3$\times$2.0 &                45 &              28 &                 4 &               4 &      3.7,2.5 &  [-310,-160] &  [-280,140] \\
 NGC 232-3 (PGC 2570) &  3.4$\times$2.2 &  3.3$\times$2.0 &                80 &              35 &                10 &               5 &      6.4,4.2 &           ND &          ND \\
              IC 1623 &  4.0$\times$2.2 &  3.5$\times$2.0 &                64 &              25 &                17 &               4 &      4.2,3.5 &   [-210,240] &  [-190,170] \\
            UGC 02982 &              -- &  3.2$\times$2.7 &                -- &              33 &                -- &               4 &       --,5.7 &           -- &  [-150,210] \\
             NGC 1614 &  3.5$\times$2.4 &  3.3$\times$2.6 &                56 &              47 &                11 &               7 &      4.5,5.4 &   [-190,200] &  [-160,200] \\
     IRAS F05189-2524 &  4.8$\times$2.2 &  3.5$\times$2.3 &                58 &              33 &                 9 &               3 &      3.5,3.6 &    [-20,220] &   [100,250] \\
      IRAS 09022-3615 &  4.7$\times$2.1 &  3.8$\times$2.1 &                31 &              36 &                15 &               8 &     7.0,15.2 &   [-450,360] &  [-220,260] \\
\enddata

\tablecomments{
(1) Target names. 
(2) and (3) The achieved synthesized beam. 
(4) and (5) The achieved sensitivity of channel maps in units of mJy~beam$^{-1}$ (*1). 
(6) and (7) The achieved sensitivity of the velocity integrated intensity map in the unit of Jy~beam$^{-1}$~km\,s$^{-1}$  (*2). 
(8) The achieved sensitivity of continuum emission in units of mJy~beam$^{-1}$ (*1). The left value is for the CO~(4--3) scheduling block and the right value is for the [\ion{C}{1}]~(1--0) scheduling block.
(9) and (10) The velocity range for the velocity-integrated intensity map. The values represent relative velocities with respect to the systematic velocity. The ``*" means that the emission is on the edge of the spectral window due to incorrect redshift. ``ND" indicates the non-detection.}
\end{deluxetable*}
               
\startlongtable
\tabletypesize{\scriptsize}
\begin{deluxetable*}{lccccccc}
\tablecaption{Line flux \label{tab:flux_line}}
\tablehead{
\colhead{Target Name} &
\colhead{$S_{\rm line}\Delta v$} &
\colhead{$S_{\rm line}\Delta v$} &
\colhead{FWHM} &
\colhead{FWHM} &
\colhead{$S_{\rm line}$d$v$}&
\colhead{size (FWHM)} &
\colhead{size (FWHM)} \\
\colhead{} &
\colhead{CO~(4--3)} &
\colhead{[\ion{C}{1}]~(1--0)} &
\colhead{CO~(4--3)} &
\colhead{[\ion{C}{1}]~(1--0)} &
\colhead{CO~(1--0)} &
\colhead{CO~(4--3)} &
\colhead{[\ion{C}{1}]~(1--0)} \\
\colhead{} &
\colhead{Jy~km~s$^{-1}$} &
\colhead{Jy~km~s$^{-1}$} &
\colhead{km~s$^{-1}$} &
\colhead{km~s$^{-1}$} &
\colhead{Jy~km~s$^{-1}$}&
\colhead{($\arcsec$,$\arcsec$,$^{\circ}$)}&
\colhead{($\arcsec$,$\arcsec$,$^{\circ}$)}
}
\decimalcolnumbers
\startdata 
                           NGC 232 &                                    905$\pm$272 &                                   461$\pm$138 &                   380 &                   390 &                            -- &                    (4.4,2.7,87) &                    (4.5,2.8,85) \\
               NGC 232-2 (NGC 235) &                                      69$\pm$22 &                                     99$\pm$30 &                   100 &                   500 &                            -- &                   (4.0,3.3,104) &                    (4.3,3.0,92) \\
              NGC 232-3 (PGC 2570) &                                          $<$28 &                                         $<$12 &                    ND &                    ND &                            -- &                              ND &                              ND \\
                           IC 1623 &                                   1655$\pm$501 &                                   632$\pm$191 &                   270 &                   270 &                    230$\pm$24 &                    (9.9,2.9,81) &                   (12.7,2.9,81) \\
                         UGC 02982 &                                             -- &                                    187$\pm$58 &                    -- &                   280 &                    289$\pm$72 &                              -- &                   (7.5,4.0,108) \\
         \shortstack{NGC 1614\\{}} &   \shortstack{{}\\1481$\pm$447\\([1710,2410])} &    \shortstack{{}\\549$\pm$167\\([505,1080])} &  \shortstack{250\\{}} &  \shortstack{250\\{}} &   \shortstack{106$\pm$26\\{}} &   \shortstack{(4.6,4.0,58)\\{}} &   \shortstack{(4.5,4.1,68)\\{}} \\
                  IRAS F05189-2524 &                                     221$\pm$67 &                                     83$\pm$25 &                   150 &                   230 &                    103$\pm$20 &                   (4.8,2.5,106) &                    (3.9,2.2,88) \\
\enddata

\tablecomments{
(1) Target names. 
(2) and (3) the velocity integrated flux. The SPIRE/FTS measurements are shown as ([A,B]), where A and B are the minimum and maximum values within the 3$\sigma$ error. The SPIRE/FTS values were obtained from the study by \citet{Kame_2016}. 
(4) and (5), the FWHM of the spectrum shown in Figure~\ref{fig:mom}.
(6) CO~(1--0) velocity integrated flux density \citep{Kame_2016}. 
(7)  and (8), the source size is base on the FWHM value of the 2D Gaussian fitting by the {\tt imfit} task. The ``*" indicates the manual aperture results (see the Section~\ref{sec:individual}).}
\end{deluxetable*}

\startlongtable
\tabletypesize{\scriptsize}
\begin{deluxetable*}{lcccccc}
\tablecaption{Continuum flux density \label{tab:flux_cont}}
\tablehead{
\colhead{Target Name} &
\colhead{$S_{\rm 650}$} &
\colhead{$S_{\rm 650}$} &
\colhead{$S_{\rm 609}$} &
\colhead{$S_{\rm 609}$} &
\colhead{size (FWHM)} &
\colhead{size (FWHM)} \\
\colhead{} &
\colhead{ACA} &
\colhead{SED} &
\colhead{ACA} &
\colhead{SED} &
\colhead{650$~\mu m$} &
\colhead{609$~\mu m$} \\
\colhead{} &
\colhead{mJy} &
\colhead{mJy} &
\colhead{mJy} &
\colhead{mJy} &
\colhead{($\arcsec$,$\arcsec$,$^{\circ}$)}&
\colhead{($\arcsec$,$\arcsec$,$^{\circ}$)}
}
\decimalcolnumbers
\startdata 
              NGC 232 &      112$\pm$35 &               -- &    153$\pm$47 &               -- &       (4.5,2.7,75) &     (4.5,2.5,78) \\
  NGC 232-2 (NGC 235) &       22$\pm$10 &               -- &     98$\pm$31 &               -- &      (3.6,3.0,155) &     (5.1,3.4,97) \\
 NGC 232-3 (PGC 2570) &         ($<$33) &               -- &       ($<$21) &               -- &                 -- &               -- \\
              IC 1623 &      109$\pm$38 &        [359,384] &    153$\pm$49 &        [447,475] &      (10.7,2.7,79) &    (10.4,2.8,79) \\
            UGC 02982 &              -- &        [340,436] &     99$\pm$38 &        [433,551] &                 -- &    (12.1,2.7,97) \\
             NGC 1614 &      148$\pm$50 &        [272,322] &    188$\pm$63 &        [337,406] &       (4.7,4.1,80) &     (4.3,3.9,52) \\
     IRAS F05189-2524 &       49$\pm$16 &          [71,89] &     49$\pm$16 &         [89,110] &      (4.9,2.3,110) &    (3.5,2.1,102) \\
      IRAS 09022-3615 &       49$\pm$18 &         [91,116] &     88$\pm$29 &        [113,142] &           (7,5,0)* &         (7,5,0)* \\
\enddata

\tablecomments{
(1) Target names. (2)-(5) The continuum flux density associated at 650 and 609~$\mu$m for ACA in this work and SED fitting from infrared data. (6)-(7) The source size was base on the FWHM value of the 2D Gaussian fitting by the {\tt imfit} task. The ``*" means the manual aperture results (see Section~\ref{sec:individual}).}
\end{deluxetable*}

\startlongtable
\tabletypesize{\scriptsize}
\begin{deluxetable*}{lccc}
\tablecaption{Luminosity \label{tab:luminosity}}
\tablehead{
\colhead{Target Name} &
\colhead{$L'_{\rm CO(4-3)}$} &
\colhead{$L'_{\rm [CI](1-0)}$} &
\colhead{$L_{\rm 630\mu m}$} \\
\colhead{} &
\colhead{$10^{6}$~${\rm K~km~s^{-1}~pc^2}$} &
\colhead{$10^{6}$~${\rm K~km~s^{-1}~pc^2}$} &
\colhead{$10^{28}~{\rm erg~s^{-1}~Hz^{-1}}$} 
}
\decimalcolnumbers
\startdata 
              NGC 232 &    1607$\pm$484 &    718$\pm$216 &    166$\pm$52 \\
  NGC 232-2 (NGC 235) &      123$\pm$40 &     154$\pm$48 &     76$\pm$26 \\
 NGC 232-3 (PGC 2570) &         ($<$49) &        ($<$19) &       ($<$34) \\
              IC 1623 &    2272$\pm$688 &    762$\pm$230 &    130$\pm$43 \\
            UGC 02982 &              -- &     172$\pm$53 &     68$\pm$27 \\
             NGC 1614 &    1249$\pm$377 &    406$\pm$124 &    104$\pm$35 \\
     IRAS F05189-2524 &    1579$\pm$481 &    519$\pm$158 &    223$\pm$73 \\
      IRAS 09022-3615 &  12132$\pm$3659 &  3744$\pm$1144 &   586$\pm$203 \\
\enddata

\tablecomments{
(1) Target names. 
(2) and (3) $L'_{\rm line}$ (equation \ref{L2}) for CO~(4--3) and[\ion{C}{1}]~(1--0), respectively.
(4) The specific luminosity in the rest frame $630~\mu$m in equation (\ref{Lcont}).
}
\end{deluxetable*}

\startlongtable
\tabletypesize{\scriptsize}
\tablewidth{1000pt}
\begin{deluxetable*}{lcccc}
\tablecaption{Ratios \label{tab:ratio_appendix}}
\tablehead{
\colhead{Target Name} &
\colhead{$L_{\rm [CI](1-0)}/L_{\rm CO(4-3)}$} &
\colhead{$L_{\rm 630\mu m}/L'_{\rm [CI](1-0)}$} &
\colhead{$L_{\rm 630\mu m}/L'_{\rm CO(4-3)}$} &
\colhead{$L'_{\rm [CI](1-0)}/L'_{\rm CO(1-0)}$} \\
\colhead{}&
\colhead{}&
\colhead{$10^{20}~{\rm erg~s^{-1}~Hz^{-1}}$~$({\rm K~km~s^{-1}~pc^2})^{-1}$}&
\colhead{$10^{20}~{\rm erg~s^{-1}~Hz^{-1}}$~$({\rm K~km~s^{-1}~pc^2})^{-1}$}
}
\decimalcolnumbers
\startdata 
              NGC 232 &    0.54$\pm$0.23 &    23$\pm$10 &       10$\pm$5 &             -- \\
  NGC 232-2 (NGC 235) &    1.51$\pm$0.69 &    49$\pm$23 &      61$\pm$30 &             -- \\
 NGC 232-3 (PGC 2570) &               ND &           ND &             ND &             -- \\
              IC 1623 &    0.41$\pm$0.17 &     17$\pm$8 &        6$\pm$3 &  0.15$\pm$0.05 \\
            UGC 02982 &               -- &    40$\pm$20 &              - &  0.04$\pm$0.01 \\
             NGC 1614 &     0.4$\pm$0.17 &    26$\pm$12 &        8$\pm$4 &  0.28$\pm$0.11 \\
     IRAS F05189-2524 &     0.4$\pm$0.17 &    43$\pm$19 &       14$\pm$6 &  0.04$\pm$0.02 \\
      IRAS 09022-3615 &    0.38$\pm$0.16 &     16$\pm$7 &        5$\pm$2 &             -- \\
\enddata

\tablecomments{
(1) Target names. (2)-(5) Ratio between luminosity. }
\end{deluxetable*}

\startlongtable
\tabletypesize{\scriptsize}
\begin{deluxetable*}{lcccc}
\tablecaption{Molecular gas mass \label{tab:Mgas}}
\tablehead{
\colhead{Target Name} &
\colhead{$M_{\rm H2}$} &
\colhead{$M_{\rm H2}$} &
\colhead{$M_{\rm H2}$} &
\colhead{$M_{\rm H2}$} \\
\colhead{} &
\colhead{[\ion{C}{1}]~(1--0)} &
\colhead{cont} &
\colhead{CO(4--3)} &
\colhead{CO(1--0)} \\
\colhead{} &
\colhead{$10^7M_{\odot}$} &
\colhead{$10^7M_{\odot}$} &
\colhead{$10^7M_{\odot}$} &
\colhead{$10^7M_{\odot}$} 
}
\decimalcolnumbers
\startdata 
              NGC 232 &   464$\pm$139 &    670$\pm$209 &   435$\pm$131 &            -- \\
  NGC 232-2 (NGC 235) &     99$\pm$31 &    306$\pm$108 &     33$\pm$10 &            -- \\
 NGC 232-3 (PGC 2570) &       ($<$12) &       ($<$137) &       ($<$13) &            -- \\
              IC 1623 &   492$\pm$148 &    523$\pm$174 &   615$\pm$186 &    404$\pm$42 \\
            UGC 02982 &    111$\pm$34 &    276$\pm$108 &            -- &    388$\pm$97 \\
             NGC 1614 &    262$\pm$80 &    419$\pm$142 &   338$\pm$102 &    114$\pm$28 \\
     IRAS F05189-2524 &   335$\pm$102 &    898$\pm$296 &   427$\pm$130 &   938$\pm$187 \\
      IRAS 09022-3615 &  2419$\pm$739 &   2363$\pm$818 &  3284$\pm$990 &            -- \\
\enddata

\tablecomments{
(1) Target names. (2)-(4) Molecular gas mass inferred from [\ion{C}{1}]~(1--0), dust continuum, CO~(4--3), and CO~(1--0), respectively (See Section~\ref{appendix:Mgas}).
}
\end{deluxetable*} 

\clearpage
\subsection{Notes for individual galaxies} \label{sec:individual}
\begin{description}
   \item[NGC~232, NGC~235, and PGC~2570] NGC~232 interacts with NGC~235, which is located at $\sim53$~kpc northeast, with a velocity offset of approximately $-123$~km\,s$^{-1}$ \citep{Larson_2016, Espada_2018}.
 In addition, NGC~235 had a minor interaction pair (PGC~2570). In NGC~232 and NGC~235, CO~(4--3), [\ion{C}{1}]~(1--0), and dust continuum were detected. In NGC~235, CO~(4--3) emission was detected only at the blue-shifted components (approximately $-250$~km\,s$^{-1}$ with a systematic velocity of  approximately $-123$~km\,s$^{-1}$). Conversely, the [\ion{C}{1}]~(1--0) emission has a double peak in both the red- and blue-shifted components.
   \item[IC~1623 (VV~114)] 
The FoV centers do not match between CO~(4--3) and [\ion{C}{1}]~(1--0) observations due to artificial mistakes when preparing the observation plans.
   \item[UGC~2982] CO~(4--3) was not fully observed owing to the ALMA schedule.
   \item[IRAS~F05189-2524] [\ion{C}{1}]~(1--0) emission was at the edge of the spectral window.
   \item[IRAS~09022-3615]  Because the continuum maps are noisy, the ``${\tt imfit}$" task does not work correctly. We manually chose a flux aperture of $7\arcsec\times5\arcsec$ for the continuum photometries.
   \item[IRAS~F09111-1007] [\ion{C}{1}]~(1--0) emission was at the edge of the spectral window.
   \item[NGC~3110] The 490GHz continuum emission associated with [\ion{C}{1}]~(1--0) was not detected due to poor observation conditions but a 460~GHz continuum associated with CO(4-3) was detected. This is due to the sky conditions during the observation.
   \item[IRAS F10565+2448] IRAS F10565+2448 consists of two colliding galaxies. In the main large galaxy,  CO~(4--3), [\ion{C}{1}]~(1--0), and dust continuum were detected. In the smaller pair galaxy (IRAS~F10565+2448-2), emission were not detected. [\ion{C}{1}]~(1--0) emission was observed at the edge of the spectral window.
   \item[NGC~4418]  The red-shifted emission associated with  the [\ion{C}{1}]~(1--0) emission line is probably fake because of the bad atmospheric window.
   \item[IC~4280] Both CO~(4--3) and [\ion{C}{1}]~(1--0) lines were detected. The continuum associated with [\ion{C}{1}]~(1--0) observation is detected, and the ``${\tt imfit}$" shows that spatially extended structure (the FWHM of $8\farcs3\times5\farcs4$) with the continuum flux density of $152\pm57$~mJy. Because the extended structure is marginal (S/N$\sim$3), this flux density may be the upper limit of the continuum flux density. However, the continuum emission associated with CO~(4--3) emission was not detected, and the upper limit was $<33$~mJy. The reason for the gap between 650$\mu$m and 609$\mu$m flux density by ACA is not clear; it is probably due to bad transmissivity around continuum emission. 
For this galaxy, the missing flux is difficult to determine.  For example, 
$S_{\rm CO(4-3)}\Delta v=361\pm74$~Jy\,km\,s$^{-1}$ 
for ACA and
 $S_{\rm CO(4-3)}\Delta v=$[531,1010]~Jy\,km\,s$^{-1}$ for SPIRE and 
 $S_{\rm [CI](1-0)}\Delta v=166\pm34$~Jy\,km\,s$^{-1}$ for ACA and 
 $S_{\rm [CI](1-0)}\Delta v=$[69,406]~Jy\,km\,s$^{-1}$ for SPIRE.
 This means that the line missions do not have a significant missing flux.
 However, in the case of continuum emission, 
 $S_{650}$=$152\pm57$~mJy for ACA but $S_{650}$=[514, 566]~mJy from SED fitting, suggesting that RF is approximately $30\%$. In the case of $S_{609}$, the situation is worsened.
This means that only continuum emission has a large missing flux in this system, probably suggesting a different spatial structure between gas and dust. However, we do not know the reason for the large difference in RF for line and continuum emission. 
   \item[IRAS~F14378-3651] Both CO~(4--3) and [\ion{C}{1}]~(1--0) lines were detected. 
    Because the continuum maps associated with CO~(4--3) emission are noisy, the ``${\tt imfit}$" task does not work correctly. We manually chose a flux aperture of $6\arcsec\times6\arcsec$ for the photometry.
   \item[CGCG~049-057]  The spurious emission is observed at the edge of the [\ion{C}{1}]~(1--0) spectral windows but it does not affect the measurements.
   \item[NGC~5936] The spectral of the frequency $>486$~GHz is significantly affected by the bad sky but it does not affect the [\ion{C}{1}]~(1--0) measurements. 
   \item[Arp 220] Owing to the strong emission line, the {\tt tclean} process cannot remove all the sidelobe components, and some components are seen in the residual maps. Therefore, moments 1 and 2 maps have uncertain structures. However, the contribution of these structures to the total integrated intensity was negligible.
   \item[NGC~5990] The [\ion{C}{1}]~(1--0) observation has issues in calibration, and the data are declared as ``QA2 semipass".  Because CO~(4--3) emission is spatially extended, single 2D Gaussian fitting by the ``${\tt imfit}$" underestimates the flux. Therefore, we used the $20\arcsec$ aperture to measure the CO~(4--3) continuum.
   \item[NGC~6052] While CO~(4--3) was clearly detected, [\ion{C}{1}]~(1--0) was not detected. The details are discussed in \citet{Michiyama_2020}.
   \item[ESO 069-IG006] ESO 069-IG00 consists of two colliding galaxies. In the main galaxy,  CO~(4--3), [\ion{C}{1}]~(1--0), and dust continuum were detected. In the smaller pair galaxy (ESO 069-IG00-2), emission were not detected. [\ion{C}{1}]~(1--0) emission was observed at the edge of the spectral window. 
   \item[IRAS~F17207-0014] [\ion{C}{1}]~(1--0) emission was observed at the edge of the spectral window.
   \item[IRAS~19542+1110] Both CO~(4--3) and [\ion{C}{1}]~(1--0) observations were done, but [\ion{C}{1}]~(1--0) emission is outside of the spectral window. The CO~(4--3) and dust continuum peak consists with optical peak.
   \item[NGC~7130] An uncertain structure is seen in the souther part of the [\ion{C}{1}]~(1--0) emission. However, we use flux derived from the ``${\tt imfit}$" which ignores the southern emission. 
   \item[ESO 467-G027] CO~(4--3) was robustly detected but the red-shifted components may have been outside of the spectral window. The [\ion{C}{1}]~(1--0) was marginally detected in the nuclear region. However, the measured flux is less than the upper limits, and we assume [\ion{C}{1}]~(1--0) to be non-detection. This galaxy is a candidate of C$^0$-poor/CO-rich galaxy but the data quality is not very good; the upper limit of  $L'_{[CI](1-0)}$/$L'_{CO(4--3)}$ is $> 0.1$. 
   \item[IRAS~F22491-1808] In CO~(4--3) observation, the spectral window does not fully cover the entire line emission. While the S/N of the spectrum spatially integrated within the FoV is not robust, a smaller aperture can boost the S/N. 
   \item[NGC~7469] No comments regarding the ACA  observations.
   \item[ESO~148-IG002] A double peak was observed in both CO~(4--3) and [\ion{C}{1}]~(1--0) emission. The uncertain north-south structure is observed in the continuum emission. This could be due to the calibration issues, but the clear reason is unknown. Further sensitive observations are required. To avoid the uncertainty, we use 20\arcsec$\times$20\arcsec aperture to measure continuum flux density.
   \item[NGC~7679] While CO~(4--3) was clearly detected, [\ion{C}{1}]~(1--0) was not detected. 
\end{description}

\subsection{Molecular gas mass}\label{appendix:Mgas}
\begin{itemize}
\item
We use the CO~(1--0) to H$_2$ conversion factor of $\alpha_{\rm CO}=0.8$ $M_{\odot}({\rm K~km~s^{-1}~pc^2})^{-1}$, which is typically applied for U/LIRGs \citep{Bolatto_2013}.
\begin{eqnarray}
&&M_{\rm H_2}^{\rm CO(1-0)} = \alpha_{\rm CO} L'_{\rm CO(1-0)} \nonumber\\
&&\alpha_{\rm CO}=0.8
\end{eqnarray}
\item
\citet{Papa_2004_ULIRG, Bothwell_2017} provided an equation to calculate the total molecular gas mass from  [\ion{C}{1}]~(1--0) luminosity:
\begin{eqnarray}
&&M_{\rm H_2}^{\rm [CI](1-0)} = \alpha_{\rm [CI]} L'_{\rm [CI](1-0)} \nonumber\\
&&\alpha_{\rm [CI]} = 6.5\left(\frac{R_{\mathrm{CI}}}{5 \times 10^{-5}}\right)^{-1}\nonumber\\
&&\times\left(\frac{A_{10}}{7.93 \times 10^{-8} \mathrm{s}^{-1}}\right)^{-1}\left(\frac{Q_{10}}{0.4}\right)^{-1},
\end{eqnarray}\label{eq:alpha_CI}
where $R_{\rm CI}$ is the C$^0$/H$_2$ abundance ratio, $A_{10}$ is the Einstein A coefficient, and $Q_{10}$ is the excitation factor.
\citet{Jiao_2017} showed that $Q_{10}$ varies from $0.35-0.45$, assuming that the molecular gas density is $10^3-10^4$~cm$^{-3}$ and the kinematic temperature is 20$-$40~K.
\citet{Papa_2004_ULIRG} considered that the typical $R_{\rm CI}$ is $3\times10^{-5}$, but a relatively higher value ($5\times10^{-5}$) was considered in starburst galaxy M82 \citep{Bothwell_2017}.
In addition, a relatively smaller value ($1.6-1.9\times10^{-5}$) was reported for main-sequence galaxies at z=1--2 \citep{Valentino_2018}.
We use $\alpha_{\rm [CI]} = 6.5$ $M_{\odot}({\rm K~km~s^{-1}~pc^2})^{-1}$, assuming that $Q_{10}=0.4$, $R_{\rm [CI]}=5\times10^{-5}$, and $A_{10}=7.93 \times 10^{-8}$. 
We note that the determination of $\alpha_{\rm [CI]}$ is uncertain because of the unknown $R_{\rm CI}$, $Q_{10}$, and $A_{10}$. 
For example, \citet{Crocker_2019} modeled the [\ion{C}{1}] and CO line emission using large-velocity gradient models and demonstrated that $\alpha_{\rm [CI]}=7.3$ $M_{\odot}({\rm K~km~s^{-1}~pc^2})^{-1}$ and  $\alpha_{\rm CO}=0.9$ $M_{\odot}({\rm K~km~s^{-1}~pc^2})^{-1}$ using the Herschel SPIRE results.
In the case of \citet{Israel_2020}, the [\ion{C}{1}] to H$_2$ conversion factor at the central molecular zone in galaxies is calculated as $X$([\ion{C}{1}]) = $(9\pm2)\times10^{19}$~cm$^{-2}$/K~km~s$^{-1}$, which corresponds to $\alpha_{\rm [CI]}\sim2$ $M_{\odot}({\rm K~km~s^{-1}~pc^2})^{-1}$ (they also calculate CO~(1--0) to H$_2$ conversion factor $X$(CO) = $(1.9\pm0.2)\times10^{19}$~cm$^{-2}$/K~km~s$^{-1}$, which is a factor of ten below the standard solar neighborhood Milky Way factor of $\sim2\times10^{20}$~cm$^{-2}$/K~km~s$^{-1}$).
Conversely, a relatively higher value of $\alpha_{\rm [CI]}\sim18.8$ $M_{\odot}({\rm K~km~s^{-1}~pc^2})^{-1}$ and $\alpha_{\rm CO}\sim3.0$ $M_{\odot}({\rm K~km~s^{-1}~pc^2})^{-1}$ was calculated by \citet{Dunne_2021}. We note that determining the conversion factor is outside the scope of this study. 
We fixed the conversion factor and attemted to understand the scatter among each method.

\item
 \citet{Scoville_2016} derived an empirical mass to luminosity ratio between 850~$\mu$m specific luminosity to molecular gas mass ($L_{\nu({\rm 850\mu m})}/M_{\rm mol} = $ 6.7e+19 $M_{\odot}^{-1}\left(\rm{erg}~\rm{s}^{-1}~~\rm{Hz}^{-1}\right) = \alpha_{\rm 850\mu m}$, where $M_{\rm mol}=1.36~M_{\rm H2}$) based on the comparison between CO~(1--0) single-dish observations, SPIRE photometry, and SCUBA 850 $\mu$m photometry. 
We note that this $L_{\nu({\rm 850\mu m})}/M_{\rm mol}$ was experimentally estimated by assuming a single standard Galactic $\alpha_{\rm CO}$.
We convert our observed continuum flux density $S_{\nu}$ into $M_{\rm H2}$ using the formulation of \citet{Scoville_2016}
\begin{eqnarray}\label{equ:MH2_cont}
M_{\rm H_2}^{\rm cont} &=& 1.78 S_{\nu} (1+z)^{-4.8} \left( \frac{\nu_{850 {\rm \mu m}}}{\nu_{\rm obs}} \right)^{3.8} \\ \nonumber
&& \times \left( \frac{6.7\times10^{19}}{\alpha_{850}} \right)
\frac{\Gamma_{\rm RJ}(T_{\rm dust},\nu_{850 {\rm \mu m}},0)}{\Gamma_{\rm RJ}(T_{\rm dust},\nu_{\rm obs},z)} \\ \nonumber
 && \times~D_{\rm L}^2~ \times \frac{M_{\rm H2}}{M_{\rm mol}}~[10^{10} M_{\odot}].
\end{eqnarray}
We note that the unit of $S_{\nu}$ is mJy and the unit of $D_{\rm L}$ is Gpc in this equation.
We assume the dust temperature of $T_{\rm dust}=25$~K for simple analyses.
 
\item
From CO~(4--3), we use
\begin{eqnarray}
&& M_{\rm H_2}^{\rm CO(4-3)}  = \alpha_{\rm CO} 
L'_{\rm CO(4-3)}
~/ 
\left(\frac{L'_{\rm CO(4-3)} }{L'_{\rm CO~(1-0)}} \right)_{\rm med}\nonumber\\
&& \left(\frac{L'_{\rm CO(4-3)}}{L'_{\rm CO(1-0)} } \right)_{\rm med} = 0.3
\end{eqnarray}
where $\left(\frac{L'_{\rm CO(4-3)} }{L'_{\rm CO(1-0)} } \right)_{\rm med}$ is the median value of $L'_{\rm CO(4-3)}/L'_{\rm CO(1-0)}$. 
However, the ratio varies for each galaxy population.
For example, $L'_{\rm CO(4-3)}/L'_{\rm CO(1-0)}$ is $\sim0.1$ in low excitation galaxies like the Milky Way \citep{Fixsen_1999}, $\sim0.8$ in starbursts like M82 \citep{Weiss_2005}, and $\sim0.8$ in some high-z SMGs  \citep{Carilli_2013, Casey_2014}.
\end{itemize}


\end{document}